\documentclass[a4paper]{article}

\newcommand{\myabstract}{{A generic search for anomalous production of events with at least three charged leptons is presented.  The data sample consists of $pp$ collisions at $\sqrt{s}=8~\textrm{Te\kern -0.1em V}$ collected in 2012 by the ATLAS experiment at the CERN Large Hadron Collider, and corresponds to an integrated luminosity of 20.3~fb$^{-1}$.  Events are required to have at least three selected lepton candidates, at least two of which must be electrons or muons, while the third may be a hadronically decaying tau.  Selected events are categorized based on their lepton flavour content and signal regions are constructed using several kinematic variables of interest.  No significant deviations from Standard Model predictions are observed.  Model-independent upper limits on contributions from beyond the Standard Model phenomena are provided for each signal region, along with prescription to re-interpret the limits for any model.  Constraints are also placed on models predicting doubly charged Higgs bosons and excited leptons.  For doubly charged Higgs bosons decaying to $e\tau$ or $\mu\tau$, lower limits on the mass are set at $400~\textrm{Ge\kern -0.1em V}$ at 95\% confidence level.  For excited leptons, constraints are provided as functions of both the mass of the excited state and the compositeness scale $\Lambda$, with the strongest mass constraints arising in regions where the mass equals $\Lambda$.  In such scenarios, lower mass limits are set at $3.0~\textrm{Te\kern -0.1em V}$ for excited electrons and muons, $2.5~\textrm{Te\kern -0.1em V}$ for excited taus, and $1.6~\textrm{Te\kern -0.1em V}$ for every excited-neutrino flavour.}}
\newcommand{\mytitle}{Search for new phenomena in events with three or more charged leptons in $pp$ collisions at $\sqrt{s}=8~\textrm{Te\kern -0.1em V}$ with the ATLAS detector}

\usepackage{preprintcover}  
\PreprintCoverPaperTitle{\mytitle}  
\PreprintIdNumber{CERN-PH-EP-2014-255}  
\PreprintCoverAbstract{\myabstract}
\PreprintJournalName{JHEP}  

\usepackage{jheppub}
\usepackage{graphicx} 
\graphicspath{{figures/}}
\usepackage{atlasphysics} 

\usepackage{rotating} 
\usepackage{multirow}
\usepackage{subfigure}
\usepackage{afterpage}
\usepackage{rotating}
\usepackage{afterpage}
\usepackage{placeins}

\newcommand{\alpgen}{{\sc alpgen}}
\newcommand{\madgraph}{{\sc MadGraph}}
\newcommand{\jimmy}{{\sc jimmy}}

\newcommand{\sherpa}{{\sc Sherpa}}

\newcommand{\powheg}{{\sc powheg-box}}
\newcommand{\herwig}{{\sc herwig}}
\newcommand{\pythia}{{\sc Pythia}}
\newcommand{\vbfnlo}{{\sc vbfnlo}}
\newcommand{\loopsim}{{\sc LoopSim}}
\newcommand{\nNLO}{\ensuremath{\mathrm{\bar{n}NLO}}}

\newcommand{\geant}{{\sc geant4}}

\newcommand{\antikt}{anti-$k_{t}$}

\newcommand{\taulep}{\ensuremath{\tau_\mathrm{lep}}}
\newcommand{\tauhad}{\ensuremath{\tau_\mathrm{had}}}

\newcommand{\tauh}    {\tauhad}

\newcommand{\threeL}  {\ensuremath{\geq3 e/\mu}}
\newcommand{\twoLtau} {\ensuremath{2 e/\mu + \geq1 \tauh}}
\newcommand{\twoLoneT}{\twoLtau}

\newcommand{\intLdt}  {\int L\mathrm{d}t}

\newcommand{\ptvis}   {\ensuremath{p_{\mathrm{T}}^{\mathrm{vis}}}}
\newcommand{\etavis}  {\ensuremath{\eta^{\mathrm{vis}}}}

\newcommand{\Ht}      {\ensuremath{H_{\mathrm{T}}^{\mathrm{jets}}}}
\newcommand{\Htjets}  {\Ht}
\newcommand{\htjets}  {\Ht}
\newcommand{\St}      {\meff}
\newcommand{\st}      {\St}
\newcommand{\htlep}   {\ensuremath{H_{\mathrm{T}}^{\mathrm{leptons}}}}
\newcommand{\Htlep}   {\htlep}

\newcommand{\mtw}     {\ensuremath{m_{\mathrm{T}}^W}}

\newcommand{\minleppt}{\ensuremath{p_{\mathrm{T}}^{\ell,\mathrm{min}}}}

\newcommand{\Etisocal}   {\ensuremath{E_{\mathrm{T,cal}}^{\mathrm{iso}}}}
\newcommand{\ptisotrack}   {\ensuremath{p_{\mathrm{T,track}}^{\mathrm{iso}}}}

\newcommand{\Etisotruth}   {\ensuremath{E_{\mathrm{T,true}}^{\mathrm{iso}}}}
\newcommand{\ptisotruth}   {\ensuremath{p_{\mathrm{T,true}}^{\mathrm{iso}}}}

\newcommand{\meff}{\ensuremath{m_{\mathrm{eff}}}}

\newcommand{\dchp}{\ensuremath{H^{\pm\pm}}}
\newcommand{\dchpl}{\ensuremath{H^{\pm\pm}_{\mathrm{L}}}}
\newcommand{\dchpr}{\ensuremath{H^{\pm\pm}_{\mathrm{R}}}}

\newcommand{\exlepmass}{\ensuremath{m_{\ell^*}}}

\newcommand{\epsfid}{\ensuremath{\epsilon_{\mathrm{fid}}}}

\newcommand{\sigmavis}{\ensuremath{\sigma_{\mathrm{95}}^{\mathrm{vis}}}}

\mathchardef\mhyphen="2D

\begin{document}

\title{\mytitle} 
\author{The ATLAS Collaboration}
\date{\today}
\abstract{\myabstract}


\maketitle

\section{Introduction}
\label{sec:Introduction}
With the delivery and exploitation of over 20~\ifb\ of integrated luminosity at a centre-of-mass energy of 8~\TeV{}
in proton--proton collisions at the CERN Large Hadron Collider, many models of new physics now face significant
constraints on their allowed parameter space.  Final states including three or more charged, prompt, and isolated leptons
have received significant attention, both in measurements of Standard Model (SM) diboson~\cite{ATLAS_WZ,ATLAS_ZZ,CMS_ZZ} and Higgs boson production~\cite{ATLAS_Higgs,CMS_Higgs}, and
in searches for new phenomena. Anomalous production of multi-lepton final states arises in many beyond the Standard Model
(BSM) scenarios, including excited-lepton models~\cite{excitednu,excitednu2}, the Zee--Babu neutrino mass model~\cite{zee1,zee2,babu},
supersymmetry~\cite{Miyazawa:1966,Ramond:1971gb,Golfand:1971iw,Neveu:1971rx,Neveu:1971iv,Gervais:1971ji,Volkov:1973ix,Wess:1973kz,Wess:1974tw},
models with pair production of vector-like quarks~\cite{VLQ}, 
and models with doubly charged Higgs bosons~\cite{DCH0,DCH1} including Higgs triplet models~\cite{DCH3,DCH2}.  
An absence of significant deviations from SM predictions in previous measurements and dedicated searches motivates 
an inclusive search strategy, sensitive to a variety of production modes and kinematic features.

In this paper, the results of a search for the anomalous production of events with at least three charged
leptons are presented.  The dataset used was collected in 2012 by the ATLAS detector at the Large Hadron Collider,
and corresponds to an integrated luminosity of 20.3~\ifb\ of $pp$ collisions at $\sqrt{s} = 8~\TeV$.
Events with at least three leptons are categorized using their flavour content, and signal regions are constructed
using several kinematic variables, to cover a wide range of different BSM scenarios.
Inspection of the signal regions reveals no significant deviations from the expected background, and model-independent upper limits on
contributions from BSM sources are evaluated.  A prescription for confronting other models with these results is also provided, along with per-lepton efficiencies
parameterized by lepton flavour and kinematics.

The model-independent limits are also used to provide constraints on two benchmark models.  The first model predicts the Drell--Yan production of
doubly charged Higgs bosons~\cite{DCH0,DCH1}, which then decay into lepton pairs.  The decays can include flavour-violating terms that can 
lead to final states such as $\ell^{\pm}\tau^{\pm}\ell^{\mp}\tau^{\mp}$, where $\ell$ denotes an electron or muon, and the tau lepton is allowed to 
decay hadronically or leptonically.  Lepton-flavor-conserving decays are not considered in this paper.  The second benchmark scenario is a composite fermion model predicting the existence of excited leptons~\cite{ExcitedLeptons}.
The excited leptons, which may be neutral ($\nu^{*}$) or charged ($\ell^{*}$), are produced in a pair or in association with a SM lepton either through contact interactions or gauge-mediated processes.
Their decay proceeds via the same mechanisms, with rates that depend on the lepton mass and a compositeness scale,~$\Lambda$.  The final states of such events often contain
three or more charged leptons with large momentum.

Related searches for new phenomena in events with multi-lepton final states have not shown any
significant deviation from SM expectations.
The CMS Collaboration has conducted a search similar to the one presented here using 5~\ifb{} of 7~\TeV\ data~\cite{cmsML7TeV} and also with 19.5~\ifb\ 
of 8~\TeV\ data~\cite{cmsML}.  
The ATLAS Collaboration has performed searches for supersymmetry
in multi-lepton final states~\cite{atlasSUSYML,atlasSUSY2L,atlasSUSY4L}, as have experiments at 
the Tevatron~\cite{cdfML,dzeroML}.  
The search presented here complements the previous searches
by providing model-independent limits and by exploring new kinematic variables.  Compared to
a similar analysis presented in ref.~\cite{atlasML2011} using 7~\TeV{} data, this search tightens the lepton 
requirements on the momentum transverse to the beamline (\pt) from 
10(15) \GeV{} to 15(20) \GeV{} for electrons and muons (hadronically decaying taus), includes
new signal regions to target models producing heavy-flavour signatures and events without $Z$ bosons, and 
tightens the requirements for previously defined signal regions to exploit the higher centre-of-mass energy and
integrated luminosity of the 2012 data sample.

\section{The ATLAS detector}
\label{sec:Detector}
The ATLAS detector~\cite{atlas} at the LHC covers nearly the entire solid angle around the collision point.\footnote{ATLAS uses a right-handed coordinate system with its
origin at the nominal interaction point (IP) in the centre of the detector and the $z$-axis along the beam pipe. The $x$-axis points from the IP to the centre of the LHC
ring, and the $y$-axis points upward. Cylindrical coordinates $(r,\phi)$ are used in the transverse plane, $\phi$ being the azimuthal angle around the beam pipe. The
pseudorapidity is defined in terms of the polar angle $\theta$ as $\eta=-\ln\tan(\theta/2)$.} It consists of an inner tracking detector surrounded by a thin superconducting
solenoid, electromagnetic and hadronic calorimeters, and a muon spectrometer incorporating three large superconducting toroid magnets with eight coils each.

The inner-detector system is immersed in a 2~T axial magnetic field and provides charged-particle tracking in the range $|\eta| < 2.5$.
A high-granularity silicon pixel detector covers the vertex region and typically provides three measurements per track, with one hit being usually registered in the innermost layer. It is followed by a silicon microstrip tracker, which usually provides four two-dimensional measurement points per track. These silicon detectors are complemented by a transition radiation tracker, which enables radially extended track reconstruction up to $|\eta| = 2.0$. The transition radiation tracker also provides electron identification information based on the fraction of hits (typically 30 in total) above a higher energy threshold corresponding to transition radiation.

The calorimeter system covers the pseudorapidity range $|\eta|< 4.9$. Within the region $|\eta|< 3.2$, electromagnetic calorimetry is provided by barrel and endcap high-granularity lead/liquid-argon (LAr) electromagnetic calorimeters, with an additional thin LAr presampler covering $|\eta| < 1.8$, to correct for energy loss in material upstream of the calorimeters. Hadronic calorimetry is provided by a steel/scintillator-tile calorimeter, segmented into three barrel structures within $|\eta| < 1.7$, and two copper/LAr hadronic endcap calorimeters. The solid angle coverage is completed with forward copper/LAr and tungsten/LAr calorimeter modules optimized for electromagnetic and hadronic measurements respectively.

The muon spectrometer comprises separate trigger and high-precision tracking chambers measuring the deflection of muons in a magnetic field generated by superconducting air-core toroids. The precision chamber system covers the region $|\eta| < 2.7$ with three layers of monitored drift tubes, complemented by cathode strip chambers in the forward region, where the background is highest. The muon trigger system covers the range $|\eta| < 2.4$ with resistive plate chambers in the barrel, and thin gap chambers in the endcap regions.

A three-level trigger system is used to select interesting events\ \cite{trigger}. The Level-1 trigger is implemented in hardware and uses a subset of detector information to reduce the event rate to a design value of at most 75~kHz. This is followed by two software-based trigger levels which together reduce the event rate to about 400~Hz.

\section{Event selection}
\label{sec:Selection}
Events are required to have fired either a single-electron or single-muon trigger.
The electron and muon triggers impose a \pt{} threshold of 24~\GeV\ along with isolation requirements on the lepton. 
To recover efficiency for higher \pt\ leptons, the isolated lepton triggers are complemented
by triggers without isolation requirements but with a higher \pt\ threshold of 60 (36) GeV for electrons (muons).
In order to ensure that the trigger has constant efficiency as a function of lepton \pt{}, the offline event selection requires 
at least one lepton (electron or muon) with $\pt > 26\GeV$
consistent with having fired the relevant single-lepton trigger.
A muon associated with the trigger must lie within $|\eta|<2.4$, while 
a triggered electron must lie within $|\eta|<2.47$, excluding the calorimeter barrel/endcap transition region ($1.37 \leq |\eta| < 1.52$).
Additional muons in the event must lie within $|\eta|<2.5$ and have $\pt > 15$~\GeV.  
Additional electrons must satisfy the same $\eta$ requirements as triggered electrons and have
$\pt > 15$~\GeV.
The third lepton in the event may be an additional electron or muon satisfying the same requirements
as the second lepton, or a hadronically decaying tau (\tauh) with 
$\ptvis > 20$~\GeV\ and $|\etavis| < 2.5$, where $\ptvis$ and $\etavis$ denote the 
\pt\ and $\eta$ of the visible products of the tau decay, with no corrections for
the momentum carried by neutrinos.  Throughout this paper, the four-momenta of tau candidates are defined
only by the visible decay products.

Events must have a reconstructed primary vertex with at
least three associated tracks with $\pt>0.4\GeV$.  
In events with multiple primary vertex candidates, the primary vertex is 
chosen to be the one  with the highest $\Sigma{p_{\mathrm{T}}^{2}}$, where the 
sum is over all reconstructed tracks associated with the vertex.  Events
with pairs of leptons that are of the same flavour but opposite sign and have an invariant
mass below 15~\GeV\ are excluded to avoid backgrounds from low-mass resonances.

The lepton selection includes requirements to reduce the contributions
from non-prompt or fake leptons.  These requirements exploit 
the transverse and longitudinal impact parameters of the tracks with respect to the primary
vertex, the isolation of the lepton candidates from nearby hadronic activity, and
in the case of electron and \tauh\ candidates, the lateral and longitudinal profiles
of the shower in the electromagnetic calorimeter.    These requirements 
are described in more detail below.  There are also requirements for electrons on the quality
of the reconstructed track and its match to the cluster in the calorimeter.

Electron candidates are required to satisfy the ``tight'' identification criteria described in ref.~\cite{elecperf}, updated
for the increased number of multiple interactions per bunch crossing (pileup)
in the 2012 dataset.  The tight criteria include requirements on the track properties and shower development of the electron candidate.  
Muons must have tracks with hits in both the inner tracking detector and muon spectrometer, and must
satisfy criteria on track quality described in ref.~\cite{muonperf}.

The transverse impact parameter significance is defined as $|d_{0}/\sigma(d_{0})|$, where 
$d_0$ is the transverse impact parameter of the reconstructed track with respect to the primary vertex and
$\sigma(d_0)$ is the estimated uncertainty on $d_0$.  This quantity must be less than 3.0 
for both the electron and muon candidates.
The longitudinal impact parameter $z_{0}$ must satisfy $|z_{0} \sin(\theta)|<0.5$~mm for both the electrons
and muons.

Electrons and muons are required to be isolated through the use of two variables sensitive to
the amount of nearby hadronic activity.  The first, \ptisotrack, is the 
scalar sum of the transverse momenta of all tracks with $\pt > 1$~\GeV\ in a cone
of $\Delta R=\sqrt{(\Delta\eta)^2 + (\Delta\phi)^2}=0.3$ around the lepton axis.  The sum excludes the track associated with the lepton candidate, 
and also excludes tracks inconsistent with originating from the primary vertex.  
The second, \Etisocal, is the sum of the transverse energy of cells in the electromagnetic
and hadronic calorimeters in a cone of size $\Delta R=0.3$ around the lepton axis.  For electron candidates, this sum excludes a rectangular 
region around the candidate axis 
of $0.125\times 0.172$ in $\eta\times\phi$ (corresponding to $5\times7$ cells in the main sampling
layer of the electromagnetic calorimeter) and is 
corrected for the incomplete containment of the electron transverse energy within the excluded region.  
For muons, the sum only includes cells above a certain threshold in order to suppress noise, and does not include
cells with energy deposits from the muon candidate.  For both the electrons and muons,
the value of \Etisocal\ is corrected for the expected effects of pileup
interactions. Electron and muon candidates are required to have $\ptisotrack/\pt < 0.1$ and
$\Etisocal/\pt < 0.1$. The isolation requirements are tightened for leptons with $\pt > 100 \gev$, 
which must satisfy $\ptisotrack < (10 \gev + 0.01\times\pt$~[\gev]) and $\Etisocal< (10 \gev + 0.01\times\pt$~[\gev]).  The tighter cut for high-\pt\ leptons 
reduces non-prompt backgrounds to negligible levels.

Jets are used as a measure of the hadronic activity within the event as well as seeds for reconstructing \tauh\ candidates.  
Jets are reconstructed using the 
\antikt\ algorithm~\cite{antikt}, with radius parameter $R=0.4$.
The jet four-momenta are corrected
for the non-compensating nature of the calorimeter, for inactive material in front of the
calorimeters, and for pileup~\cite{jetcorr,JER}.
Jets used in this analysis are required to have $\pt > 30$~\GeV\ and lie within $|\eta|<4.9$.  
Jets within the acceptance of the inner tracking detector must fulfil a requirement, based on tracking information, that they
originate from the primary vertex. Jets containing $b$-hadrons are identified using a multivariate technique~\cite{ATLAS-CONF-2014-046} based on quantities
such as the impact parameters of the tracks associated with the jet. The working point of the identification algorithm used in this analysis 
has an efficiency for tagging $b$-jets of 80\%, with corresponding rejection factors of approximately
30 for light-jets and 3 for charm-jets, as determined for jets with $\pt > 20$ GeV within the inner tracker's acceptance in simulated $t\bar{t}$ events.

Tau leptons decaying to an electron (muon) and neutrinos are selected with the electron (muon) identification criteria described above, and are 
classified as electrons (muons).
Hadronically decaying tau candidates are seeded by reconstructed jets and are selected using an identification algorithm based on a
boosted decision tree (BDT)
trained to distinguish hadronically decaying tau leptons from quark-  and gluon-initiated jets~\cite{tauid}.
The BDT uses track and calorimeter quantities associated with the tau candidate, including the properties of nearby tracks
and the shower development in the calorimeter.
It is trained separately for tau candidates with one and three charged decay products, referred to as ``one-prong'' and ``three-prong'' taus, respectively.
In this analysis, only one-prong $\tauh$ candidates satisfying the criteria for the ``tight'' working point~\cite{tauid} are considered.
This working point is roughly 40\% efficient for one-prong $\tauh$ candidates originating 
from $W$ or $Z$ boson decays, and has a jet rejection factor of roughly 300 in multi-jet topologies.
Additional requirements to remove \tauh\ candidates initiated by prompt electrons or muons are also imposed.  

To further ensure the prompt nature of our lepton candidates, and to resolve ambiguities in cases where tracks and clusters of energy deposited in the calorimeter are reconstructed as multiple physics objects, the following logic is applied.  Muon candidates with a jet within $\Delta R<0.4$ are neglected.  If a reconstructed jet lies within $\Delta R<0.2$ of an electron or \tauh\ candidate, this object is considered to be a lepton and the jet is neglected. If the separation of the jet axis from an electron candidate satisfies $0.2<\Delta R<0.4$, the electron is considered non-isolated due to the nearby hadronic activity and is neglected.  Jets within $0.2<\Delta R<0.4$ of \tauh\ candidates are considered as separate objects within the \tauh\ reconstruction algorithm, and are not explicitly treated here.  Electrons within $\Delta R<0.1$ of a muon candidate are also neglected, as are \tauh\ candidates within $\Delta R<0.2$ of electron or muon candidates.  Finally, if two electrons are separated by $\Delta R < 0.1$, the candidate with lower \pt{} is neglected.

The missing transverse momentum is defined as the negative vector sum of the transverse momenta 
of reconstructed jets and leptons, using the energy calibration appropriate for each object~\cite{atlasMET}.
Any remaining calorimeter energy deposits unassociated with reconstructed objects are also
included in the sum. The magnitude of the missing transverse momentum is denoted \met.

\section{Signal regions}
\label{sec:SignalRegions}
Events satisfying all selection criteria are classified into one of two channels.  Events in which at least three of the lepton candidates 
are electrons or muons are selected first, followed by events with two electrons or muons (or one of each)
and at least one \tauh\ candidate.  These two channels are referred to as \threeL\ and \twoLtau\ respectively.

Next, events are further divided into three categories.  The first category includes events that contain
at least one opposite-sign, same-flavour (OSSF) pair of leptons with an invariant mass within 20~\GeV\ of the $Z$ boson mass.  This category
also includes events in which an OSSF pair can combine with a third lepton to satisfy the same invariant mass requirement, allowing this category to capture
events in which a $Z$ boson decays to four leptons (e.g. via $Z\rightarrow\ell\ell\rightarrow\ell\ell\gamma^*\rightarrow\ell\ell\ell'\ell'$) or has some significant 
final-state radiation that is reconstructed as a prompt electron.
This category is referred to as ``on-$Z$''.  
The second category is composed of events that contain an OSSF pair of leptons that do not satisfy the on-$Z$ requirements; this category is
labelled ``off-$Z$, OSSF''.  The final category is composed of all remaining events, and is labelled ``no-OSSF''.  The wide dilepton mass window used to
define the on-$Z$ category is chosen to reduce the leakage of events with real $Z$ bosons into the off-$Z$ categories, which would otherwise see
larger backgrounds from SM production of $ZZ$, $WZ$, and $Z$+jets events.
In \threeL\ events, the categorization is performed using only the three leading leptons (ordered by lepton \pt).  In \twoLtau\ events, the categorization
is performed using the two light-flavour leptons and the \tauh\ candidate with the highest \pt.  The categorization always ignores any additional leptons.

Several kinematic variables are used to characterize events that satisfy all selection criteria.
The variable \Htlep\ is defined as the scalar sum of the \pt, or \ptvis\ for $\tauh$ candidates,
of the three leptons used to categorize the event. The variable \minleppt\ is defined as the minimum \pt\ of the three leptons used to categorize the event.
The variable \Htjets\ is defined as the scalar sum of the \pt{} of all selected jets in the event.  
The ``effective mass'', \St, is the scalar sum of \met, \Htjets, and the
\pt{} of all identified leptons in the event.  For events classified as on-$Z$, the 
transverse mass (\mtw) is constructed using the \met\ and the highest-\pt\ lepton not associated with a $Z$ boson candidate.  It is defined as $\mtw = \sqrt{2p_{\mathrm{T}}^\ell\met(1-\cos(\Delta\phi))}$
where $\Delta\phi$ is the azimuthal angle between the lepton and the missing transverse momentum.  In on-$Z$ events where a triplet of leptons forms the $Z$-boson
candidate, another $Z$ boson is defined using the OSSF pair of leptons with the largest invariant mass, and \mtw\ is constructed using the third lepton.
In events in which two $Z$ boson candidates can be formed from the three leading leptons, the candidate with mass closer to the pole mass is defined
as the $Z$ boson.

Signal regions are defined in each channel and category by requiring one or more variables to exceed minimum values.  
Signal regions based on \Htlep\ are made without requirements on other variables, as are regions based on \minleppt\ and the number 
of $b$-tagged jets.
Signal regions based on \met\ are defined 
separately for events with \Htjets\ below and above 150~\GeV, which serves to distinguish weak production (e.g. $pp\rightarrow W^*\rightarrow \ell^*\nu^*$) from strong production (e.g. $pp\rightarrow Q\bar{Q}' \rightarrow W\bar{q}Zq'$, where $Q$ is some new heavy quark).
Signal regions based on \St\ are constructed with and without additional requirements of $\met \geq 100$~\GeV\ and $\mtw \geq 100$~\GeV. 
The definitions of all 138 signal regions are given in table~\ref{tab:signal_regions}.

\begin{table*}[tbp]
  \begin{center}
    \begin{tabular}{l r r r r l}
      \hline
      \hline
      Variable     &\multicolumn{4}{c}{Lower Bounds [GeV]}  &Additional Requirements\\
      \hline       
      \htlep       &200 &500 &800 &       &\\
      \minleppt    &50  &100 &150 &       &\\
      \met         &0   &100 &200 &300    &$\Ht < 150$ \GeV\\
      \met         &0   &100 &200 &300    &$\Ht \geq 150$ \GeV\\
      \st          &600 &1000&1500&       &\\
      \st          &0   &600 &1200&       &$\met{}\geq100$ \GeV\\
      \st          &0   &600 &1200&       &$\mtw{}\geq100$ \GeV, on-$Z$\\
      \hline
      \hline
      Variable     &\multicolumn{4}{c}{Multiplicity}\\
      \hline
      $b$-tags     &$\geq 1$&$\geq 2$&&                           &\\
      \hline
    \end{tabular}
    \caption{Kinematic requirements for the signal regions defined in the analysis. The signal regions are constructed
    by combining these criteria with the six exclusive event categories.  The regions with combined requirements on \st{}
    and \mtw{} are an exception as they are only defined for the on-$Z$ category. }
    \label{tab:signal_regions}
  \end{center}
\end{table*}

Several of the categories and signal regions described above are new with respect to the analysis performed using the 7~\TeV{} dataset~\cite{atlasML2011}.
The distinction between the off-$Z$, OSSF and the off-$Z$, no-OSSF categories is introduced, as are the signal regions defined using the variables \minleppt, \mtw, 
and the number of $b$-tagged jets.  As mentioned earlier, thresholds that define signal regions in the 7~\TeV{} analysis are also raised to exploit the higher
centre-of-mass energy and larger dataset at 8~\TeV.

\section{Simulation}
\label{sec:Samples}
Simulated samples are used to estimate backgrounds from events with three or more
prompt leptons, where prompt leptons are those originating in the hard scattering process or from the decays of gauge bosons. 
The response of the ATLAS detector is modelled~\cite{atlassim} using the \geant~\cite{geant}\ toolkit,
and simulated events are reconstructed using the same software as used for collision data.
Small post-reconstruction corrections are applied to account for differences
in reconstruction and trigger efficiency, energy resolution, and energy scale between
data and simulation~\cite{Aad:2014fxa,ATLAS-CONF-2014-032,muonperf}.  
Additional $pp$ interactions (pileup) in the same or nearby bunch crossings are modelled 
with \pythia\ 6.425~\cite{pythia}.  Simulated events are reweighted to reproduce the distribution of the average number of $pp$ interactions per
crossing observed in data over the course of the 2012 run.  

The largest SM backgrounds with at least three prompt leptons are $WZ$ and $ZZ$
production where the bosons decay leptonically.  These processes are modelled with \sherpa~\cite{sherpa} using version 1.4.3 (1.4.5) for $WZ$ ($ZZ$).
These samples include the continuum Drell--Yan processes ($\gamma^{*}$), where the boson has an invariant mass above twice the muon (tau) mass for 
decays to muons (taus), and above 100 MeV for decays to electrons.
Diagrams where a $\gamma^{*}$ is produced as radiation from a final-state lepton and decays to additional leptons, i.e. $W\to\ell^{*}\nu\to\ell\gamma^{*}\nu\to\ell\ell'\ell'\nu$ and $Z\to\ell\ell^{*}\to\ell\ell\gamma^{*}\to\ell\ell\ell'\ell'$, where $\ell$ and $\ell'$ need not have the same flavour, are also included.
Simulated samples of SM $Z\gamma^*\rightarrow\ell^+\ell^-e^+e^-$ events generated with \madgraph\ 5.1.3.28~\cite{madgraph} are used to verify that this analysis has negligible acceptance for $Z\gamma^*$ events when the mass of the $\gamma^*$ is less than $100~\MeV$.  The simulation and reconstruction efficiency of such events was probed in an analysis of Dalitz decays~\cite{ATLAS-CONF-2010-007}, where good agreement of simulation and data was observed.
The leading-order predictions from \sherpa\ are cross-checked with next-to-leading-order (NLO) calculations
from \vbfnlo-2.6.2~\cite{vbfnlo}.  Diagrams including a SM Higgs boson give negligible contributions
compared to other diboson backgrounds in all signal regions under study.

The production of $\ttbar+W/Z$ processes (also denoted $\ttbar+V$) is simulated with 
\alpgen\ 2.13~\cite{alpgen} for the hard scattering, \herwig\ 6.520~\cite{herwig} for the 
parton shower and hadronization, and \jimmy\ 4.31~\cite{jimmy} for the underlying event.
Single-top production in association with a $Z$ boson ($tZ$) is simulated with \madgraph\ 5.1.3.28~\cite{madgraph}.
Both the $\ttbar+V$ and $tZ$ samples use \pythia\ 6.425 for the parton shower and hadronization.
These samples also include production of $\ttbar\gamma^*$ and $t\gamma^*$, with the mass of the generated $\gamma^*$ required to be above 5~\GeV.
As for $Z\gamma^*$, cross checks with dedicated \madgraph\ samples in which the mass of the $\gamma^*$ is allowed to drop to twice the electron mass
show that the contributions from such events are negligible in this search.
Corrections to the normalization from higher-order effects for these samples are 30\%~\cite{ttW,ttZ}. 
Leptons from Drell--Yan processes produced in association with a photon that converts 
in the detector (denoted $Z+\gamma$ in the following) are modelled with \sherpa\ 1.4.1.
Additional samples are used to model dilepton backgrounds for 
control regions with fewer than three leptons.
Events from $\ttbar$ production are generated using \powheg~\cite{powheg} with \pythia~6.425 used for the parton shower and hadronization.
Production of $Z$+jets is performed      
with \alpgen\ 2.13~\cite{alpgen} for the hard scattering and \pythia 6.425 for the parton shower.

Samples of doubly charged Higgs bosons, generated with \pythia~8.170~\cite{pythia8}, are normalized to NLO cross sections.
The samples include events with pair-produced doubly charged Higgs bosons mediated by a $Z/\gamma^*$, and do not include
single-production or associated production with a singly charged state. Samples of excited charged leptons and excited
neutrinos are generated with \pythia~8.175 using the effective Lagrangian described in ref.~\cite{ExcitedLeptons}.

The CT10~\cite{ct10} parton distribution functions (PDFs) are used for the \sherpa\ and \powheg\ samples.  MRST2007 LO$^{**}$~\cite{mrst} 
PDFs are used for the \pythia\ and \herwig\ samples. For \powheg, \madgraph\ and \alpgen, the CTEQ6L1~\cite{cteq6l1} PDFs are used.  The underlying event tune 
for \powheg{} and \pythia~8.175 is the ATLAS Underlying Event Tune 2 (AUET2)~\cite{AUET2}, while for the \pythia~6.425 and \madgraph{} samples the tune is AUET2B~\cite{AUET2B}.
The \alpgen{} $ttV$ samples use AUET2B, while the \alpgen{} $Z$+jets samples use P2011C~\cite{P2011C}.

\section{Background estimation}
\label{sec:Backgrounds}
Standard Model processes that produce events with three or more lepton candidates fall into three classes.  The first 
consists of events in which prompt leptons are produced in the hard
interaction or in the decays of gauge bosons.
A second class of events includes Drell--Yan production in association
with an energetic $\gamma$, which then converts in the detector to produce
a single reconstructed electron.  A third class of events
includes events with at least one non-prompt, non-isolated, or fake lepton candidate satisfying the identification
criteria described above.

The first class of backgrounds is dominated by 
$WZ\to\ell\nu\ell'\ell'$ and 
$ZZ\to\ell\ell\ell'\ell'$ events.  Smaller contributions
come from $\ttbar+W$, $\ttbar+Z$, and $t+Z$ events, where the vector bosons, including those from top quark decays, decay leptonically.  Contributions
from triboson events, such as $WWW$, and events containing a Higgs boson, are negligible.  All 
processes in this class of backgrounds are modelled with the dedicated simulated samples described above.
Reconstructed leptons in the simulated samples are required
to be consistent with the decay of a vector boson or tau lepton using generator-level information.

The second class of backgrounds, from Drell--Yan production in association
with a hard photon, is also modelled with simulation.  Prompt electrons reconstructed with incorrect charge (charge-flips) are modelled in simulation, 
with correction factors derived using $Z\rightarrow ee$ events in data.  Similar corrections are applied to photons reconstructed as prompt
electrons.

The class of events that includes non-prompt or fake leptons, referred to here
as the reducible background, is estimated
using {\it in situ} techniques that rely minimally on simulation.  
Such backgrounds for muons arise from semileptonic $b$- or $c$-hadron decays, from in-flight decays 
of pions or kaons, and from energetic particles that reach the muon spectrometer.
Non-prompt or fake electrons can also arise from misidentified hadrons or jets.
Hadronically decaying taus have large backgrounds from narrow, low-track-multiplicity jets that mimic $\tauh$ signatures.

The reducible background is estimated by reweighting events with one or more leptons that do not satisfy the nominal 
identification criteria, but satisfy a set of relaxed criteria, defined separately for each lepton flavour.  To define the relaxed
criteria for electrons, 
the identification working point is changed from tight to loose~\cite{elecperf}.  For muons, the $|d_{0}/\sigma(d_{0})|$ 
and isolation cuts are loosened.  For taus, the BDT working point is changed from tight to loose.  The reweighting
factors are defined as the ratio of fake or non-prompt leptons that satisfy the nominal criteria to those which only fulfil the
relaxed criteria.  These factors are measured as a function of the candidate $\pt{}$ and $\eta{}$ in samples of data that are
enriched in non-prompt and fake leptons.  Corrections for the contributions from prompt leptons in the background-enriched
samples are taken from simulation.

The background estimates and lepton modelling are tested in several validation regions.
The \tauh\ modelling and background estimation are tested in a region enriched in $Z\rightarrow\tau\tau\rightarrow\mu\tauh$ events.
This region is constructed by placing requirements on the invariant mass of the muon and $\tauh$ pair, on the angles between the muon, $\tauh$
and missing transverse momentum, and on the muon and $\met$ transverse mass.  These requirements were optimized to suppress the contribution from $W\rightarrow\mu\nu$
+ jets events. The \tauh\ \pt\ distribution in this validation region is shown in figure~\ref{fig:tauVR_pt}.

\begin{figure}[tbp]
  \subfigure[Prompt \tauh\ validation region]{\includegraphics[width=0.48\columnwidth]{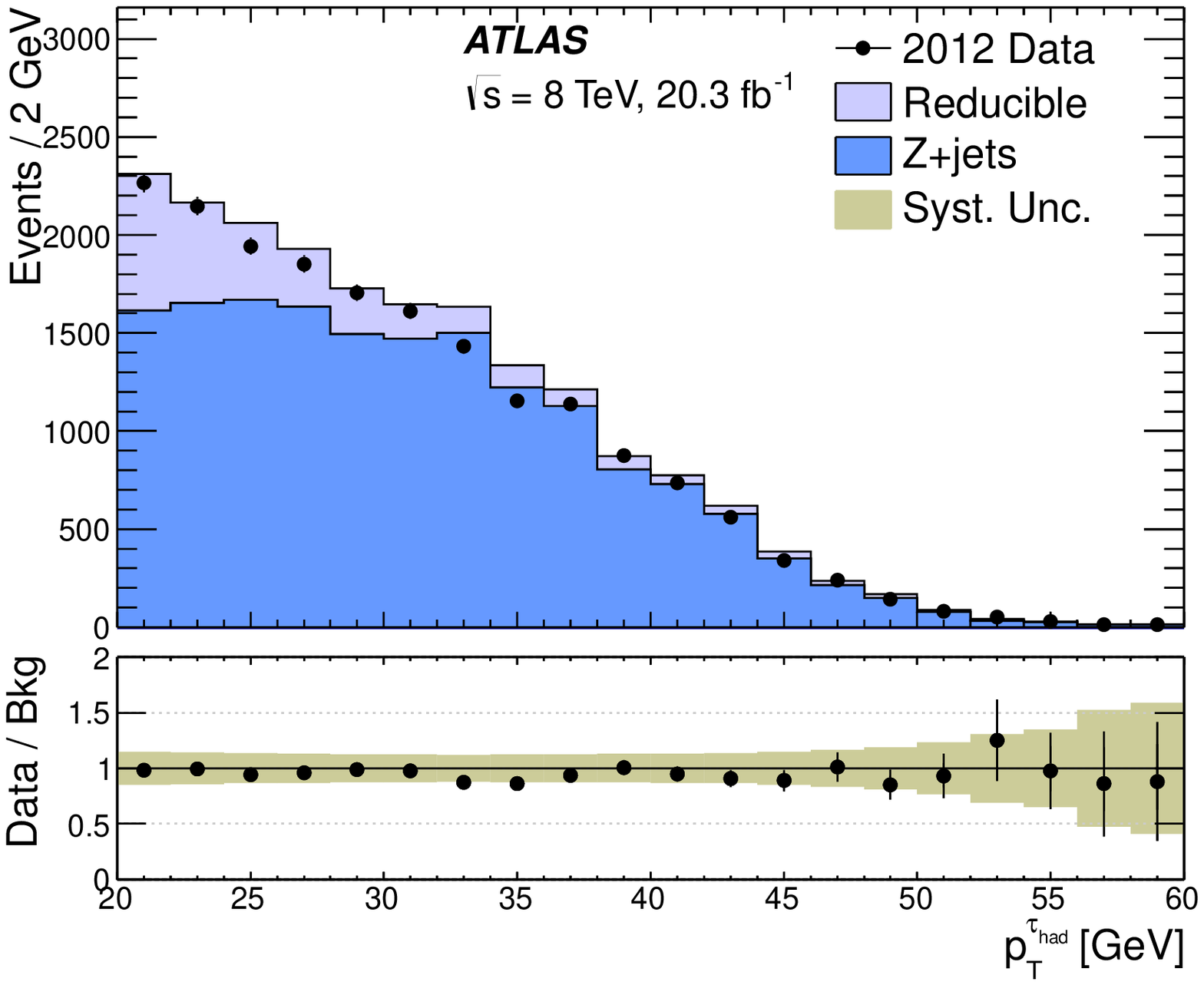}\label{fig:tauVR_pt}}
  \subfigure[Electron/muon validation region]{\includegraphics[width=0.48\columnwidth]{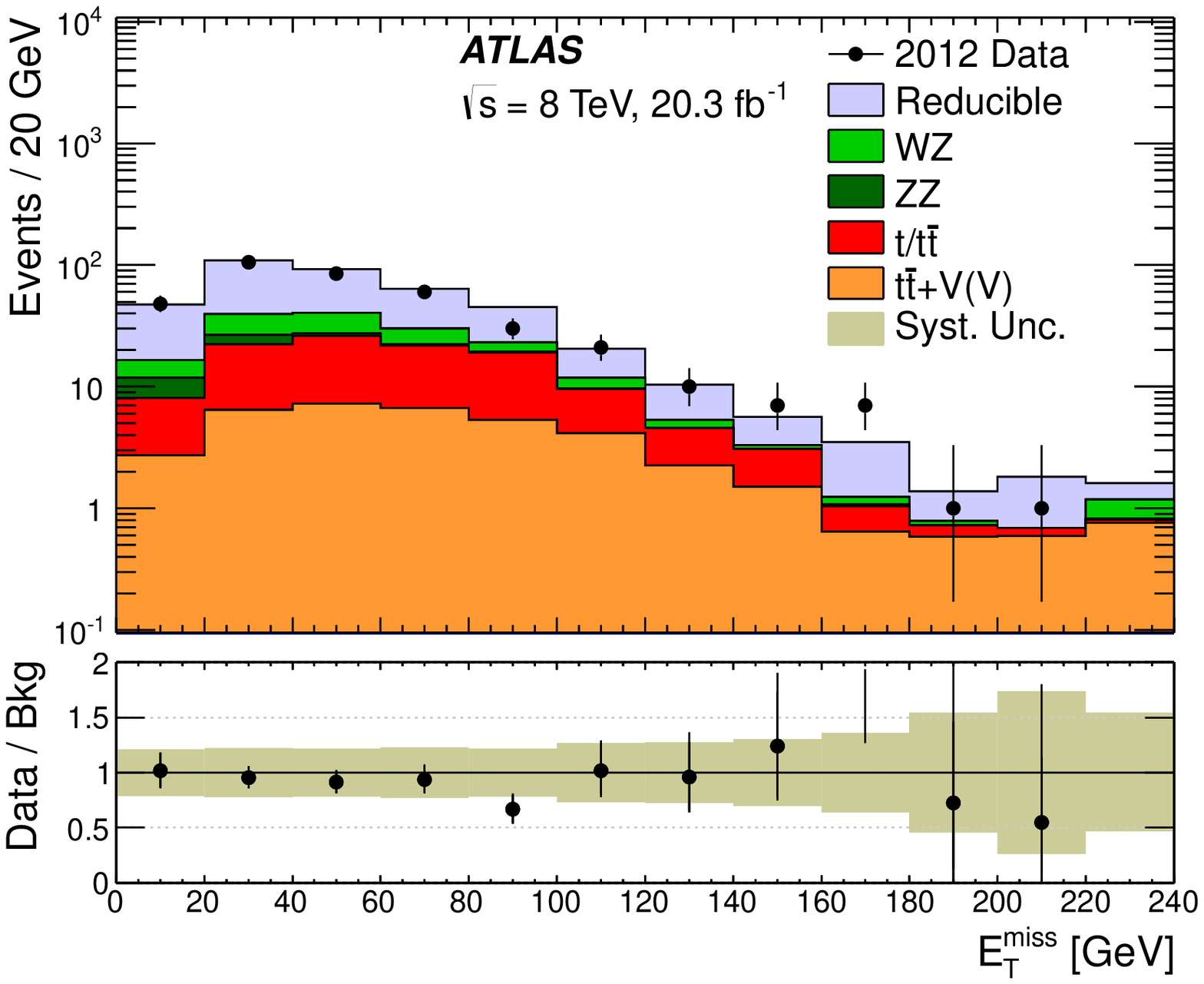}\label{fig:ttbar_emu_met}}\\
  \subfigure[Intermediate-\tauh\ validation region]{\includegraphics[width=0.48\columnwidth]{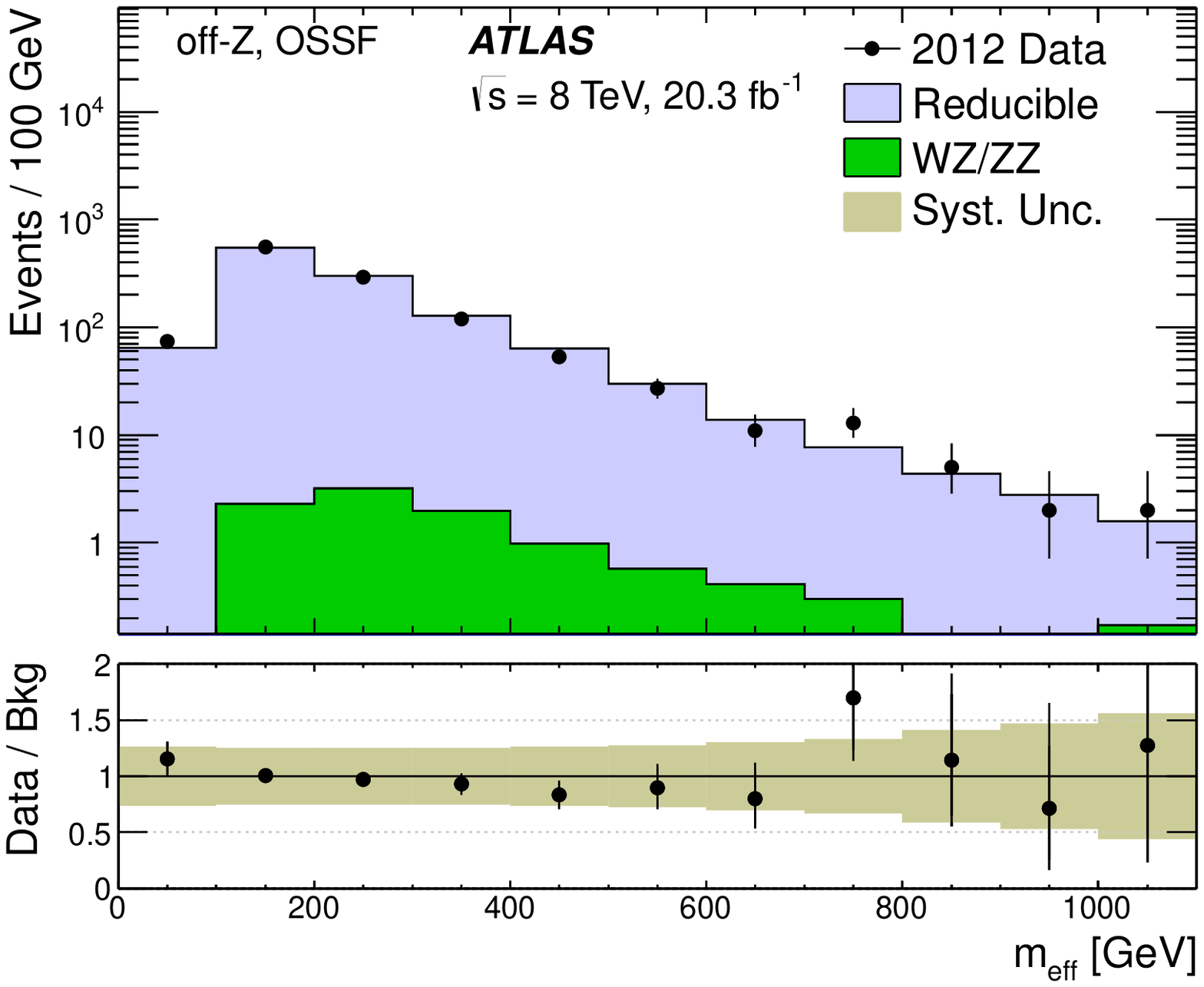}\label{fig:int_tau_ST}}
  \subfigure[Intermediate-muon validation region]{\includegraphics[width=0.48\columnwidth]{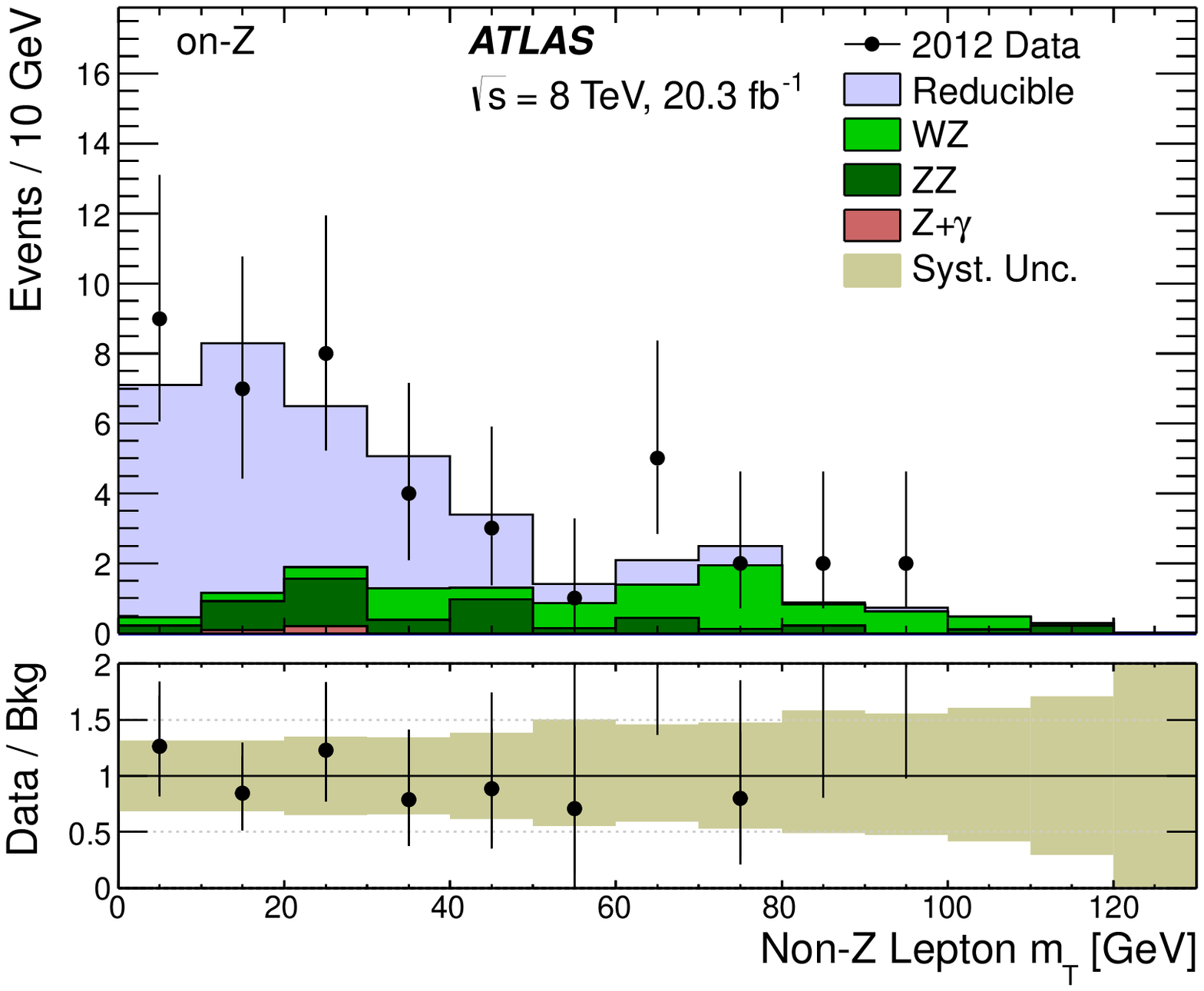}\label{fig:int_mu_mt}}
  \caption{
    (a) Tau \pt\ distribution for \tauh\ candidates in the enriched \tauh\ validation region. 
    (b) Missing transverse momentum distribution in the $\ttbar$ validation region for electrons and muons.  
    (c) Effective mass distribution in the intermediate \tauh\ validation region, in the off-$Z$, OSSF category.
    (d) Distribution of the transverse mass of the missing transverse momentum and the muon not associated with the $Z$-boson candidate in the intermediate-muon validation region.
    Signal contamination from doubly charged Higgs bosons and excited leptons in all validation regions is negligible.
    The lower panel shows the ratio of data to the expected SM backgrounds in each bin.  The last bin in all figures includes overflows.
  }
  \label{fig:vrs}
\end{figure}

A validation region rich in $\ttbar$ events is defined to test the estimates of the reducible background.  
Events in this region have exactly two identified lepton candidates with the same charge (but any flavour combination), at least one $b$-tag, and $\Htjets \le 500\gev$.
This sample is estimated to be primarily composed of lepton+jets
$\ttbar$ events.   The same-sign requirement suppresses events where both $W$ bosons decay leptonically,
and enhances the contributions from events where one lepton candidate originates from semileptonic $b$-decay.
The upper limit on \Htjets\ of 500~\GeV\ reduces potential contamination from hypothesized signals.  An example of the \met\ distribution in the $\ttbar$~region enriched in reducible backgrounds from the same-sign electrons and/or muons is shown in figure~\ref{fig:ttbar_emu_met}.

Additional validation regions that test the estimation of reducible backgrounds lepton identification criteria tighter than those used in
the background-enriched samples but looser than and orthogonal to those used in the signal regions.  This
set of identification criteria is referred to as the ``intermediate'' selection,
and leptons satisfying the intermediate selection are referred to as intermediately identified leptons, or simply intermediate leptons.
The reweighting factors are remeasured for the intermediate selection and used in the validation region.
Events are selected as in the analysis, with the intermediate selection used for a single lepton flavour.  For intermediate electrons and muons, only events in the on-$Z$ channel are considered, and intermediate leptons
are required to have a flavour different from that of the OSSF pair forming the $Z$ boson candidate.  
For intermediate taus, all channels are considered.  
An example of the \meff\ distribution for the intermediate tau selection is shown in figure~\ref{fig:int_tau_ST}.
For the intermediate muon validation region, the transverse mass distribution for intermediate muons combined with \met\ is shown in figure~\ref{fig:int_mu_mt}.

Good agreement between the expected and observed event yields is seen in all validation regions.  A summary of expected and observed event yields
for all validation regions is shown in table~\ref{tab:vrs}.

\begin{table}[tbp]
  \centering
  \begin{tabular}{l r@{ $\pm$ }l r@{ $\pm$ }l r@{ $\pm$ }l c}
    \hline
    \hline
    Region                      &\multicolumn{2}{c}{Prompt}  &\multicolumn{2}{c}{Fake}  &\multicolumn{2}{c}{Total Expected}  &Observed\\
    \hline
    $Z\rightarrow\taulep\tauh$          &16400&800                   &2900&700                  &19300&1100                 &18323\\
    \hline
    \ttbar: $\ell\ell$                  &130&40                      &230&60                    &360&70                     &375\\
    \ttbar: $\ell\tauh$                 &37&3                        &1700&400                  &1700&400                   &1469\\
    \hline
    Intermediate electron               &130&70                      &53&17                     &180&80                     &207\\
    Intermediate muon                   &13&2                        &26&8                      &39&8                       &43 \\
    Intermediate tau, on-$Z$            &74&7                        &19000&5000                &19000&5000                 &17361\\
    Intermediate tau, off-$Z$, OSSF     &11&2                        &1160&290                  &1170&290                   &1155\\
    Intermediate tau, off-$Z$, no-OSSF  &21&3                        &320&80                    &340&80                     &340\\
    \hline
    \hline
  \end{tabular}
  \caption{Expected and observed event yields for all validation regions.  The expected contributions from signal processes such as excited leptons
  or doubly charged Higgs bosons are negligible in all validation regions.}
  \label{tab:vrs}
\end{table}

\section{Systematic uncertainties}
\label{sec:Systematics}
The backgrounds modelled with simulated samples have systematic uncertainties related to the trigger, selection efficiency, 
momentum scale and resolution, \met, and luminosity.  These uncertainties, when evaluated as fractions of the total background
estimate, are usually small, and are summarized in table~\ref{tab:systematics}.
Predictions from simulations are normalized to the integrated luminosity collected in 2012.  The uncertainty on the luminosity is 
2.8\% and is obtained following the same methodology as that detailed in ref.~\cite{luminositypaper}.

\begin{table}
  \centering
  \begin{tabular}{l c}
    \hline
    \hline
    Source         &Uncertainty [\%]\\
    \hline
    Luminosity               &2.8\\
    \hline
    Trigger efficiency       &1 \\
    Lepton momentum scale/resolution  &1\\
    Lepton identification    &2 \\
    Jet energy resolution    &2 \\
    Jet energy scale         &5 \\
    $b$-tagging efficiency   &5 \\
    \met\ scale/resolution   &4 \\
    \hline
    $\ttbar+V$ cross section &30 \\
    $WZ/ZZ$ cross section    &7 \\
    $WZ/ZZ$ shape            &20--50 \\
    \hline
    Charge misidentification &8\\
    Non-prompt and fake \tauh               &25\\
    Non-prompt and fake $e/\mu$       &40\\
    \hline
    \hline
  \end{tabular}
  \caption{Typical systematic uncertainties from various sources, in signal regions where the uncertainty is relevant.
  The uncertainties on the backgrounds are presented as the percent uncertainty on the total background estimate. }
  \label{tab:systematics}
\end{table}

Uncertainties on the cross sections of SM processes modelled by simulation are also considered.  
The normalization of the $\ttbar+W$ and $\ttbar+Z$ backgrounds have an uncertainty of 30\%\ based on PDF and scale variations~\cite{ttW,ttZ}.
The \sherpa\ predictions~\cite{sherpa} of the $WZ$ and $ZZ$ processes are cross-checked with next-to-leading-order 
predictions from \vbfnlo.  
Scale uncertainties are evaluated by varying the factorization and renormalization scales up and down
by a factor of two, and range from 3.5\% for the inclusive prediction to 6.6\% for events with at least
one additional parton.  PDF uncertainties are evaluated by taking the envelope of predictions from all
PDF error sets for CT10-NLO, MSTW2008-NLO, and NNPDF-2.3-NLO, and are between 3\% and 4\%.

An additional uncertainty on the \sherpa\ predictions is applied to cover possible mismodelling of events with
significant jet activity.  This shape uncertainty is evaluated using \loopsim+\vbfnlo~\cite{loopsim}, which makes ``beyond-NLO'' 
predictions (denoted \nNLO) for high-\pt\ observables, and is based on the study presented in ref.~\cite{loopsim2}.
Predictions of \htjets\ and \meff\ at $\nNLO$ are compared with those from \sherpa\ in a phase space similar to
that used in this analysis.
Good agreement between \sherpa\ and the \nNLO\ predictions is observed across the full range of \htjets\ and \meff. 
The uncertainty on the \nNLO\ prediction is evaluated by changing the renormalization and factorization scales used in the \nNLO\ calculation 
by factors of two.  These uncertainties increase linearly with event activity with a slope of (50\%)$\times$(\htjets~[\TeV]) and are
applied to the \sherpa\ predictions.  A study of $Z$+jets events at $\sqrt{s}=7~\TeV$~\cite{ZplusjetsSM} shows good agreement of
\sherpa\ predictions with data in events with significant transverse activity, showing deviations of data from predictions within the uncertainties used here.

The estimates of the reducible background carry large uncertainties from several sources. 
These uncertainties are determined in dedicated studies using a combination of simulation
and data.  They account for potential biases in the methods used to extract the
reweighting factors, and for the dependency of the reweighting factors on the event topology.
The electron reweighting factors have uncertainties that range from 24\%
to 30\% as a function of the electron \pt, while for muons the uncertainties range
from 25\% to 50\%.  
For the estimates of fake $\tauh$ candidates, the \pt-dependent uncertainty on the reweighting factors is approximately 25\%.
In signal regions where the relaxed
samples are poorly populated, statistical uncertainties on the estimates of the reducible background
become significant, especially in regions with high \met\ or \Htjets\ requirements.

The relative uncertainty on the correction factors for electron charge-flip modelling in simulation is estimated to be
40\%, resulting in a maximum uncertainty on the total background yield in any signal region of 11\%.  Studies of simulated data show that
the majority of charge-flip electrons are due to bremmstrahlung photons that interact with detector material and convert to an electron-positron
pair, yielding an energetic secondary lepton with the opposite sign of the prompt lepton.  As this is the same process by which prompt photons
mimic prompt leptons, the same 40\% uncertainty is assigned to the modelling of prompt photons reconstructed as electrons.

In all signal regions, the dominant systematic uncertainty is either the uncertainty on the reducible background or the shape uncertainty on the diboson samples.  
In \twoLoneT\ channels,
the uncertainty on the reducible background always dominates.  In \threeL\ channels, the $WZ$ theory uncertainties dominate in most regions 
except in the no-OSSF categories, 
where the uncertainties on the reducible background are dominant.  The uncertainties on $\ttbar+V$ are large in regions requiring two $b$-tagged jets.
The uncertainties on  the trigger, selection efficiency, momentum scale and resolution, and \met\ are always subdominant.

\section{Results}
\label{sec:Results}
Expected and observed event yields for the most inclusive signal regions are summarized in table~\ref{tab:srs}.
Results of the search in all signal regions are summarized in figure~\ref{fig:summary_plot}, which shows the deviation of the observed event yields from the expected yields, 
divided by the total uncertainty on the expected yield, for all signal regions.  The total uncertainty on the expected yield includes statistical uncertainties on the background estimate
as well as the systematic uncertainties discussed in the previous section.  There are no signal regions in which the observed event yield exceeds
the expected yield by more than three times the uncertainty on the expectation, and only one region in which the observed event yield is lower than expected
by more than three times the uncertainty, i.e. the \threeL, off-$Z$ no-OSSF category, with $\htjets < 150\GeV$ and $\met>100\GeV$.  
The smallest $p$-value is 0.05, which corresponds to a $1.7\sigma$ deviation, and is observed in the $\meff>1000~\GeV$ region in the \twoLtau, on-$Z$ channel.
Examples of kinematic distributions for all channels and categories are shown in figure~\ref{fig:example_SRs}.

\begin{table}[tbp]
  \centering
  \begin{tabular}{c r@{ $\pm$ }l r@{ $\pm$ }l r@{ $\pm$ }l c}
    \hline
    \hline
    Channel                   &\multicolumn{2}{c}{Prompt}  &\multicolumn{2}{c}{Fake}  &\multicolumn{2}{c}{Total Expected}  &Observed\\
    \hline
    \multicolumn{8}{c}{off-$Z$, no-OSSF}\\
    \hline
$\geq 3 e/\mu$            &13&2	 &18&5  	 &30&5	        &36\\
$2 e/\mu + \geq 1 \tau$   &26&3	 &180&40	 &200&40	&208\\
\hline
\multicolumn{8}{c}{off-$Z$, OSSF}\\
\hline
$\geq 3 e/\mu$           &206&23	 &33&9	         &239&25	&221\\
$2 e/\mu + \geq 1 \tau$  &15&2	 &630&170	 &640&170	&622\\
\hline
\multicolumn{8}{c}{on-$Z$}\\
\hline
$\geq 3 e/\mu$           &2900&340	 &180&40	 &3080&350	&2985\\
$2 e/\mu + \geq 1 \tau$  &141&13	 &10300&2800	 &10400&2800	&9703\\
    \hline
    \hline
  \end{tabular}
  \caption{Expected and observed event yields for the most inclusive signal regions.}
  \label{tab:srs}
\end{table}

\afterpage{\clearpage}
\begin{sidewaysfigure}
  \includegraphics[width=\textwidth]{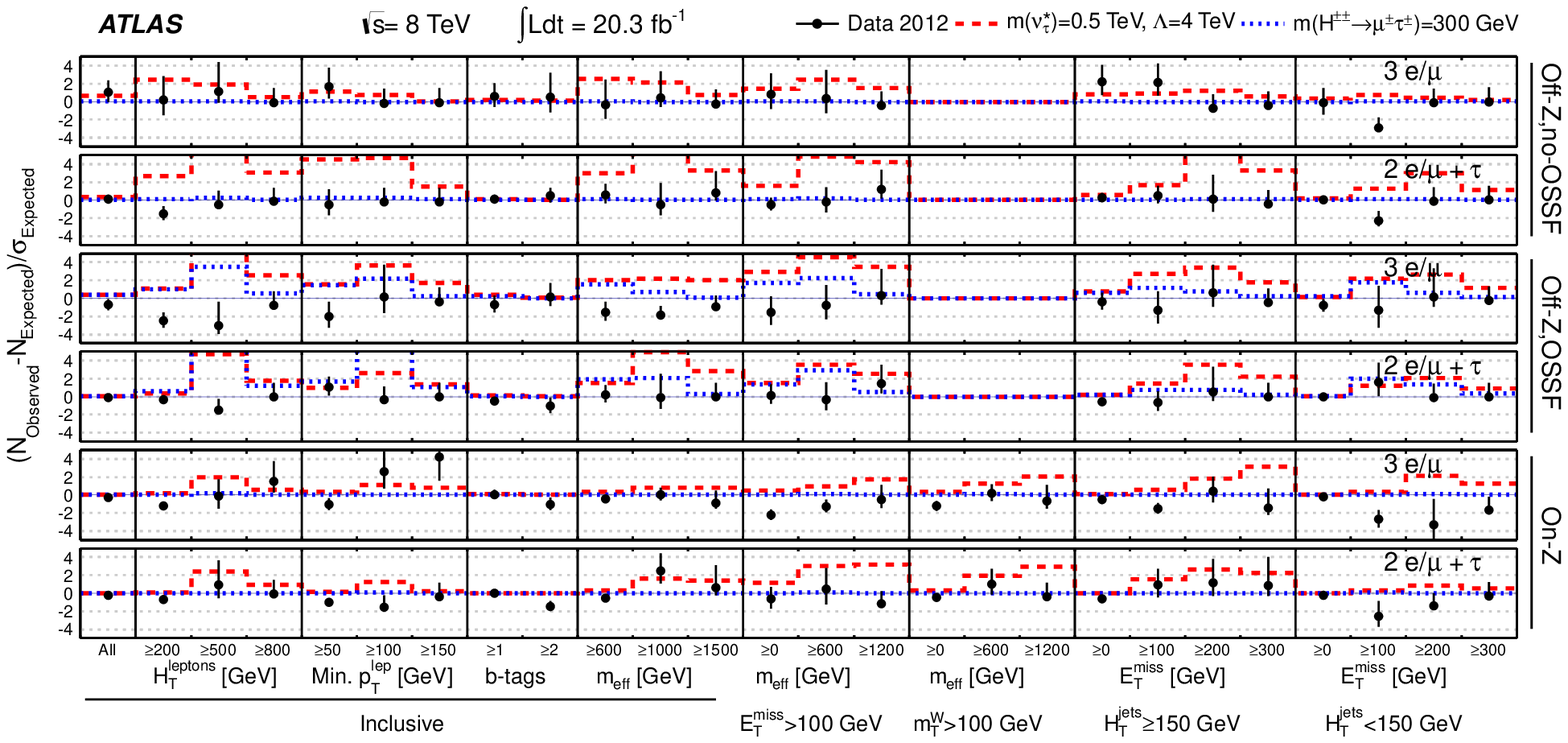}
  \caption{Deviations of observed event yields from expected yields, divided by the total uncertainty on the expected yield, for all signal regions under study.  The total uncertainty on the expected yield includes statistical uncertainties on the background estimate as well as the systematic uncertainties discussed in the previous section.   The error bars on
the data points show Poisson uncertainties with 68\% coverage.  Expected yields for two benchmark BSM scenarios, a model with excited tau neutrinos with mass 500~\GeV\ and 
compositeness scale 4~\TeV, and a model with pair-produced doubly charged Higgs bosons with a mass 300~\GeV\ decaying to $\mu\tau$, are shown by red and blue dashed lines, respectively.}
  \label{fig:summary_plot}
\end{sidewaysfigure}

\begin{figure*}[tbp]
  \subfigure[\threeL, off-$Z$, no-OSSF]{\includegraphics[width=.48\textwidth]{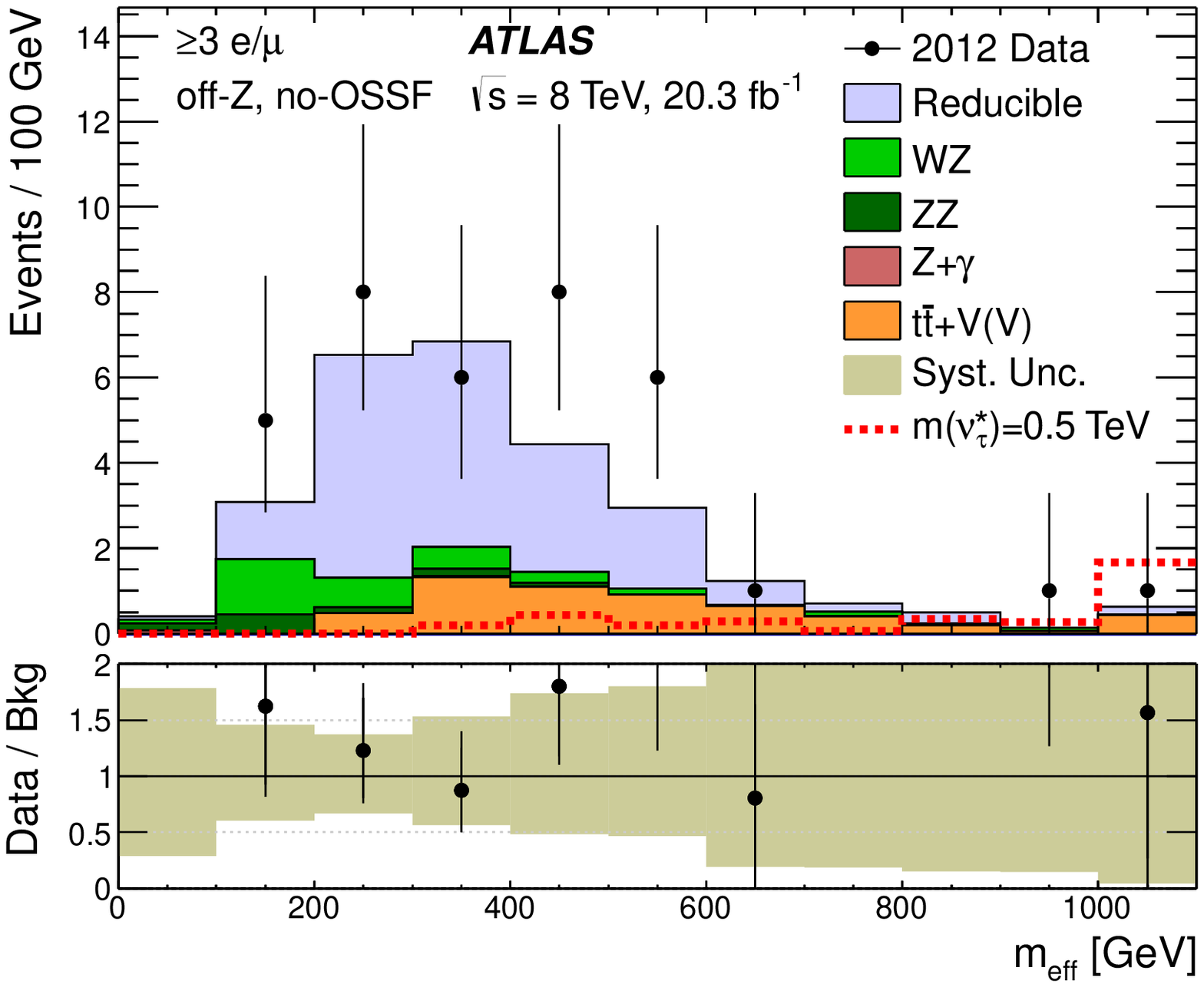}}
  \subfigure[\twoLoneT, off-$Z$, no-OSSF]{\includegraphics[width=.48\textwidth]{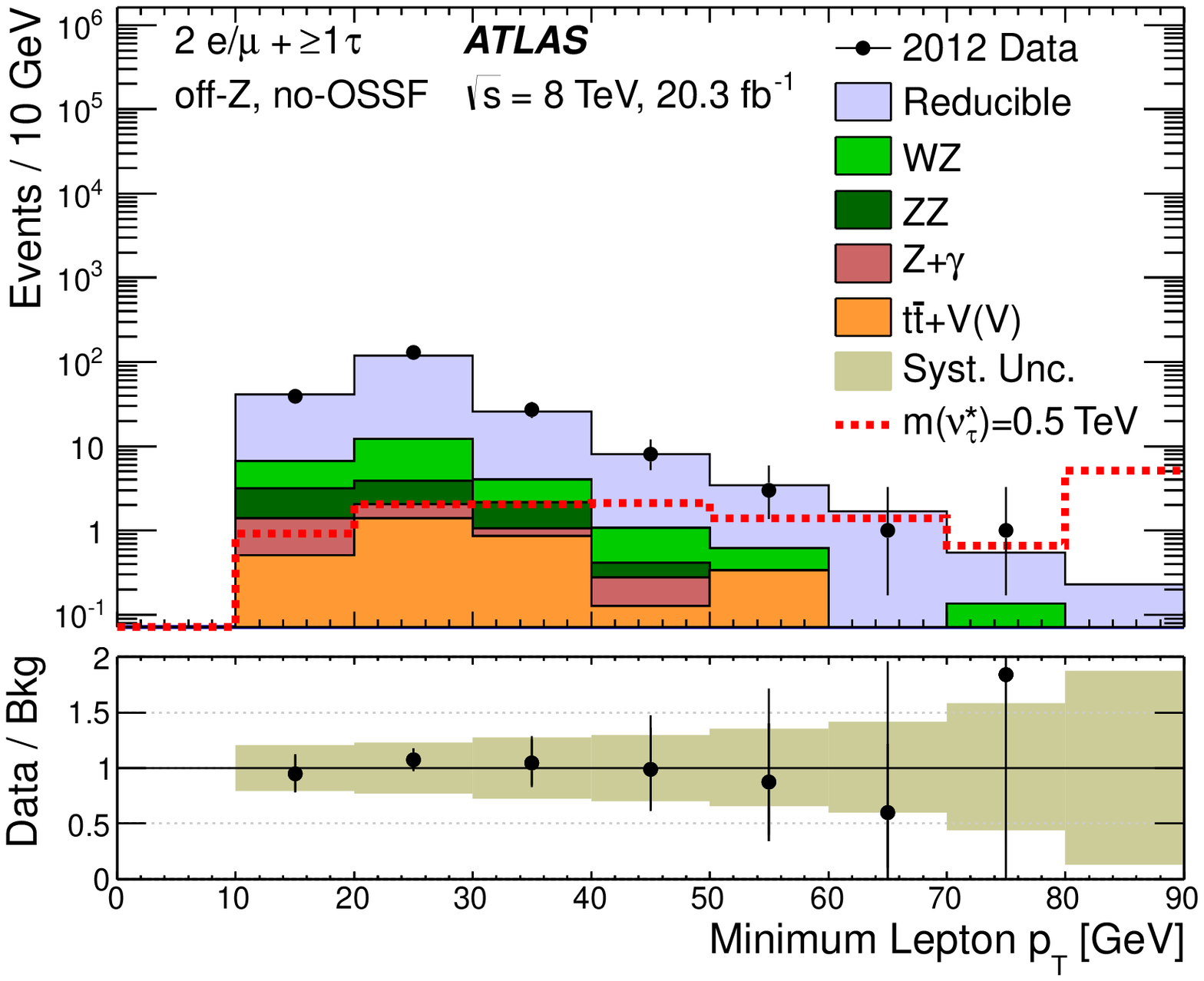}}\\
  \subfigure[\threeL, off-$Z$, OSSF]{\includegraphics[width=.48\textwidth]{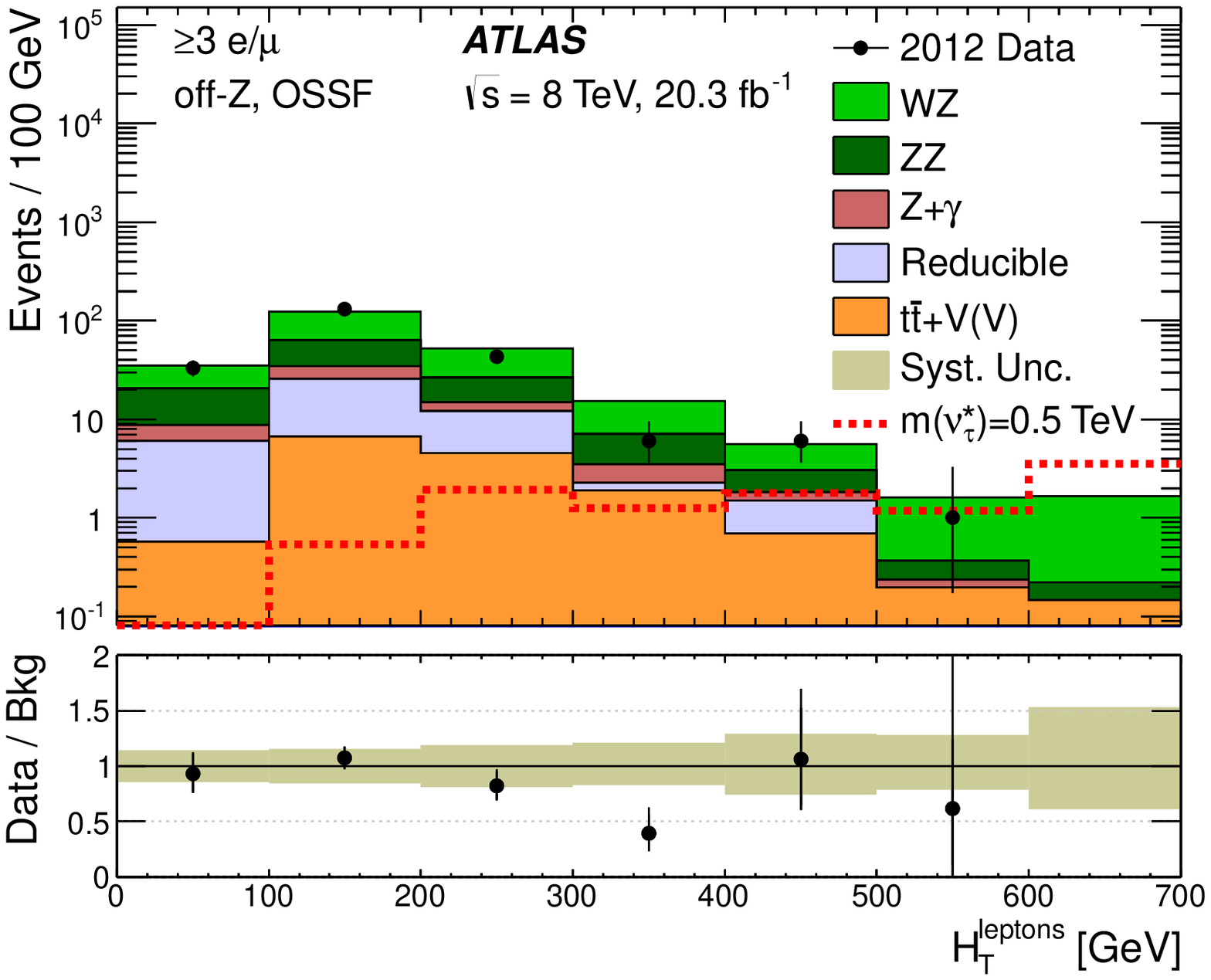}}
  \subfigure[\twoLoneT, off-$Z$, OSSF]{\includegraphics[width=.48\textwidth]{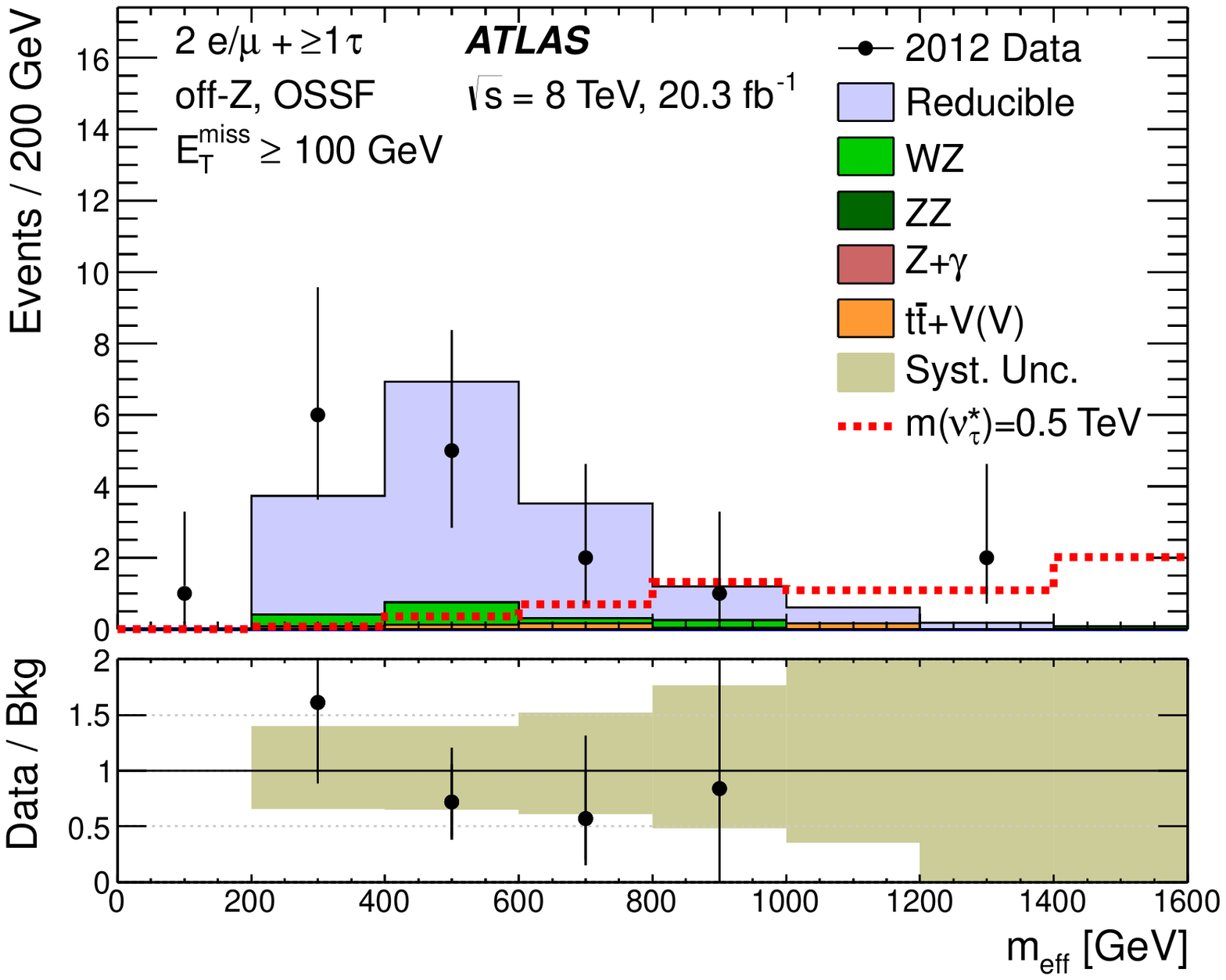}}\\
  \subfigure[\threeL, on-$Z$]{\includegraphics[width=.48\textwidth]{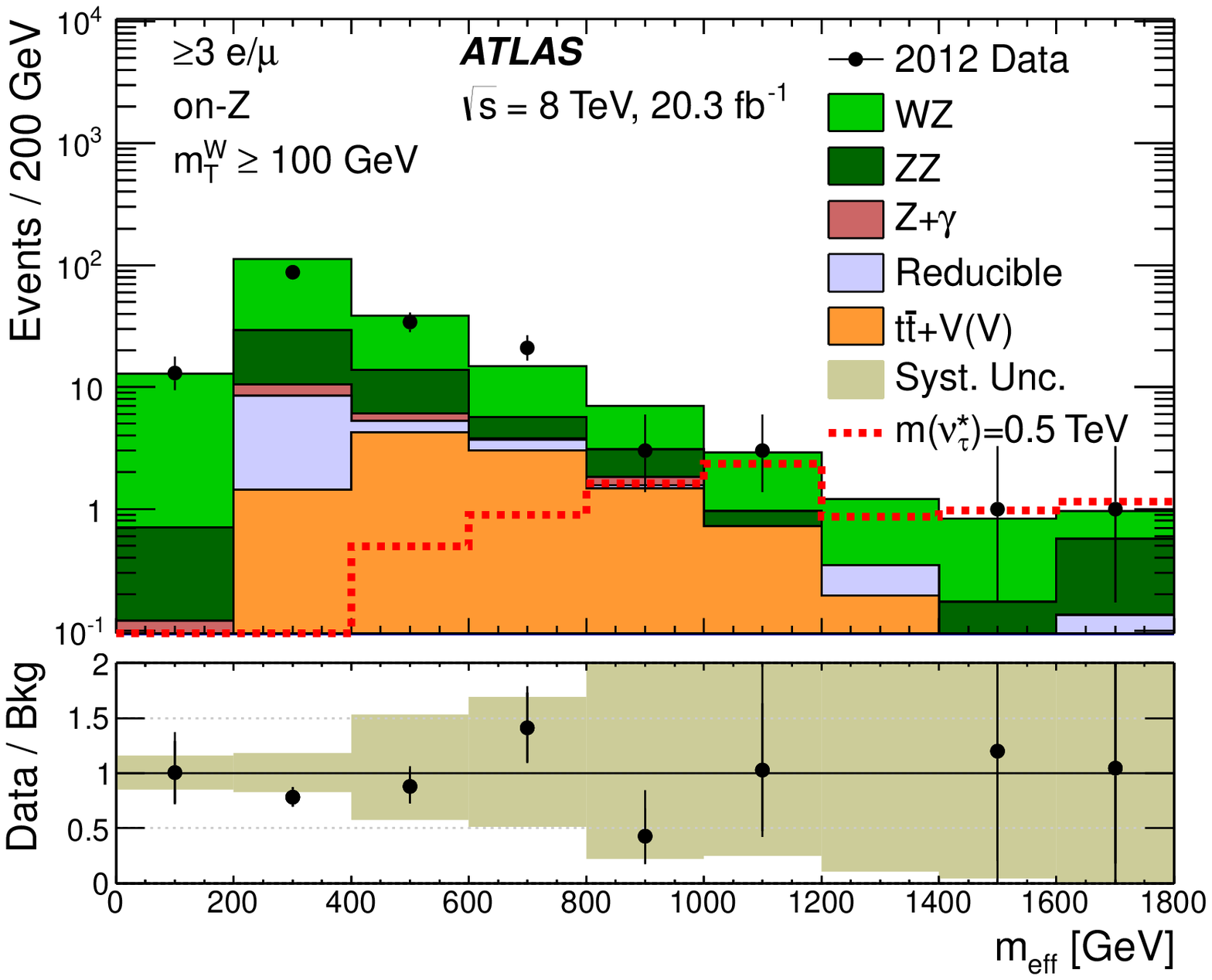}}
  \subfigure[\twoLoneT, on-$Z$]{\includegraphics[width=.48\textwidth]{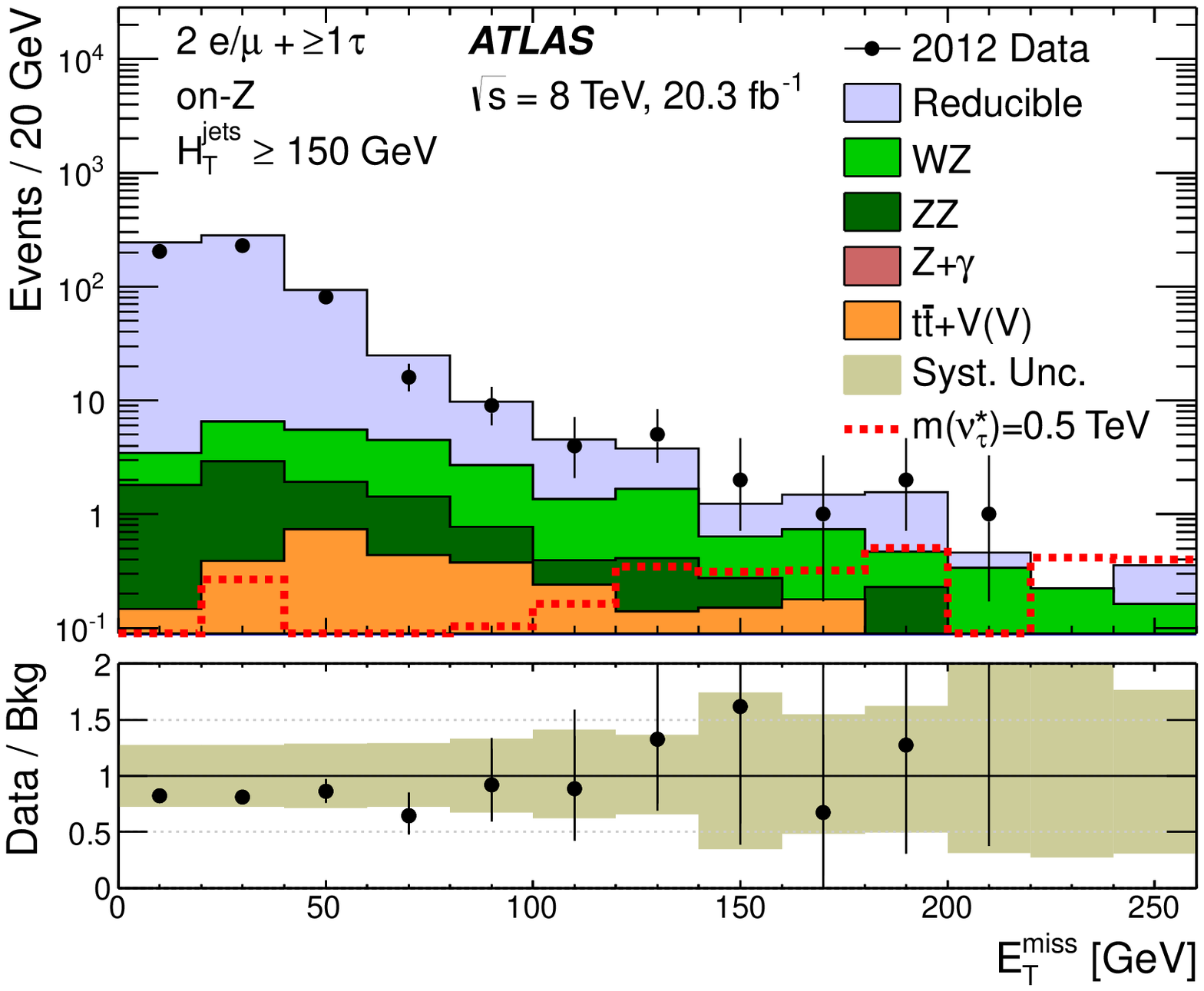}}
  \caption{Sample results for all categories: (a) \threeL, off-$Z$, no-OSSF, (b) \twoLoneT, off-$Z$, no-OSSF, (c) \threeL, off-$Z$, OSSF, (d) \twoLoneT, off-$Z$, OSSF, (e) \threeL, on-$Z$ and (f) \twoLoneT, on-$Z$.  A predicted signal of excited tau neutrinos is overlaid to illustrate the sensitivity of the different signal regions; the compositeness scale $\Lambda$ of this signal scenario is 4~\TeV.  The lower panel shows the ratio of data to the expected SM backgrounds in each bin.  The last bin in all figures includes overflows.}  
  \label{fig:example_SRs}
\end{figure*}

Since the data are in good agreement with SM predictions, the observed event yields are used to constrain contributions from new phenomena.
The 95\% confidence level (CL) upper limits on the number of events from non-SM sources ($N_{95}$) are calculated
using the modified Frequentist $CL_{s}$ prescription~\cite{cls}.  All statistical and systematic uncertainties on estimated backgrounds are
incorporated into the limit-setting procedure, with correlations taken into account where appropriate.  
The $N_{95}$ limits are then converted into limits on the ``visible cross section'' (\sigmavis) 
using the relationship $\sigmavis=N_{95}/\intLdt$, where $\intLdt$ is the integrated luminosity of the data sample.

Figure~\ref{fig:summary_limits} shows the 
resulting observed limits, along with the median expected limits with $\pm1\sigma$ and $\pm2\sigma$ uncertainties.  Table~\ref{tab:limits_all} shows the expected and observed limits for the most
inclusive signal regions.

\afterpage{\clearpage}
\begin{sidewaysfigure*}[tbp]
  \includegraphics[width=\textwidth]{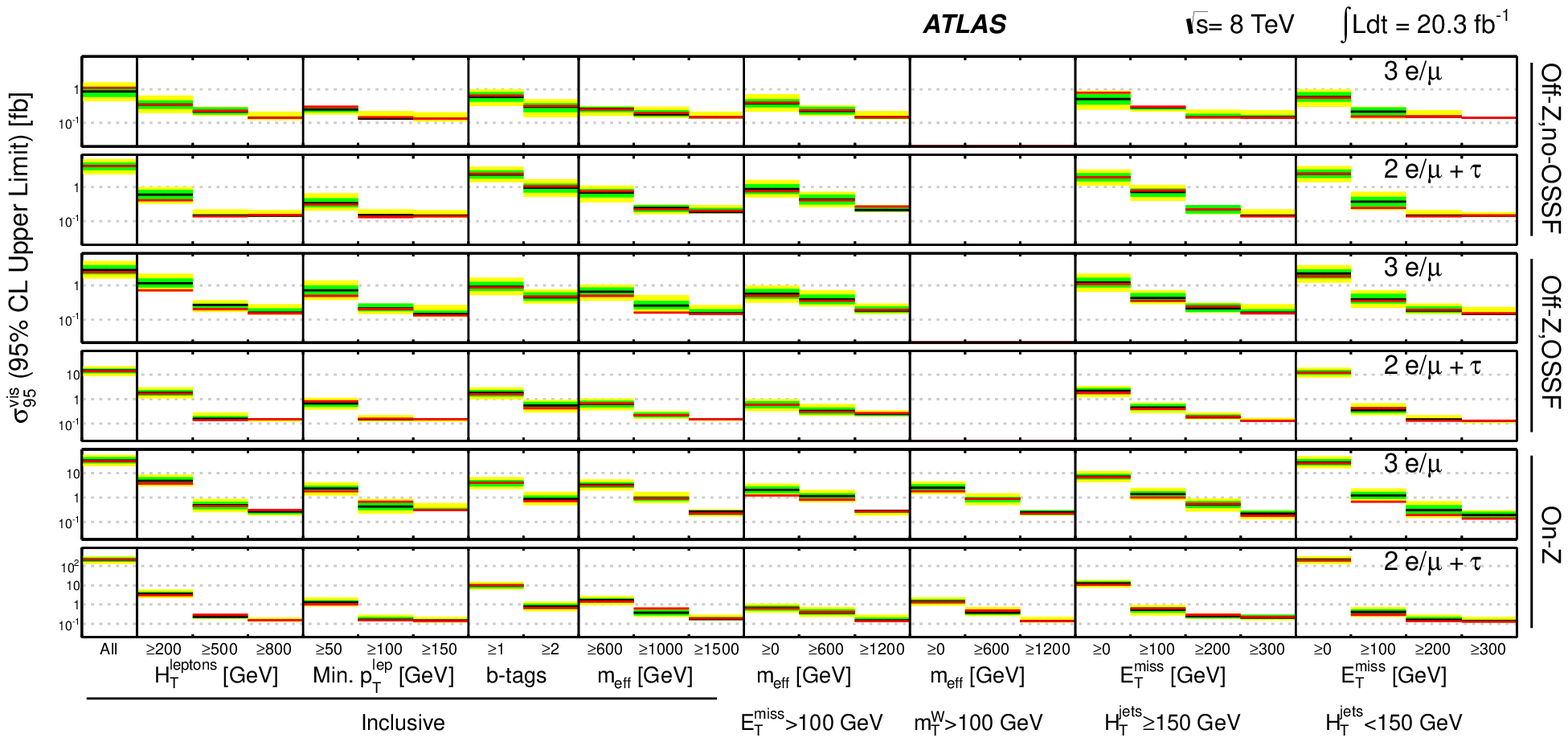}
  \caption{Expected (black) and observed (red) 95\% CL upper limits on the visible cross section from BSM sources for all signal regions.  Confidence intervals of one and two standard deviations on the expected limits are shown as green and yellow bands.}
  \label{fig:summary_limits}
\end{sidewaysfigure*}

\renewcommand{\arraystretch}{1.3}
\begin{table}[tbp]
  \centering
  \begin{tabular}{c c l l c}
    \hline
    \hline
    Channel  &Expected &$\pm 1\sigma$ &$\pm 2\sigma$    &Observed    \\
             &{[fb]}   &[fb]          &[fb]             &[fb]        \\
    \hline
    \multicolumn{5}{c}{off-$Z$, no-OSSF} \\
    \hline
    $\geq 3 e/\mu$            &0.82 &$^{+0.19}_{-0.22}$ &$^{+0.56}_{-0.38}$  &0.89  \\ 
    $2 e/\mu + \geq 1 \tau$   &4.2 &$^{+1.2}_{-1.0}$   &$^{+2.1}_{-1.7}$    &4.3  \\ 
    \hline
    \multicolumn{5}{c}{off-$Z$, OSSF} \\
    \hline
    $\geq 3 e/\mu$            &3.0 &$^{+1.1}_{-0.8}$ &$^{+2.4}_{-1.3}$  &2.5  \\ 
    $2 e/\mu + \geq 1 \tau$   &14.4&$^{+3.2}_{-3.3}$ &$^{+6.2}_{-5.7}$  &14.0  \\ 
    \hline
    \multicolumn{5}{c}{on-$Z$} \\
    \hline
    $\geq 3 e/\mu$            &33  &$^{+11}_{-9}$  &$^{+24}_{-15}$  &31  \\ 
    $2 e/\mu + \geq 1 \tau$   &220 &$^{+50}_{-50}$ &$^{+90}_{-90}$  &207  \\ 
    \hline\hline
  \end{tabular}
\caption{Expected and observed limits on $\sigmavis$ for inclusive signal regions, along with confidence intervals of one and two standard deviations on the expected limits.}
\label{tab:limits_all}
\end{table}
\renewcommand{\arraystretch}{1.0}

\section{Model testing}
\label{sec:ModelTesting}
The model-independent exclusion limits presented in section~\ref{sec:Results} can be re-interpreted in
the scope of any model of new phenomena predicting final states with three or more leptons.  
This section provides a prescription for such re-interpretations.
In order to convert the $\sigmavis$
limits into upper limits on the cross section in a specific model, the fiducial acceptance ($\mathcal{A}$) 
must be known.  The efficiency 
to select signal events within the fiducial volume (fiducial efficiency, or $\epsfid$) is also needed.  
The 95\% CL upper limit on the cross section $\sigma_{95}$ is then given by

\begin{equation}
  \sigma_{95} = \frac{\sigmavis}{\mathcal{A}\times\epsfid}.
\end{equation}

Both $\mathcal{A}$ and \epsfid\ are determined using simulated events at the particle level, i.e. using all particles
after the parton shower and hadronization with mean lifetimes longer than $10^{-11}~\mathrm{s}$.
Event selection proceeds as described in
section~\ref{sec:Selection}, with minor modifications detailed below.  The acceptance is determined by selecting
trilepton events, categorizing them, applying the signal region requirements, and dividing the resulting event yield
by the signal yield before any selection.  The fiducial efficiency is then determined using parameterized
efficiencies provided below.  Events should be generated without pileup -- the effects of pileup are small, and are handled
in the parameterized efficiencies.

Electron and muons are selected using the same $|\eta|$ requirements described in section~\ref{sec:Selection}, but
with a lower $\pt$ requirement of 10~\GeV.  Electrons or muons from tau decays must satisfy the same requirements
as prompt leptons.  
The tau four-momentum at the particle level is defined using only the visible decay products, which include all 
particles except neutrinos.
Hadronically decaying taus are required to have $\ptvis\geq 15\GeV$ and $|\etavis|<2.5$.

Generated electrons and muons are required to be isolated.  A track isolation energy
at the particle level corresponding to \ptisotrack, denoted \ptisotruth, is defined as the scalar sum of transverse momenta of 
charged particles within a cone of $\Delta R = 0.3$ around the lepton axis.  
Particles used in the sum are included after hadronization and must have $\pt>1\GeV$.
A fiducial isolation energy corresponding to \Etisocal, denoted
\Etisotruth, is defined as the sum of all particles inside the annulus $0.1 < \Delta R < 0.3$ around the lepton axis.  
Neutrinos and other stable, weakly interacting particles produced in models of new phenomena are excluded from both \ptisotruth\ and \Etisotruth; muons are excluded
from \Etisotruth.  Electrons and muons must satisfy $\ptisotruth/\pt < 0.15$ and $\Etisotruth/\pt < 0.15$.

A simulated sample of $WZ$ events is used to extract 
the per-lepton efficiencies $\epsilon_{\ell}$.  Generated leptons are matched to reconstructed lepton candidates that satisfy
the selection criteria defined in section~\ref{sec:Selection} by requiring their $\Delta R$ separation be less than 0.1 for prompt electrons and muons, and less than 0.2 for taus.
Reconstructed electrons and muons originating from true tau decays are also required to be within $\Delta R$ of 0.2 of the true lepton from the tau decay.
The per-lepton fiducial efficiency, $\epsilon_{\ell}$, is defined as the ratio of 
the number of reconstructed leptons satisfying all selection criteria to the number of generated leptons within acceptance.
Separate values of $\epsilon_{\ell}$ are measured for each lepton flavour, and $\epsilon_{\ell}$
is determined separately for leptons from tau decays.  The effects of the trigger requirements are folded into the per-lepton efficiencies;
for SM $WZ$ events with both bosons on-shell, the trigger efficiency is over 95\% when all offline selection criteria are applied.

The efficiencies as functions of $\pt$ are shown in table~\ref{t:fideff_num1}, and efficiencies as functions of $|\eta|$ for electrons and taus are shown in table~\ref{t:fideff_num2}.
For empty bins, the value from the preceding filled bin is the suggested central value.
For electrons and taus, the final per-lepton efficiency 
is given as $\epsilon_{\ell} = \epsilon(\pt)\cdot\epsilon(\eta) / \langle \epsilon \rangle$, where
$\langle \epsilon \rangle$ is the inclusive efficiency of the full sample, and is $0.66$ for prompt electrons, $0.39$ for electrons from tau decays, and
$0.26$ for hadronically decaying taus.
The $\eta$ dependence of the muon efficiencies is treated by separate \pt\ efficiency measurements for muons with
$|\eta|<0.1$ and those with $|\eta|\geq0.1$. 

\begin{table*}[tbp]
\begin{center}
\resizebox{\textwidth}{!}{%
 \begin{tabular}{cccccccc} 

 \hline 
 \hline 
\multicolumn{1}{c}{\pt} & Prompt $e$             &\multicolumn{2}{c}{Prompt $\mu$}           & $\tau\rightarrow e$   & \multicolumn{2}{c}{$\tau\rightarrow\mu$} & $\tauh$       \\
$[$\GeV$]$              &                        & $|\eta| > 0.1$       &$|\eta| < 0.1$      &                       &$|\eta| > 0.1$      &$|\eta| < 0.1$       &               \\ 
\hline
 10--15                 & 0.0256 $\pm$  0.0003   & 0.0224 $\pm$ 0.0002& 0.0071 $\pm$ 0.0003  &  0.0086 $\pm$ 0.0006  &                     &                    &                    \\ 
 15--20                 &  0.522 $\pm$  0.005    &  0.839 $\pm$ 0.008 &  0.402 $\pm$ 0.015   &   0.409 $\pm$ 0.029   &   0.62 $\pm$  0.04  &  0.66 $\pm$  0.19  & 0.0311$\pm$  0.0021\\ 
 20--25                 &  0.607 $\pm$  0.005    &  0.887 $\pm$ 0.007 &  0.478 $\pm$ 0.017   &   0.44  $\pm$ 0.04    &   0.66 $\pm$  0.06  &  0.12 $\pm$  0.04  & 0.148 $\pm$  0.012 \\ 
 25--30                 &  0.654 $\pm$  0.005    &  0.910 $\pm$ 0.007 &  0.490 $\pm$ 0.016   &   0.55  $\pm$ 0.04    &   0.68 $\pm$  0.05  &  0.13 $\pm$  0.03  & 0.229 $\pm$  0.018 \\ 
 30--40                 &  0.708 $\pm$  0.004    &  0.919 $\pm$ 0.005 &  0.492 $\pm$ 0.011   &   0.63  $\pm$ 0.04    &   0.71 $\pm$  0.04  &  0.53 $\pm$  0.13  & 0.217 $\pm$  0.013 \\ 
 40--50                 &  0.737 $\pm$  0.005    &  0.923 $\pm$ 0.005 &  0.499 $\pm$ 0.012   &   0.62  $\pm$ 0.05    &   0.74 $\pm$  0.06  &  0.28 $\pm$  0.11  & 0.292 $\pm$  0.025 \\ 
 50--60                 &  0.761 $\pm$  0.005    &  0.925 $\pm$ 0.006 &  0.527 $\pm$ 0.016   &   0.62  $\pm$ 0.06    &   0.71 $\pm$  0.07  &  0.50 $\pm$  0.20  & 0.245 $\pm$  0.026 \\ 
 60--80                 &  0.784 $\pm$  0.005    &  0.925 $\pm$ 0.006 &  0.512 $\pm$ 0.013   &   0.64  $\pm$ 0.07    &   0.78 $\pm$  0.08  &  0.25 $\pm$  0.13  & 0.307 $\pm$  0.032 \\ 
 80--100                &  0.815 $\pm$  0.008    &  0.922 $\pm$ 0.008 &  0.530 $\pm$ 0.020   &   0.72  $\pm$ 0.13    &   0.65 $\pm$  0.10  &  0.50 $\pm$  0.25  & 0.227 $\pm$  0.033 \\ 
100--200                &  0.835 $\pm$  0.008    &  0.918 $\pm$ 0.008 &  0.528 $\pm$ 0.018   &   0.62  $\pm$ 0.11    &   0.75 $\pm$  0.13  &  0.33 $\pm$  0.19  & 0.28  $\pm$  0.04\\ 
200--400                &  0.851 $\pm$  0.021    &  0.884 $\pm$ 0.022 &  0.465 $\pm$ 0.041    \\
400--600                &  0.84  $\pm$  0.10     &  0.83  $\pm$ 0.10  &  0.17  $\pm$ 0.07     \\
$\ge 600$               &  0.90  $\pm$  0.26      \\                                                
 \hline \hline 
 \end{tabular} 
}
\caption{The fiducial efficiency for electrons, muons, and taus in different $\pt$ ranges ($\epsfid(\pt)$).  For electrons and muons from tau decays, the $\pt$ is that of the electron or muon, not the tau.  The uncertainties shown reflect the statistical uncertainties of the simulated samples only. }
 \label{t:fideff_num1} 
 \end{center} 
 \end{table*} 

Table~\ref{t:fideff_num1} includes entries to cover cases where leptons with true $\pt$ below the nominal $\pt$
threshold of 15 (20) \GeV\ for electrons and muons (taus) are reconstructed with $\pt$ above threshold.  These efficiencies
are typically small, but are needed for proper modelling of events with low-$\pt$ leptons.

\begin{table}[tbp]
\begin{center}
  {\small
 \begin{tabular}{cccc} 
 \hline \hline 
 $|\eta|$ & Prompt $e$           & $\tau\rightarrow e$    & $\tauh$ \\
 \hline 
 0.0--0.1 &   0.650 $\pm$ 0.006  &    0.55  $\pm$  0.06   &  0.166 $\pm$  0.017\\ 
 0.1--0.5 &   0.714 $\pm$ 0.004  &    0.500 $\pm$  0.026  &  0.150 $\pm$  0.009\\ 
 0.5--1.0 &   0.722 $\pm$ 0.004  &    0.513 $\pm$  0.026  &  0.188 $\pm$  0.010\\ 
 1.0--1.5 &   0.689 $\pm$ 0.004  &    0.421 $\pm$  0.026  &  0.175 $\pm$  0.010\\ 
 1.5--2.0 &   0.635 $\pm$ 0.004  &    0.470 $\pm$  0.030  &  0.142 $\pm$  0.009\\ 
 2.0--2.5 &   0.615 $\pm$ 0.004  &    0.433 $\pm$  0.032  &  0.109 $\pm$  0.008\\
 \hline \hline 
 \end{tabular} 
}
\caption{The fiducial efficiency for electrons and taus in different $\eta$ ranges ($\epsfid(\eta)$).  For electrons from tau decays, the $\eta$ is that of the electron, not the tau.  The uncertainties shown reflect the statistical uncertainties of the simulated samples only.}

 \label{t:fideff_num2} 
 \end{center} 
 \end{table}

The resulting per-lepton efficiencies are then combined to yield a selection efficiency for a given event satisfying
the fiducial acceptance criteria.  For events with exactly three leptons,
the total efficiency for the event is the product of the individual lepton efficiencies.  For events
with more than three leptons, the additional leptons in order of descending \pt\ only contribute to 
the total efficiency when a lepton with higher \pt\ is not selected, leading to terms such as 
$\epsilon_1\epsilon_2\epsilon_4(1-\epsilon_3)$, where $\epsilon_i$ denotes the fiducial efficiency for
the $i^{th}$ \pt-ordered lepton.  The method can be extended to cover the number of leptons
expected in the model under consideration.

Jets at the particle level are reconstructed from all stable particles, excluding muons and neutrinos, 
with the \antikt\ algorithm using a radius parameter $R=0.4$.  
Overlaps between jets and leptons are removed as described in section~\ref{sec:Selection}.
\met\ is defined as the magnitude of the vector sum of the transverse momenta of all neutrinos and any stable, non-interacting particles
produced in models of new phenomena.  The kinematic variables used to define signal regions are defined as in section~\ref{sec:Selection}.

Predictions of both the rates and kinematic properties of doubly charged Higgs and excited-lepton events, 
when made with the method described above,
agree well with the same quantities after detector simulation.
Uncertainties, based on the level of agreement seen across the studied models, are estimated 
at 10\% for the \threeL\  channels, and 20\% for the \twoLtau\  channels.  
When calculating limits on specific models, these uncertainties must be applied to the
estimated signal yields after selection to take into account the limited precision of
the fiducial efficiency approach.

\section{Interpretation}
\label{sec:Interpretation}
The results of the model-independent search are interpreted in the context of two specific models
of new phenomena: a model with pair-produced doubly charged Higgs bosons, and a
model with excited, non-elementary leptons.

Doubly charged Higgs bosons
can be either pair-produced or produced in association with a singly charged state.  
In this paper, the \dchp\ are assumed to be pair-produced, with decays to charged leptons.  One
feature of most models with \dchp\ is the presence of lepton-flavour-violating terms, leading to decays
such as $\dchp\rightarrow e^{\pm}\mu^\pm$ in addition to $\dchp\rightarrow e^\pm e^\pm$ or $\dchp\rightarrow\mu^{\pm}\mu^{\pm}$.
Decays to electrons and/or muons have been probed at $\sqrt{s}=8$~\TeV{} in ref.~\cite{atlasDCH}, while decays to all flavours of leptons are probed
at $\sqrt{s}=$7~\TeV{} in ref.~\cite{cmsDCH}.  In this paper, only the lepton-flavour-violating decays $\dchp\to e^{\pm}\tau^{\pm}$ 
and $\dchp\to\mu^{\pm}\tau^{\pm}$ are considered.

The visible cross-section limits presented above are used to constrain this model.  The off-$Z$, OSSF category provides
the largest acceptance for the lepton-flavour-violating decays; contributions from the remaining categories are small and
have a negligible impact on the sensitivity.  The signal regions based on \htlep\ provide the best expected sensitivity,
followed by limits based on \minleppt; here only limits based on \htlep\ are used.  For \dchp\ masses up to 200 GeV, the
signal region defined by $\htlep>200\GeV$ is used; for higher masses the requirement is $\htlep>500\GeV$.  Finally, both the
\threeL\ and \twoLoneT\ channels are used to maximize the total acceptance.

Table~\ref{tab:dch} summarizes the expected acceptance, efficiency, and cross-section limit for several mass values,
channels, and decay scenarios.  The \threeL\ and \twoLoneT\ channels have comparable sensitivity for high masses, and
are therefore combined when setting the final limits to improve the overall constraint on this model.  
The $\dchp$ can couple preferentially to left-handed ($\dchpl$) or right-handed ($\dchpr$) leptons, with the production cross
section for the right-handed coupling scenario being roughly half that for the left-handed coupling scenario.  
The acceptance and efficiency are the same for both couplings.  The final limits
on $\dchp\rightarrow e^{\pm}\tau^{\pm}$ and $\dchp\rightarrow \mu^{\pm}\tau^{\pm}$ for both scenarios are shown in figure~\ref{fig:dch}.
In both cases, a branching ratio of 100\% is assumed for the chosen decay.
For $\dchp\rightarrow e^{\pm}\tau^{\pm}$, the expected mass limit for left-handed couplings is 350$\pm$50~\GeV, with an observed limit of 400~\GeV.
For $\dchp\rightarrow \mu^{\pm}\tau^{\pm}$, the expected mass limit for left-handed couplings is 370$^{+20}_{-40}$~\GeV, with an observed limit at 400~\GeV.
The expected (observed) limit on $\dchp\rightarrow \mu^{\pm}\tau^{\pm}$ from the 7~\TeV\ ATLAS analysis~\cite{atlasML2011} 
is 229 (237) ~\GeV, which only uses the \threeL\ channel.  The corresponding observed limits from the 7~\TeV\ CMS analysis~\cite{cmsDCH}
are 293~\GeV{} for $\dchp\rightarrow e^{\pm}\tau^{\pm}$ and 300~\GeV{} for $\dchp\rightarrow \mu^{\pm}\tau^{\pm}$.

\begin{table*}
\renewcommand{\arraystretch}{1.25}
\resizebox{\textwidth}{!}{%
\begin{tabular}{| c | c | c c | c c | c c |}
  \hline
  \hline
                                &           &\multicolumn{6}{c|}{\dchp\ mass and decay mode}\\
  \cline{3-8}
                                &           &\multicolumn{2}{c|}{100 GeV} &\multicolumn{2}{c|}{300 GeV} &\multicolumn{2}{c|}{500 GeV}\\
                                &Channel    &\multicolumn{1}{c}{$e\tau$} &\multicolumn{1}{c|}{$\mu\tau$} &\multicolumn{1}{c}{$e\tau$} &\multicolumn{1}{c|}{$\mu\tau$} &\multicolumn{1}{c}{$e\tau$} &\multicolumn{1}{c|}{$\mu\tau$}\\
  \hline
  \hline
  $\sigma$ [fb]                 & Combined       &\multicolumn{2}{c|}{504}  &\multicolumn{2}{c|}{5.55} &\multicolumn{2}{c|}{0.396}\\
  \hline
  \multirow{2}{*}{$\mathcal{A}$}
                                &\threeL    & 0.14  $\pm$   0.01 & 0.15  $\pm$   0.01 &0.35  $\pm$    0.01 &0.38  $\pm$    0.02 &0.39  $\pm$    0.02 &0.45  $\pm$     0.02\\ 
                                &\twoLoneT  & 0.33  $\pm$    0.01 &0.36  $\pm$    0.01 &0.48  $\pm$    0.02 &0.49  $\pm$    0.02 &0.49  $\pm$    0.02 &0.47  $\pm$    0.02\\
  \hline
  \multirow{2}{*}{$\epsfid$}
                                &\threeL    & 0.24  $\pm$    0.02 &0.26  $\pm$    0.02 &0.36  $\pm$     0.02 &0.37  $\pm$    0.01 &  0.40  $\pm$    0.02 &0.37  $\pm$    0.01\\
                                &\twoLoneT  & 0.21  $\pm$   0.01 &0.24  $\pm$   0.01 &0.29  $\pm$    0.01 &0.31  $\pm$    0.01 &0.32  $\pm$    0.01 &0.31  $\pm$    0.01\\
  \hline
  \multirow{2}{*}{$\mathcal{A}\times\epsfid$}
                                &\threeL    & 0.034  $\pm$   0.002 & 0.039  $\pm$   0.003 &  0.12  $\pm$   0.01 &  0.14  $\pm$   0.01 &  0.16  $\pm$   0.01 &  0.17  $\pm$    0.01\\

                                &\twoLoneT  & 0.071  $\pm$   0.003 &  0.087  $\pm$   0.004 &  0.14  $\pm$   0.01 &  0.15  $\pm$   0.01 &  0.15  $\pm$   0.01 &  0.14  $\pm$   0.01\\

  \hline
  \multirow{2}{*}{Rec. $\mathcal{A}\times\epsilon$}
                                &\threeL    &0.034  $\pm$   0.004 &0.046  $\pm$   0.005 & 0.12  $\pm$   0.01 & 0.12  $\pm$   0.01 & 0.13  $\pm$   0.01 & 0.14  $\pm$   0.01\\

                                &\twoLoneT  &0.062  $\pm$   0.006 &0.083  $\pm$   0.007 & 0.14  $\pm$      0.01 & 0.16  $\pm$    0.01 & 0.16  $\pm$    0.01 & 0.18  $\pm$    0.01\\
  \hline
  \multirow{3}{*}{Exp. Limit [fb]}   &\threeL    &\multicolumn{1}{c}{$53^{+26}_{-17}$}  &\multicolumn{1}{c|}{$54^{+25}_{-17}$}  &\multicolumn{1}{c}{$ 5.0^{+ 2.6}_{- 0.9}$}  &\multicolumn{1}{c|}{$ 6.7^{+ 3.0}_{- 1.9}$}  &\multicolumn{1}{c}{$ 2.7^{+ 1.4}_{- 0.7}$}  &\multicolumn{1}{c|}{$ 2.3^{+ 1.0}_{- 0.6}$}\\
                                     &\twoLoneT  &\multicolumn{1}{c}{$54^{+21}_{-14}$}  &\multicolumn{1}{c|}{$38^{+14}_{-10}$}  &\multicolumn{1}{c}{$ 2.6^{+ 0.4}_{- 0.2}$}  &\multicolumn{1}{c|}{$ 2.4^{+ 0.4}_{- 0.2}$}  &\multicolumn{1}{c}{$ 1.3^{+ 0.5}_{- 0.2}$}  &\multicolumn{1}{c|}{$ 1.1^{+ 0.5}_{- 0.2}$}\\
                                     &Combined   &\multicolumn{1}{c}{$42^{+18}_{-12}$}  &\multicolumn{1}{c|}{$34^{+14}_{-~9}$}  &\multicolumn{1}{c}{$ 2.6^{+ 0.4}_{- 0.2}$}  &\multicolumn{1}{c|}{$ 2.6^{+ 1.0}_{- 0.4}$}  &\multicolumn{1}{c}{$ 1.2^{+ 0.5}_{- 0.2}$}  &\multicolumn{1}{c|}{$ 1.1^{+ 0.4}_{- 0.2}$}\\
  \hline
  \multirow{3}{*}{Obs. Limit [fb]}   &\threeL    &\multicolumn{1}{c}{32}              &\multicolumn{1}{c|}{32}               &\multicolumn{1}{c}{ 3.2}                  &\multicolumn{1}{c|}{ 4.2}                   &\multicolumn{1}{c}{ 1.7}                  &\multicolumn{1}{c|}{ 1.5}\\
                                     &\twoLoneT  &\multicolumn{1}{c}{51}              &\multicolumn{1}{c|}{36}               &\multicolumn{1}{c}{ 2.4}                  &\multicolumn{1}{c|}{ 2.2}                   &\multicolumn{1}{c}{ 1.2}                  &\multicolumn{1}{c|}{ 1.0}\\
                                     &Combined   &\multicolumn{1}{c}{28}              &\multicolumn{1}{c|}{24}               &\multicolumn{1}{c}{ 2.4}                  &\multicolumn{1}{c|}{ 1.9}                   &\multicolumn{1}{c}{ 0.8}                  &\multicolumn{1}{c|}{ 0.7}\\
  \hline
  \hline
\end{tabular}
}
\caption{Theoretical cross section and the acceptances, efficiencies and 95\% CL upper limits on the cross section for pair-produced \dchp\ decaying
to $e^{\pm}\tau^{\pm}$ and $\mu^{\pm}\tau^{\pm}$.  Rec. $\mathcal{A}\times\epsilon$ represents the fraction of signal events passing
all analysis cuts after detector-level simulation and event reconstruction.}
\label{tab:dch}
\end{table*}

\begin{figure*}
\subfigure[$\dchp\to e^{\pm}\tau^{\pm}$]{\includegraphics[width=0.48\columnwidth]{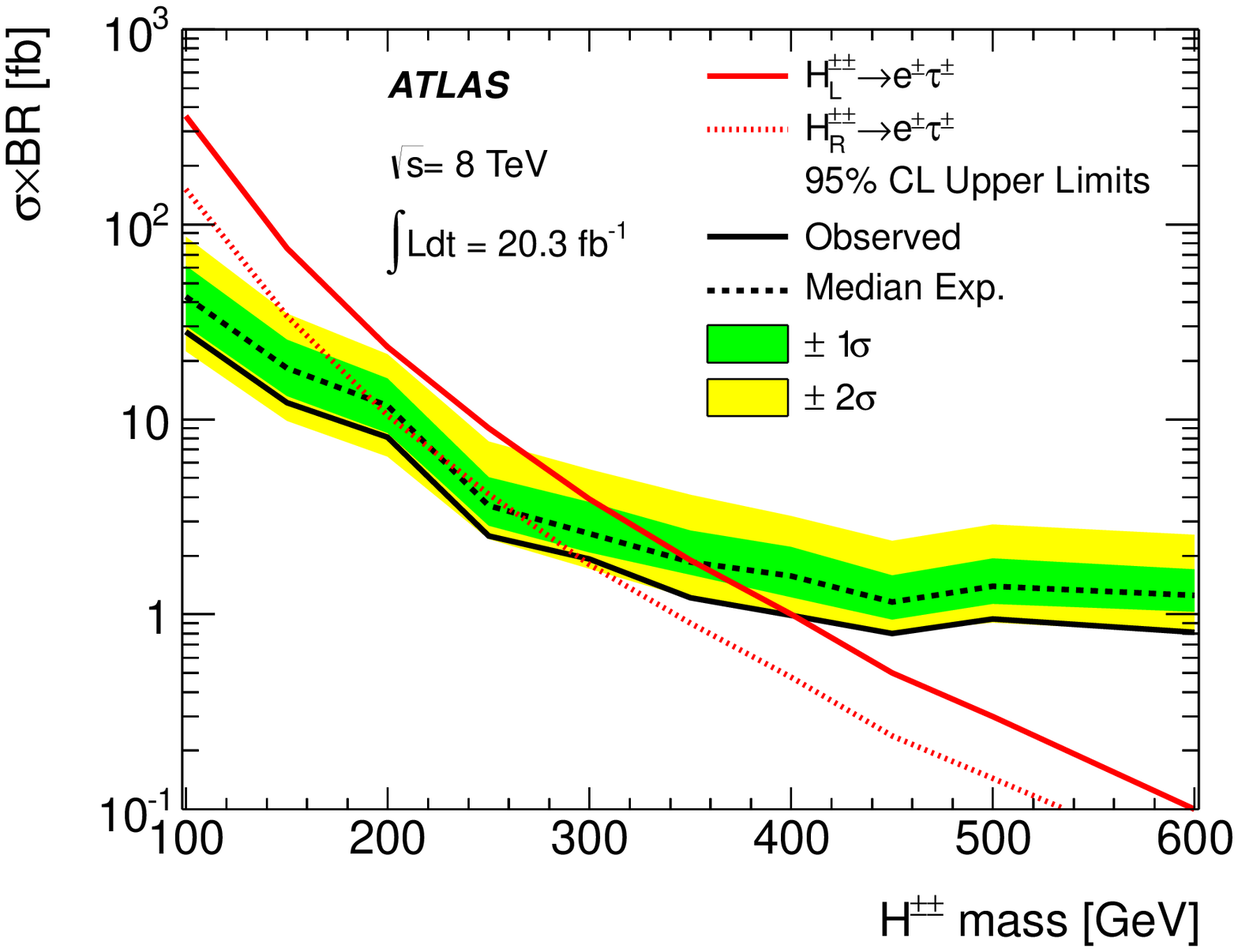}}
\subfigure[$\dchp\to\mu^{\pm}\tau^{\pm}$]{\includegraphics[width=0.48\columnwidth]{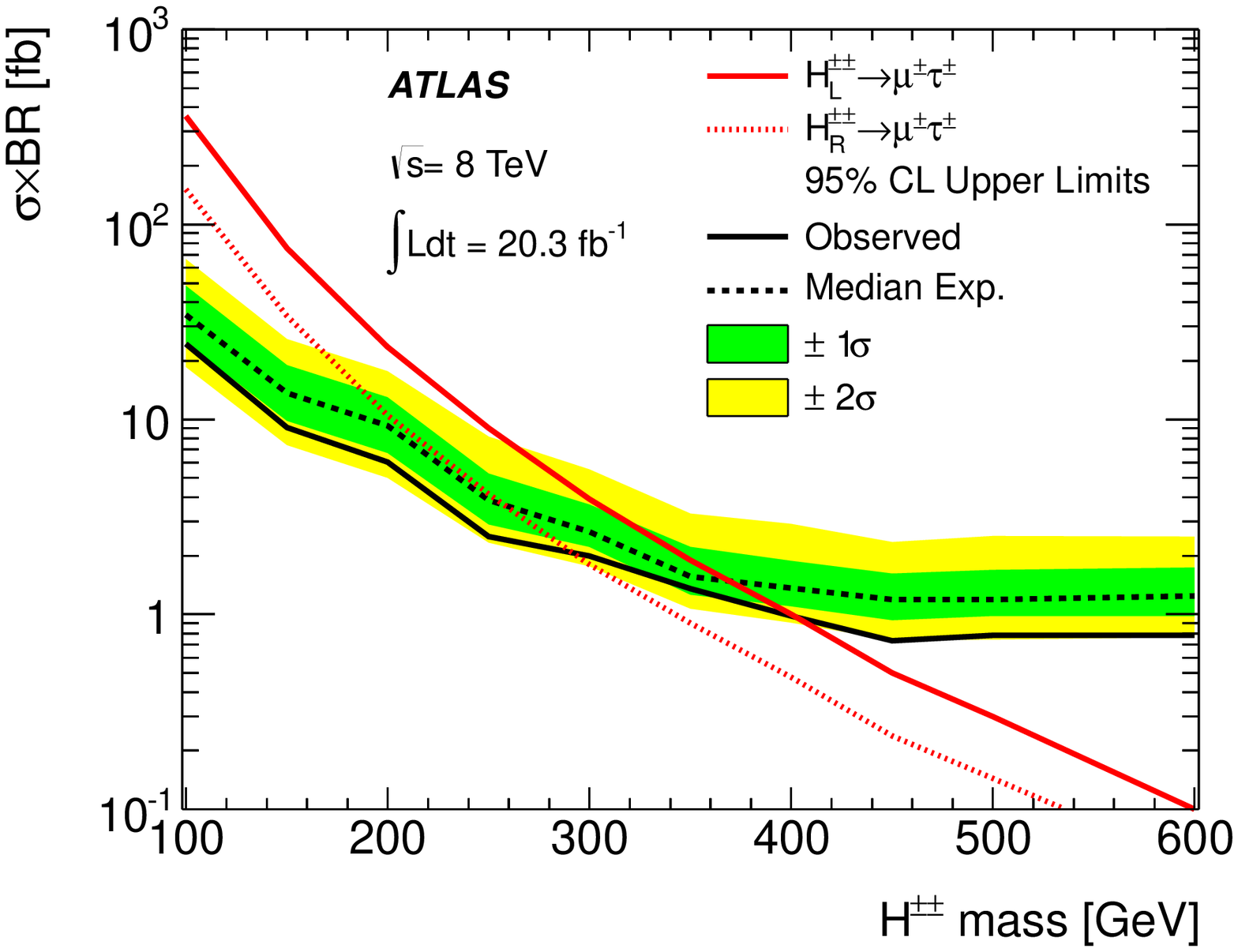}}
\caption{Observed and expected 95\% upper limits on the cross section times branching ratio for $\dchp$ decaying to (a) $e^{\pm}\tau^{\pm}$ and (b) $\mu^{\pm}\tau^{\pm}$.  Separate mass constraints are extracted for $\dchp$ coupling to left- and right-handed fermions from the intersections with the predicted cross sections shown by the dotted and solid red curves.}
\label{fig:dch}
\end{figure*}

Composite fermion models often imply the existence of excited-lepton states~\cite{ExcitedLeptons}.
Excited leptons are either pair-produced, produced in association with another excited lepton of a different flavour,
 or produced in association with a SM lepton~\cite{excitednu,excitednu2}.
The production is mediated either by gauge bosons (gauge-mediated, GM) or by auxiliary, massive fields that can be
approximated as a four-fermion contact interaction (CI) vertex.
The scales of the CI and GM processes are assumed to be identical and called $\Lambda$,
while the masses of the excited leptons are referred to as \exlepmass{}.
The CI process dominates the production and decay of excited leptons
for $\exlepmass/\Lambda > 0.3$, while for lower values the GM process becomes important. 
Additionally, the parameters $f_{\mathrm{s}}$, $f$ and $f^{\prime}$, corresponding to the SU(3), SU(2) and U(1)
couplings of the model respectively, can be chosen arbitrarily and dictate the dynamics of the model.  For this study, all coupling
parameters are set to unity, as used in ref.~\cite{ExcitedLeptons}. This specific choice of $f = f^{\prime}$ forbids
the radiative decays of excited neutrinos.  

Searches for excited electrons and muons have been performed using a similar benchmark model by CMS~\cite{cmsexcited}, at $\sqrt{s}=7~\TeV$, and by ATLAS~\cite{ATLASExcitedL}, with 13~\ifb{} at $\sqrt{s}=8$~\TeV{}.  The most stringent lower limits on \exlepmass{} from these searches are at 2.2~\TeV{} for $\Lambda=\exlepmass$.
Lower limits on the mass of excited leptons were set by the L3 experiment.  These limits, which are independent of $\Lambda$, range from 91~\GeV{} to 102~\GeV{}, with limits on excited 
taus and excited tau neutrinos being somewhat weaker than those for other flavours~\cite{L3result}.  

The decay products for each excited neutrino are a neutrino (or charged lepton) of the same generation and a $Z$ ($W$) boson, or a fermion pair.
Similarly, excited charged leptons can decay into a charged lepton (or neutrino) of the same generation and a $\gamma/Z$ (or $W$) boson,
or into a fermion pair. For excited neutrinos, only the pair production of two excited neutrinos $\nu$*$\bar{\nu}$* is taken into account;
single production of excited neutrinos producing final states with three or more leptons is suppressed and its contribution is negligible. For
the excited charged leptons, both single and pair production of excited states are taken into account.

The upper limits on the visible cross section can be used to constrain $\exlepmass$ and $\Lambda$.
In all cases, the signal region with the best expected sensitivity is used to constrain each scenario.
In the cases where the excited charged lepton or neutrino masses are large, the decay products typically carry a
large amount of momentum. This leads to signal events with large $\htlep$.  Additionally, in this regime,
the GM decay through $Z$ bosons is disfavoured compared to the CI decay.  Consequently, for such scenarios,
the off-$Z$ channel provides better sensitivity due to lower background rates.

The production of excited electrons, excited muons, and excited electron and muon neutrinos is constrained using the $\geq 3 e/\mu$, off-$Z$, OSSF region requiring $\htlep>800\GeV$ ($\htlep>500\GeV$) for masses above (below) 600 GeV. 
Excited tau neutrinos with high values of $\exlepmass/\Lambda$ are constrained using the $\geq 3 e/\mu$, off-$Z$, OSSF region requiring $\meff>1.5\TeV$.
The only excited tau neutrino decays that preferentially produce final states with taus are the GM decays via a $W$ boson, which
become significant at lower values of $\exlepmass/\Lambda$.  For such cases, the $2 e/\mu + \geq 1 \tau_{\mathrm{had}}$, off-$Z$,
no-OSSF region requiring $\minleppt > 100 \GeV$ is used.

For excited taus, the $\geq 3 e/\mu$, off-$Z$, OSSF region requiring $\meff>1.5\TeV$ is used for masses above 1~\TeV.
For masses between 500~\GeV{} and 1~\TeV{}, the $\geq 3 e/\mu$, off-$Z$, OSSF regions requiring $\meff>1\TeV$ is used.
For masses below 500~\GeV, where the GM decay through $Z$ bosons again becomes significant, the $2 e/\mu + \geq 1 \tau_{\mathrm{had}}$,
on-$Z$ region requiring $\minleppt>100\GeV$ is most sensitive.

Table~\ref{tab:exlep} summarizes the expected acceptance and efficiency for several flavours, mass values and $\Lambda$ values for the most sensitive signal region.
Figure~\ref{Excited_neutrinos_Limits_nu} shows the excluded regions of the mass parameter and the scale $\Lambda$ for all lepton flavours extracted from the expected 
and observed upper limits on the visible cross section.  Exclusion regions are also shown for the case where excited leptons are only produced via the CI process.

\begin{table*}
\renewcommand{\arraystretch}{1.25}
\resizebox{\textwidth}{!}{%
\begin{tabular}{| c | r | r | r@{ $\pm$ }l | r@{ $\pm$ }l | r@{ $\pm$ }l | r@{ $\pm$ }l | c |}
  \hline
  \hline
              &\multicolumn{1}{|c|}{\exlepmass}                 &\multicolumn{1}{|c|}{$\sigma$}   &\multicolumn{2}{|c|}{\multirow{2}{*}{$\mathcal{A}$}}  &  \multicolumn{2}{|c|}{\multirow{2}{*}{$\epsfid$}}  &  \multicolumn{2}{|c|}{\multirow{2}{*}{$\mathcal{A}\times\epsfid$}}  &\multicolumn{2}{|c|}{\multirow{2}{*}{Rec. $\mathcal{A}\times\epsilon$}} &Limit\\
              &\multicolumn{1}{|c|}{ [GeV] }                        &\multicolumn{1}{|c|}{ [fb] }          &\multicolumn{2}{|c|}{}  &\multicolumn{2}{|c|}{} &\multicolumn{2}{|c|}{} &\multicolumn{2}{|c|}{} & [fb]\\
  \hline 
  \multicolumn{12}{c}{$\Lambda = 4$ \TeV}\\
  
  \hline                             
                                
   $\nu_e^{*}\bar{\nu}_e^{*}$           &  500   &  127     & 0.036  &   0.001 & 0.63  &    0.07 & 0.023  &   0.003  &  0.023  &   0.001           &6.5  \\
   $\nu_e^{*}\bar{\nu}_e^{*}$           &  1500  &   0.562  & 0.041 &   0.001 & 0.66 &    0.07 & 0.027 &   0.003 &   0.027 &   0.001            &5.6 \\
   $\nu_{\mu}^{*}\bar{\nu}_{\mu}^{*}$    &  500   &  127     & 0.036  &    0.001 & 0.51  &     0.06 & 0.018  &   0.003 &    0.022  &   0.001        &6.8 \\
   $\nu_{\mu}^{*}\bar{\nu}_{\mu}^{*}$    &  1500  &  0.562   & 0.039  &    0.001 & 0.52  &    0.06 & 0.020  &   0.004 &    0.025  &   0.001       &6.0 \\
   $\nu_{\tau}^{*}\bar{\nu}_{\tau}^{*}$  &  500   &  127     & 0.0022  & 0.0003 & 0.43 &   0.05 &  0.0009  &   0.0003 &   0.0010  &   0.0002        &150  \\
   $\nu_{\tau}^{*}\bar{\nu}_{\tau}^{*}$  &  1500  &  0.562   & 0.014  &     0.001 & 0.52  &    0.06 & 0.007  &    0.002 &   0.008  &    0.001     &19  \\
   $\tau^{*}\bar{\tau}^{*}$            &  500   &   127    & 0.0011  &     0.0002 & 0.40 &   0.04 &  0.0004  &    0.0001 &    0.0002  &    0.0001 &750 \\
   $\tau^{*}\bar{\tau}^{*}$            &  1500  &  0.562   & 0.027  &     0.001 & 0.29  &    0.03 &  0.008  &    0.002  &   0.006  &    0.001    &25 \\
   $\tau^{*}\bar{\tau}$                &  500   &   276   & 0.0012  &     0.0002 & 0.47 &   0.05 &  0.0006  &    0.0002 &   0.0007  &    0.0002  &210 \\
   $\tau^{*}\bar{\tau}$                &  1500  &  1.41   & 0.032  &     0.001 & 0.48  &   0.05 &  0.015  &    0.002 &    0.015  &    0.001     &10  \\
     
  \hline  
   \multicolumn{12}{c}{$\Lambda = 10$ \TeV} \\
   \hline
   $\nu_e^{*}\bar{\nu}_e^{*}$           &  500   &  3.24    & 0.044   &   0.001 & 0.61  &    0.07 & 0.027  &   0.004  &  0.030  &   0.001  &5.0\\
   $\nu_e^{*}\bar{\nu}_e^{*}$           &  1500  &   0.015  & 0.088   &   0.002 & 0.66 &    0.07 & 0.058 &   0.007 &   0.056 &   0.002   &2.7\\
   $\nu_{\mu}^{*}\bar{\nu}_{\mu}^{*}$    &  500   &  3.24    & 0.041   &    0.001 & 0.54  &     0.06 & 0.022  &   0.003 &    0.028  &   0.001  &5.4\\
   $\nu_{\mu}^{*}\bar{\nu}_{\mu}^{*}$    &  1500  &  0.015   & 0.084   &    0.002 & 0.50  &    0.05 & 0.042  &   0.006 &    0.052  &   0.002  &2.9\\
   $\nu_{\tau}^{*}\bar{\nu}_{\tau}^{*}$  &  500   &  3.24    & 0.0020  &    0.0006 & 0.19 &   0.02 &  0.0004  &   0.0002 &   0.0005  &   0.0001  &300\\
   $\nu_{\tau}^{*}\bar{\nu}_{\tau}^{*}$  &  1500  &  0.015   & 0.012   &     0.002 & 0.36  &    0.04 & 0.0043  &    0.0008 &   0.0045  &    0.0004  &33\\
   $\tau^{*}\bar{\tau}^{*}$            &  500   &   3.24   & 0.0002  &     0.0001 & 0.33 &   0.04 &  0.0001  &    0.0001 &    0.0001  &    0.0001 &1500\\
   $\tau^{*}\bar{\tau}^{*}$            &  1500  &  0.015   & 0.0070  &     0.0001 & 0.17  &    0.02 &  0.0012  &    0.0007  &   0.0022  &    0.0003 &68\\
   $\tau^{*}\bar{\tau}$               &  500    &   3.81   & 0.0003  &     0.0001 & 0.53 &   0.06 &  0.0002  &    0.0002 &   0.0002  &    0.0002 &750\\
   $\tau^{*}\bar{\tau}$               &  1500   &  0.022   & 0.012   &     0.001 & 0.48  &   0.05 &  0.0056  &    0.0015 &    0.0048  &    0.0004 &31\\
  \hline
  \hline
\end{tabular}}
\caption{Cross section, acceptances, efficiencies, and 95\% CL upper limits on the cross section for various excited-lepton flavours and mass values using the $\geq 3 e/\mu$, off-$Z$, OSSF region requiring $\htlep > 800 \GeV$.  The observed limit is equal to the expected limit in this signal region.  Rec. $\mathcal{A}\times\epsilon$ represents the fraction of signal events passing
all analysis cuts after detector-level simulation and event reconstruction.}
\label{tab:exlep}
\end{table*} 

\begin{figure}[tbp]
  \centering    
  \subfigure[$e^*$]{\includegraphics[width=0.48\columnwidth]{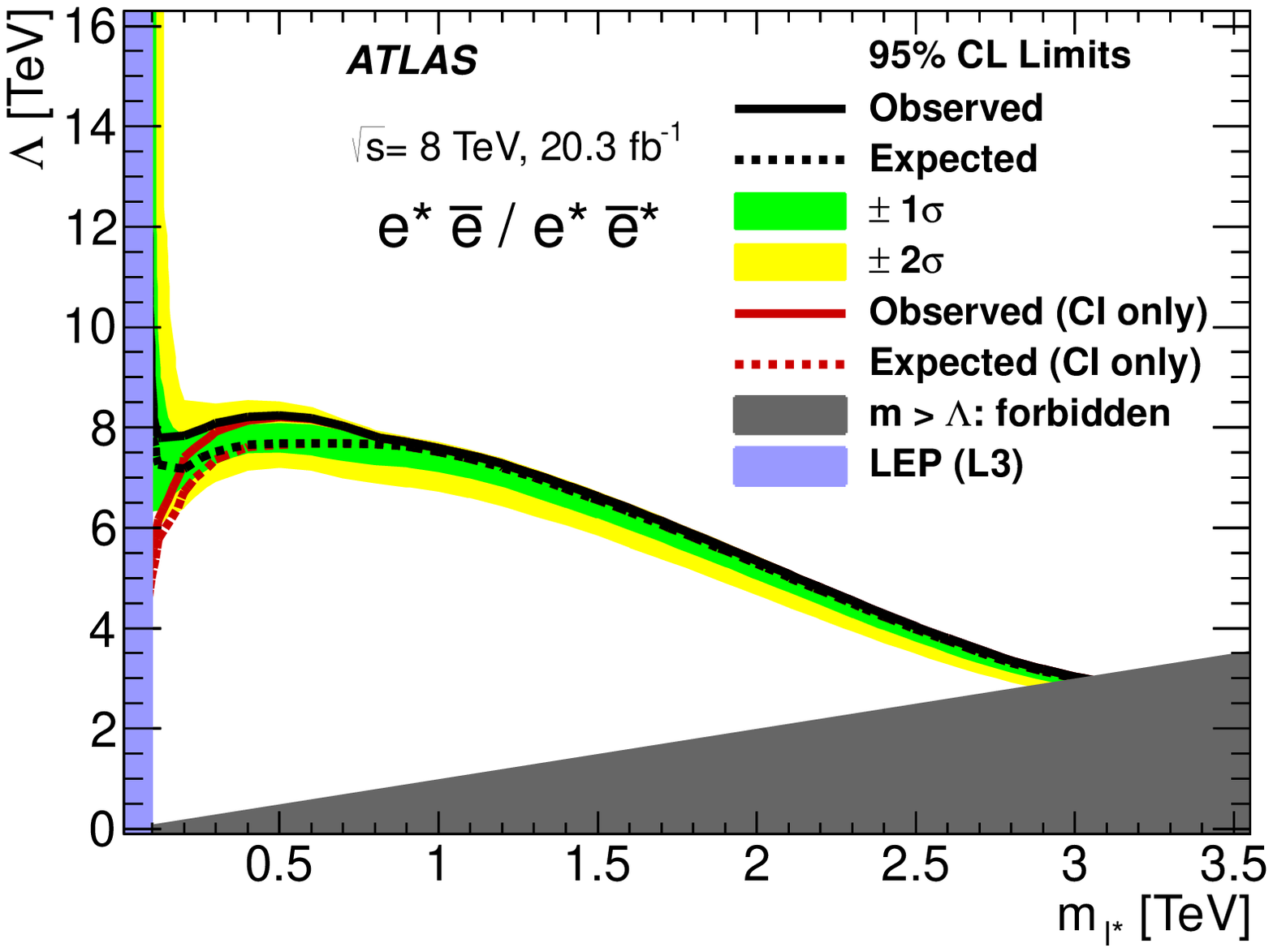}}
  \subfigure[$\nu_e^*$]{\includegraphics[width=0.48\columnwidth]{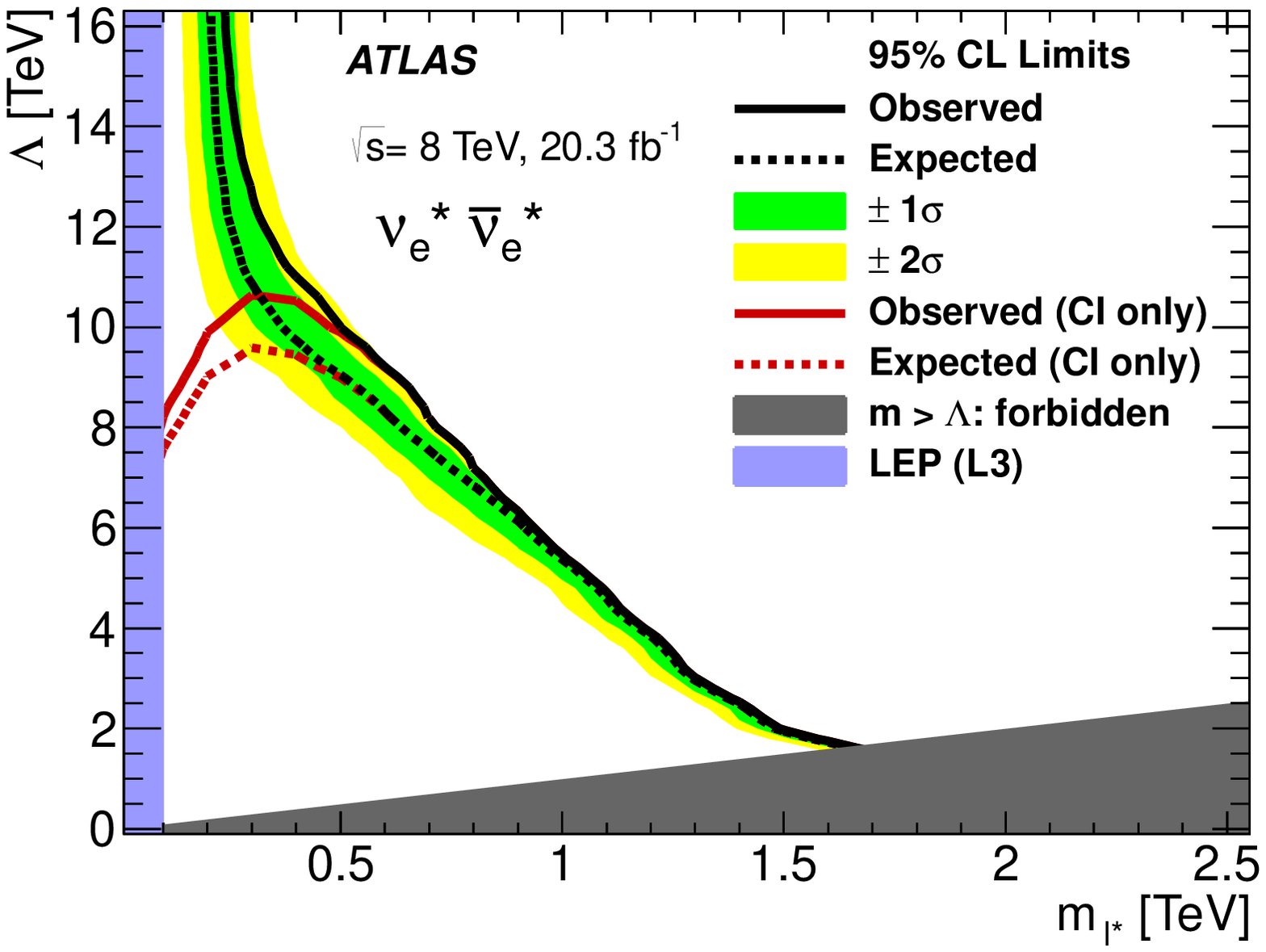}}
  \subfigure[$\mu^*$]{\includegraphics[width=0.48\columnwidth]{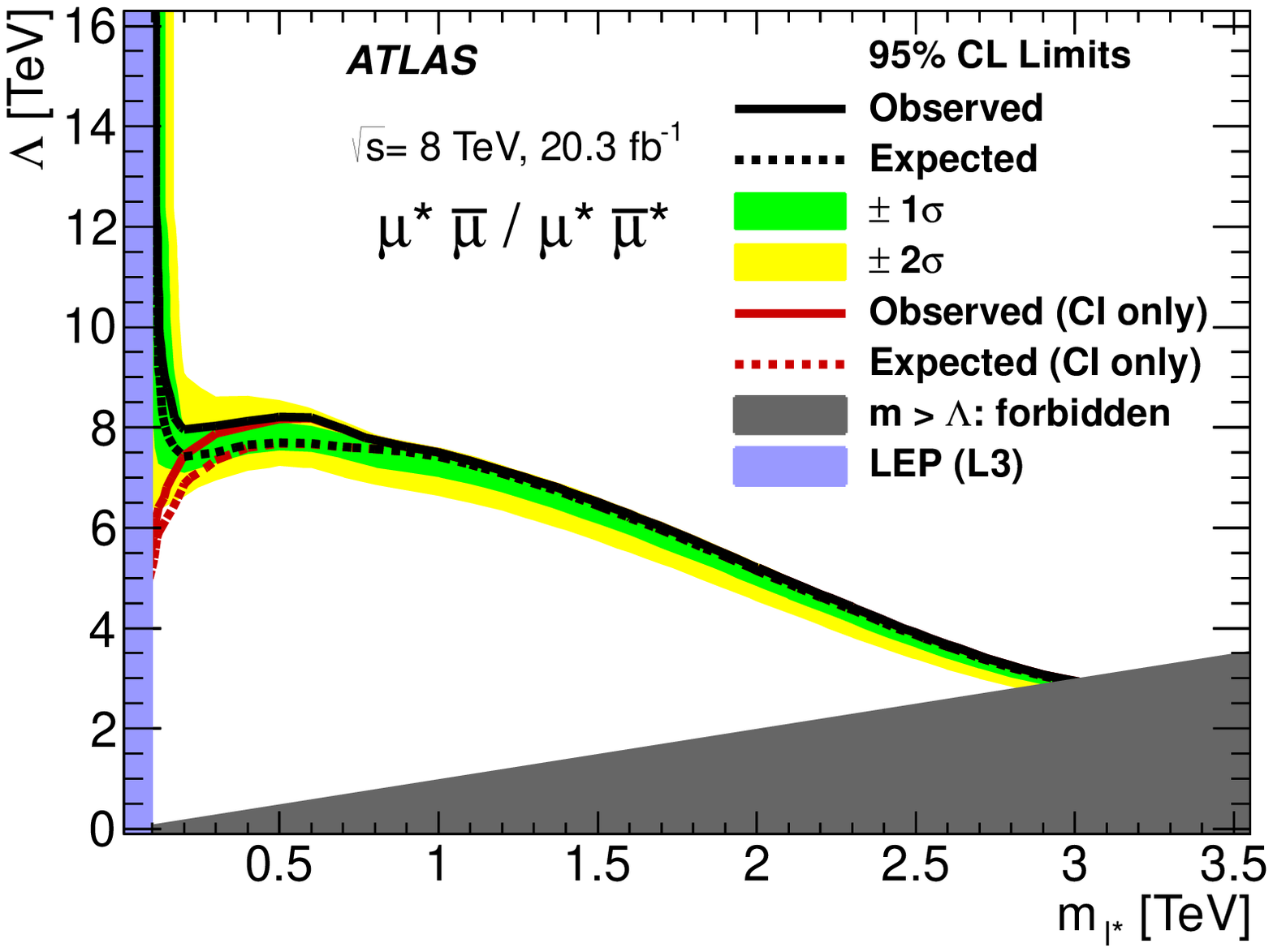}}
  \subfigure[$\nu_\mu^*$]{\includegraphics[width=0.48\columnwidth]{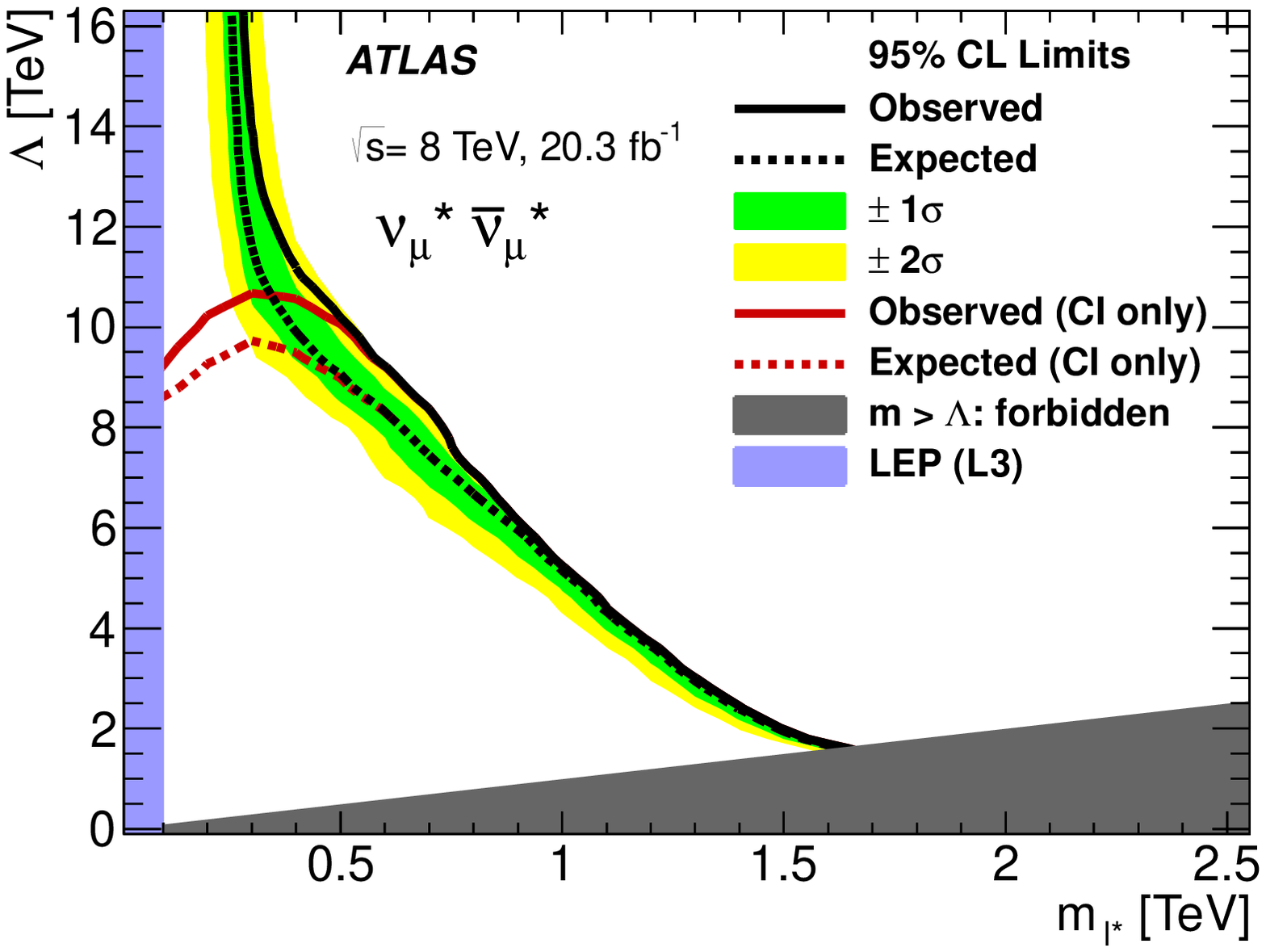}}
  \subfigure[$\tau^*$]{\includegraphics[width=0.48\columnwidth]{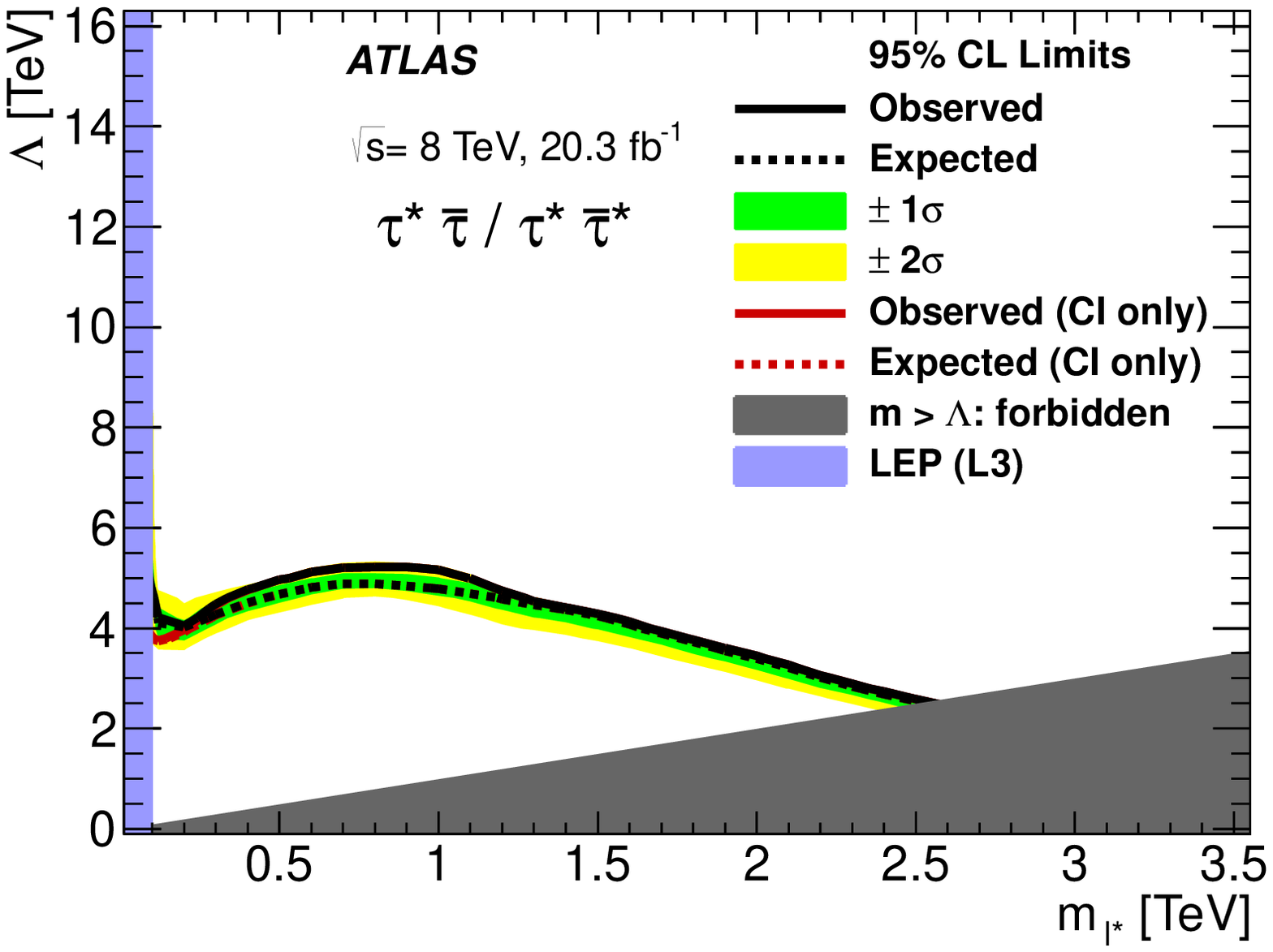}}
  \subfigure[$\nu_\tau^*$]{\includegraphics[width=0.48\columnwidth]{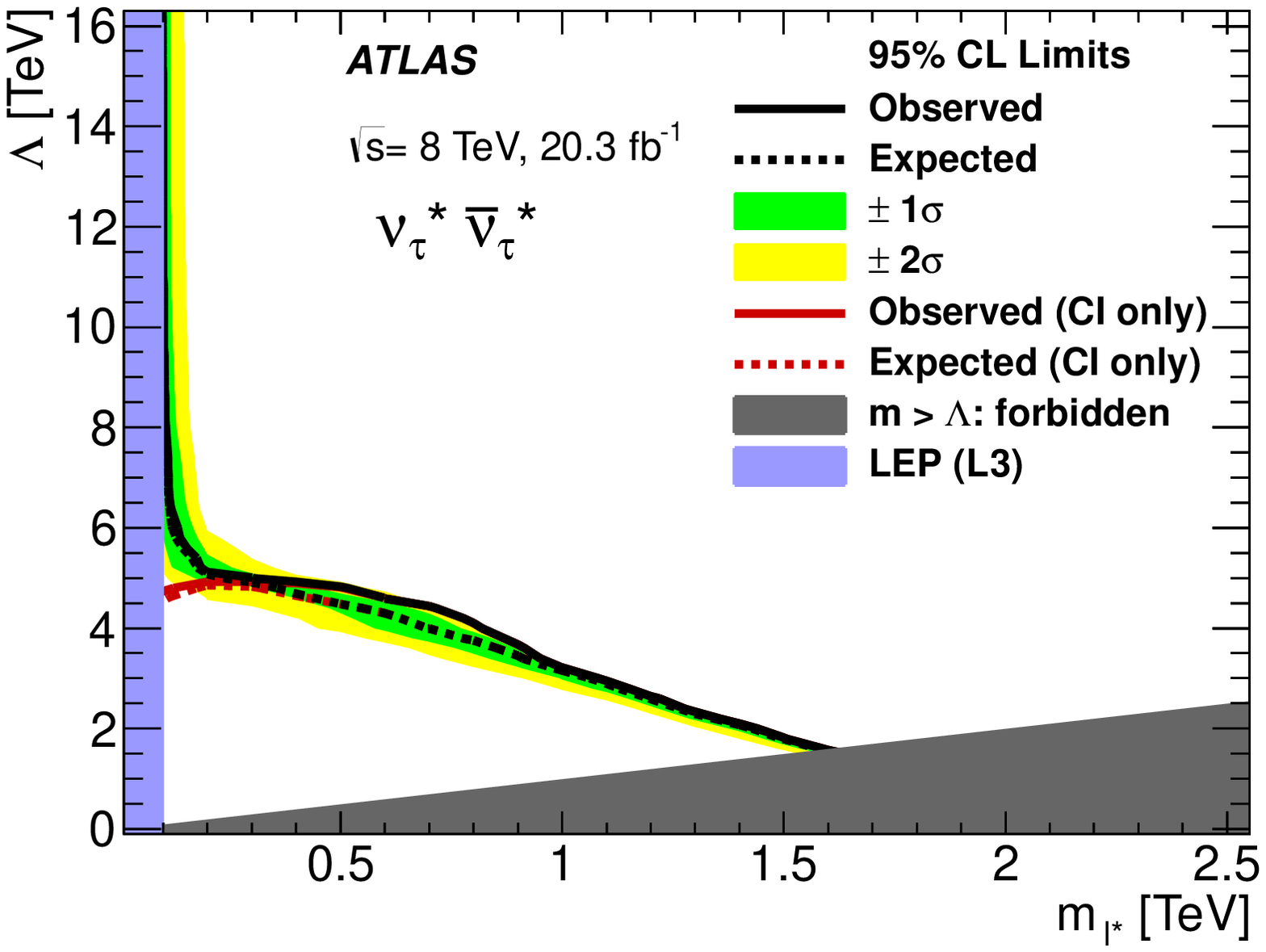}}
  
  \caption{Observed and expected 95\% CL limits on the mass parameter and the
    compositeness scale $\Lambda$ for excited leptons. The region under the curve is excluded by this analysis, the blue region is excluded by LEP, and the gray region represents $\exlepmass>\Lambda$ and is unphysical.  The red line shows the limits taking only CI production into account.}
  \label{Excited_neutrinos_Limits_nu}
\end{figure}

For low $\Lambda$-values, a broad range of masses up to 2~\TeV{} can be excluded, while for higher $\Lambda$-values, 
only low masses are excluded. In the low-mass region, $\nu_{\ell}^*\rightarrow \ell + W$ is the main decay mode for excited neutrinos, while 
$\ell^*\rightarrow \ell + \gamma$ is the main decay mode for charged leptons. Therefore, pair-produced $\nu_{e}^*$ and $\nu_{\mu}^*$ have the 
highest acceptance due to their final states with at least three leptons, and thus they have the most stringent limits. 

The production cross section of pair-produced excited leptons via the GM process is independent of $\Lambda$, which leads to improved sensitivity 
at low excited-lepton masses.  
The low efficiency for reconstructing tau leptons leads to a relatively small gain in sensitivity for $\nu_\tau^*$ from GM production.

For $\nu_e^*$ ($\nu_\mu^*$), the expected $\Lambda$-independent mass limit is $210\pm25$~\GeV{} ($225\pm25$~\GeV), with an observed limit of 230~\GeV{} (250~\GeV).  
For masses higher than 300~\GeV, the limits for these two particles follow approximately a line of: $\Lambda + 8.3\times m_{\nu_{\ell}^*}=14500$~\GeV.
The most stringent upper limits on the mass of the excited leptons are found when $\exlepmass = \Lambda$.  In this case, the resulting limits are
3.0~\TeV{} for excited electrons and muons, 2.5~\TeV{} for excited taus, and 1.6~\TeV{} for every excited-neutrino flavour.

\FloatBarrier

\section{Conclusion}
\label{sec:Conclusion}
A search for anomalous production of events with at least three charged leptons is
been presented, using 20.3~\ifb{} of $pp$ collisions at $\sqrt{s}=8\TeV$ recorded by the ATLAS detector at the CERN Large Hadron Collider. 
Data distributions are compared to SM predictions in a variety of observables and final
states, designed to probe a large range of BSM scenarios.  Good agreement between the
data and SM predictions is observed. Model-independent exclusion limits on visible cross sections are derived,
and a prescription to re-interpret such limits for any model is presented.  Additionally, limits
are set on specific models predicting doubly charged Higgs bosons and
excited leptons.  Doubly charged Higgs bosons coupling to left-handed fermions and decaying exclusively to $e\tau$ or $\mu\tau$
pairs are constrained to have mass above 400~\GeV{} at 95\% confidence level.
For excited leptons, the mass constraints depend on the compositeness scale, with the strongest mass constraints reached where 
the mass of the excited state and the compositeness scale are the same; the lower limits on the mass extend to 3.0~\TeV{} for excited electrons and excited muons, 
2.5~\TeV{} for excited taus, and 1.6~\TeV{} for every excited-neutrino flavour.

\section*{Acknowledgements}
\label{sec:Acknowledgements}

We thank CERN for the very successful operation of the LHC, as well as the
support staff from our institutions without whom ATLAS could not be
operated efficiently.

We acknowledge the support of ANPCyT, Argentina; YerPhI, Armenia; ARC,
Australia; BMWFW and FWF, Austria; ANAS, Azerbaijan; SSTC, Belarus; CNPq and FAPESP,
Brazil; NSERC, NRC and CFI, Canada; CERN; CONICYT, Chile; CAS, MOST and NSFC,
China; COLCIENCIAS, Colombia; MSMT CR, MPO CR and VSC CR, Czech Republic;
DNRF, DNSRC and Lundbeck Foundation, Denmark; EPLANET, ERC and NSRF, European Union;
IN2P3-CNRS, CEA-DSM/IRFU, France; GNSF, Georgia; BMBF, DFG, HGF, MPG and AvH
Foundation, Germany; GSRT and NSRF, Greece; RGC, Hong Kong SAR, China; ISF, MINERVA, GIF, I-CORE and Benoziyo Center, Israel; INFN, Italy; MEXT and JSPS, Japan; CNRST, Morocco; FOM and NWO, Netherlands; BRF and RCN, Norway; MNiSW and NCN, Poland; GRICES and FCT, Portugal; MNE/IFA, Romania; MES of Russia and NRC KI, Russian Federation; JINR; MSTD,
Serbia; MSSR, Slovakia; ARRS and MIZ\v{S}, Slovenia; DST/NRF, South Africa;
MINECO, Spain; SRC and Wallenberg Foundation, Sweden; SER, SNSF and Cantons of
Bern and Geneva, Switzerland; NSC, Taiwan; TAEK, Turkey; STFC, the Royal
Society and Leverhulme Trust, United Kingdom; DOE and NSF, United States of
America.

The crucial computing support from all WLCG partners is acknowledged
gratefully, in particular from CERN and the ATLAS Tier-1 facilities at
TRIUMF (Canada), NDGF (Denmark, Norway, Sweden), CC-IN2P3 (France),
KIT/GridKA (Germany), INFN-CNAF (Italy), NL-T1 (Netherlands), PIC (Spain),
ASGC (Taiwan), RAL (UK) and BNL (USA) and in the Tier-2 facilities
worldwide.

\bibliographystyle{JHEP}
\bibliography{EXOT-2012-20}

\clearpage

\appendix
\section{Yields and cross-section limits}
\label{app:yields}
Expected and observed event yields for all signal regions are provided in tables~\ref{t:Result_All_HTLep}--\ref{t:Result_Weak_MET}.

\begin{table*}[hp]\footnotesize
 \begin{center}
  \begin{tabular}{r r@{\ $\pm$\ }r r@{\ $\pm$\ }r r@{\ $\pm$\ }r r@{\ $\pm$\ }r c}
\hline
$\Htlep \geq$ &\multicolumn{2}{c}{$t\bar{t}+V(V)$}	 &\multicolumn{2}{c}{$VV(V)$}	 &\multicolumn{2}{c}{Reducible}	 &\multicolumn{2}{c}{Total}	&Observed\\\hline
\multicolumn{10}{c}{$\geq 3 e/\mu$, off-$Z$, no-OSSF} \\
\hline
       200 \GeV	&2.0&0.6	 &1.6&0.4	 &2.1&1.1	 &5.7&1.4	&6\\
       500 \GeV	&0.13&0.06	 &0.09&0.03	 &\multicolumn{1}{r@{\ $^+_-$\ }}{0}&{$^{0.7}_{0}$}	 &\multicolumn{1}{r@{\ $^+_-$\ }}{0.22}&{$^{0.70}_{0.22}$}	&1\\
       800 \GeV	&0.06&0.04	 &\multicolumn{1}{r@{\ $^+_-$\ }}{0}&{$^{0.03}_{0}$}	 &\multicolumn{1}{r@{\ $^+_-$\ }}{0}&{$^{0.7}_{0}$}	 &\multicolumn{1}{r@{\ $^+_-$\ }}{0.06}&{$^{0.70}_{0.06}$}	&0\\
\hline
\multicolumn{10}{c}{$2 e/\mu + \geq 1 \tau$, off-$Z$, no-OSSF} \\
\hline
       200 \GeV	&1.2&0.4	 &2.3&0.5	 &19&6	 &22&6	&14\\
       500 \GeV	&0.01&0.01	 &0.03&0.01	 &\multicolumn{1}{r@{\ $^+_-$\ }}{0.32}&{$^{0.73}_{0.32}$}	 &\multicolumn{1}{r@{\ $^+_-$\ }}{0.36}&{$^{0.73}_{0.36}$}	&0\\
       800 \GeV	&\multicolumn{1}{r@{\ $^+_-$\ }}{0}&{$^{0.003}_{0}$}	 &0.01&0.01	 &\multicolumn{1}{r@{\ $^+_-$\ }}{0.11}&{$^{0.71}_{0.11}$}	 &\multicolumn{1}{r@{\ $^+_-$\ }}{0.12}&{$^{0.71}_{0.12}$}	&0\\
\hline
\multicolumn{10}{c}{$\geq 3 e/\mu$, off-$Z$, OSSF} \\
\hline
       200 \GeV	&7.5&2.3	 &63&8	 &9&4	 &78&9	&56\\
       500 \GeV	&0.34&0.12	 &3.3&0.5	 &\multicolumn{1}{r@{\ $^+_-$\ }}{0}&{$^{0.7}_{0}$}	 &3.7&0.9	&1\\
       800 \GeV	&{0.01}&{0.01}	 &0.54&0.12	 &\multicolumn{1}{r@{\ $^+_-$\ }}{0}&{$^{0.7}_{0}$}	 &\multicolumn{1}{r@{\ $^+_-$\ }}{0.5}&{$^{0.7}_{0.5}$}	&0\\
\hline
\multicolumn{10}{c}{$2 e/\mu + \geq 1 \tau$, off-$Z$, OSSF} \\
\hline
       200 \GeV	&0.64&0.21	 &4.4&0.6	 &68&20	 &73&20	&67\\
       500 \GeV	&0.06&0.03	 &0.17&0.04	 &1.1&0.9	 &1.3&0.9	&0\\
       800 \GeV	&\multicolumn{1}{r@{\ $^+_-$\ }}{0}&{$^{0.003}_{0}$}	 &\multicolumn{1}{r@{\ $^+_-$\ }}{0}&{$^{0.03}_{0}$}	 &\multicolumn{1}{r@{\ $^+_-$\ }}{0}&{$^{0.7}_{0}$}	 &\multicolumn{1}{r@{\ $^+_-$\ }}{0}&{$^{0.7}_{0}$}	&0\\
\hline
\multicolumn{10}{c}{$\geq 3 e/\mu$, on-$Z$} \\
\hline
       200 \GeV	&23&7	 &410&50	 &18&8	 &450&50	&387\\
       500 \GeV	&0.82&0.25	 &10.9&2.3	 &\multicolumn{1}{r@{\ $^+_-$\ }}{0.6}&{$^{0.8}_{0.6}$}	 &12.3&2.4	&12\\
       800 \GeV	&0.05&0.03	 &0.92&0.23	 &\multicolumn{1}{r@{\ $^+_-$\ }}{0.10}&{$^{0.70}_{0.10}$}	 &1.1&0.7	&3\\
\hline
\multicolumn{10}{c}{$2 e/\mu + \geq 1 \tau$, on-$Z$} \\
\hline
       200 \GeV	&1.1&0.4	 &20.7&2.7	 &160&50	 &180&50	&148\\
       500 \GeV	&0.02&0.01	 &0.82&0.23	 &1.2&0.9	 &2.0&1.0	&3\\
       800 \GeV	&\multicolumn{1}{r@{\ $^+_-$\ }}{0}&{$^{0.003}_{0}$}	 &0.04&0.02	 &\multicolumn{1}{r@{\ $^+_-$\ }}{0}&{$^{0.71}_{0}$}	 &\multicolumn{1}{r@{\ $^+_-$\ }}{0.04}&{$^{0.71}_{0.04}$}	&0\\
\hline
\end{tabular}
   \caption{Expected and observed event yields for the $\Htlep$~signal regions.}
   \label{t:Result_All_HTLep}
\end{center}
\end{table*}

\begin{table*}[tbp]\footnotesize
 \begin{center}
  \begin{tabular}{r r@{\ $\pm$\ }r r@{\ $\pm$\ }r r@{\ $\pm$\ }r r@{\ $\pm$\ }r c}
\hline
$\minleppt \geq$ &\multicolumn{2}{c}{$t\bar{t}+V(V)$}	 &\multicolumn{2}{c}{$VV(V)$}	 &\multicolumn{2}{c}{Reducible}	 &\multicolumn{2}{c}{Total}	&Observed\\\hline
\multicolumn{10}{c}{$\geq 3 e/\mu$, off-$Z$, no-OSSF} \\
\hline
        50 \GeV	&0.83&0.27	 &0.45&0.13	 &\multicolumn{1}{r@{\ $^+_-$\ }}{0.28}&{$^{0.70}_{0.28}$}	 &1.6&0.8	&4\\
       100 \GeV	&0.09&0.04	 &0.03&0.01	 &\multicolumn{1}{r@{\ $^+_-$\ }}{0}&{$^{0.7}_{0}$}	 &\multicolumn{1}{r@{\ $^+_-$\ }}{0.12}&{$^{0.69}_{0.12}$}	&0\\
       150 \GeV	&0.06&0.04	 &\multicolumn{1}{r@{\ $^+_-$\ }}{0}&{$^{0.03}_{0}$}	 &\multicolumn{1}{r@{\ $^+_-$\ }}{0}&{$^{0.7}_{0}$}	 &\multicolumn{1}{r@{\ $^+_-$\ }}{0.06}&{$^{0.70}_{0.06}$}	&0\\
\hline
\multicolumn{10}{c}{$2 e/\mu + \geq 1 \tau$, off-$Z$, no-OSSF} \\
\hline
        50 \GeV	&0.43&0.15	 &0.62&0.13	 &5.0&1.9	 &6.1&1.9	&5\\
       100 \GeV	&0.01&0.01	 &\multicolumn{1}{r@{\ $^+_-$\ }}{0.03}&{$^{0.05}_{0.03}$}	 &\multicolumn{1}{r@{\ $^+_-$\ }}{0.11}&{$^{0.71}_{0.11}$}	 &\multicolumn{1}{r@{\ $^+_-$\ }}{0.15}&{$^{0.72}_{0.15}$}	&0\\
       150 \GeV	&\multicolumn{1}{r@{\ $^+_-$\ }}{0}&{$^{0.005}_{0}$}	 &0.01&0.00	 &\multicolumn{1}{r@{\ $^+_-$\ }}{0.11}&{$^{0.71}_{0.11}$}	 &\multicolumn{1}{r@{\ $^+_-$\ }}{0.12}&{$^{0.71}_{0.12}$}	&0\\
\hline
\multicolumn{10}{c}{$\geq 3 e/\mu$, off-$Z$, OSSF} \\
\hline
        50 \GeV	&3.7&1.1	 &27.1&3.4	 &1.8&1.0	 &33&4	&25\\
       100 \GeV	&0.17&0.07	 &1.73&0.25	 &\multicolumn{1}{r@{\ $^+_-$\ }}{0}&{$^{0.7}_{0}$}	 &1.9&0.7	&2\\
       150 \GeV	&0.01&0.01	 &0.24&0.05	 &\multicolumn{1}{r@{\ $^+_-$\ }}{0}&{$^{0.7}_{0}$}	 &\multicolumn{1}{r@{\ $^+_-$\ }}{0.25}&{$^{0.69}_{0.25}$}	&0\\
\hline
\multicolumn{10}{c}{$2 e/\mu + \geq 1 \tau$, off-$Z$, OSSF} \\
\hline
        50 \GeV	&0.33&0.11	 &1.72&0.25	 &15&5	 &18&5	&23\\
       100 \GeV	&\multicolumn{1}{r@{\ $^+_-$\ }}{0}&{$^{0.003}_{0}$}	 &0.14&0.06	 &\multicolumn{1}{r@{\ $^+_-$\ }}{0.12}&{$^{0.72}_{0.12}$}	 &\multicolumn{1}{r@{\ $^+_-$\ }}{0.26}&{$^{0.72}_{0.26}$}	&0\\
       150 \GeV	&\multicolumn{1}{r@{\ $^+_-$\ }}{0}&{$^{0.003}_{0}$}	 &\multicolumn{1}{r@{\ $^+_-$\ }}{0}&{$^{0.03}_{0}$}	 &\multicolumn{1}{r@{\ $^+_-$\ }}{0}&{$^{0.7}_{0}$}	 &\multicolumn{1}{r@{\ $^+_-$\ }}{0}&{$^{0.7}_{0}$}	&0\\
\hline
\multicolumn{10}{c}{$\geq 3 e/\mu$, on-$Z$} \\
\hline
        50 \GeV	&8.7&2.6	 &168&19	 &6.3&3.1	 &183&19	&163\\
       100 \GeV	&0.54&0.17	 &9.6&1.6	 &\multicolumn{1}{r@{\ $^+_-$\ }}{0.22}&{$^{0.72}_{0.22}$}	 &10.4&1.7	&16\\
       150 \GeV	&0.05&0.02	 &0.88&0.21	 &\multicolumn{1}{r@{\ $^+_-$\ }}{0}&{$^{0.7}_{0}$}	 &0.9&0.7	&4\\
\hline
\multicolumn{10}{c}{$2 e/\mu + \geq 1 \tau$, on-$Z$} \\
\hline
        50 \GeV	&0.31&0.11	 &8.0&1.2	 &54&18	 &62&18	&45\\
       100 \GeV	&0.01&0.01	 &0.53&0.22	 &\multicolumn{1}{r@{\ $^+_-$\ }}{0.8}&{$^{0.8}_{0.8}$}	 &1.3&0.9	&0\\
       150 \GeV	&\multicolumn{1}{r@{\ $^+_-$\ }}{0}&{$^{0.003}_{0}$}	 &0.09&0.07	 &\multicolumn{1}{r@{\ $^+_-$\ }}{0.16}&{$^{0.72}_{0.16}$}	 &\multicolumn{1}{r@{\ $^+_-$\ }}{0.25}&{$^{0.73}_{0.25}$}	&0\\
\hline
\end{tabular}
   \caption{Expected and observed event yields for the minimum $p_{\mathrm{T}}^{\ell}$~signal regions.}
   \label{t:Result_All_MinLepPt}
\end{center}
\end{table*}

\begin{table*}[tbp]\footnotesize
 \begin{center}
  \begin{tabular}{r r@{\ $\pm$\ }r r@{\ $\pm$\ }r r@{\ $\pm$\ }r r@{\ $\pm$\ }r c}
\hline
$b$-tags $\geq$ &\multicolumn{2}{c}{$t\bar{t}+V(V)$}	 &\multicolumn{2}{c}{$VV(V)$}	 &\multicolumn{2}{c}{Reducible}	 &\multicolumn{2}{c}{Total}	&Observed\\\hline
\multicolumn{10}{c}{$\geq 3 e/\mu$, off-$Z$, no-OSSF} \\
\hline
         1 	&5.3&1.7	 &0.37&0.12	 &11.1&3.3	 &17&4	&19\\
         2 	&2.2&0.7	 &\multicolumn{1}{r@{\ $^+_-$\ }}{0}&{$^{0.03}_{0}$}	 &2.2&1.0	 &4.4&1.2	&5\\
\hline
\multicolumn{10}{c}{$2 e/\mu + \geq 1 \tau$, off-$Z$, no-OSSF} \\
\hline
         1 	&3.1&1.0	 &0.9&0.4	 &91&24	         &95&24	        &98\\
         2 	&1.3&0.5	 &0.05&0.03	 &29&8	 &30&8	&34\\
\hline
\multicolumn{10}{c}{$\geq 3 e/\mu$, off-$Z$, OSSF} \\
\hline
         1 	&13&4   	 &11.4&2.0	 &15&5  	 &39&7  	&34\\
         2 	&5.5&1.8	 &0.32&0.17	 &2.7&2.1	 &8.5&2.8	&9\\
\hline
\multicolumn{10}{c}{$2 e/\mu + \geq 1 \tau$, off-$Z$, OSSF} \\
\hline
         1 	&1.09&0.35	 &0.88&0.28	 &74&21	 &76&21	&65\\
         2 	&0.38&0.19	 &\multicolumn{1}{r@{\ $^+_-$\ }}{0.05}&{$^{0.06}_{0.05}$}	 &17&5	 &17&5	&12\\
\hline
\multicolumn{10}{c}{$\geq 3 e/\mu$, on-$Z$} \\
\hline
         1 	&51&16  	 &144&23	 &41&11 	 &235&32	&237\\
         2 	&23&8   	 &8.0&2.1	 &4.6&1.5	 &36&8  	&27\\
\hline
\multicolumn{10}{c}{$2 e/\mu + \geq 1 \tau$, on-$Z$} \\
\hline
         1 	&2.9&0.9	 &8.8&1.7	 &398&11	 &410&11	&409\\
         2 	&1.3&0.4	 &0.26&0.17	 &33&9  	 &34&9	        &21\\
\hline
\end{tabular}
   \caption{Expected and observed event yields for the $b$-tag~signal regions.}
   \label{t:Result_All_nBtags}
\end{center}
\end{table*}

\begin{table*}[tbp]\footnotesize
 \begin{center}
  \begin{tabular}{r r@{\ $\pm$\ }r r@{\ $\pm$\ }r r@{\ $\pm$\ }r r@{\ $\pm$\ }r c}
\hline
$\St \geq$ &\multicolumn{2}{c}{$t\bar{t}+V(V)$}	 &\multicolumn{2}{c}{$VV(V)$}	 &\multicolumn{2}{c}{Reducible}	 &\multicolumn{2}{c}{Total}	&Observed\\\hline
\multicolumn{10}{c}{$\geq 3 e/\mu$, off-$Z$, no-OSSF} \\
\hline
       600 \GeV	&1.7&0.6	 &0.40&0.13	 &1.2&0.8	 &3.4&1.0	&3\\
      1000 \GeV	&0.44&0.16	 &\multicolumn{1}{r@{\ $^+_-$\ }}{0.01}&{$^{0.03}_{0.01}$}	 &\multicolumn{1}{r@{\ $^+_-$\ }}{0.20}&{$^{0.70}_{0.20}$}	 &{0.7}&{0.7}	&1\\
      1500 \GeV	&\multicolumn{1}{r@{\ $^+_-$\ }}{0.07}&{$^{0.09}_{0.07}$}	 &\multicolumn{1}{r@{\ $^+_-$\ }}{0}&{$^{0.03}_{0}$}	 &\multicolumn{1}{r@{\ $^+_-$\ }}{0.08}&{$^{0.69}_{0.08}$}	 &\multicolumn{1}{r@{\ $^+_-$\ }}{0.16}&{$^{0.70}_{0.16}$}	&0\\
\hline
\multicolumn{10}{c}{$2 e/\mu + \geq 1 \tau$, off-$Z$, no-OSSF} \\
\hline
       600 \GeV	&1.4&0.5	 &1.23&0.32	 &17&5  	 &19&5  	&22\\
      1000 \GeV	&0.26&0.22	 &0.07&0.05	 &2.2&1.1	 &2.6&1.1	&2\\
      1500 \GeV	&0.03&0.02	 &0.01&0.01	 &\multicolumn{1}{r@{\ $^+_-$\ }}{0.17}&{$^{0.73}_{0.17}$}	 &\multicolumn{1}{r@{\ $^+_-$\ }}{0.20}&{$^{0.73}_{0.20}$}	&1\\
\hline
\multicolumn{10}{c}{$\geq 3 e/\mu$, off-$Z$, OSSF} \\
\hline
       600 \GeV	&6.7&2.1	 &13.4&2.9	 &3.7&2.5	 &24&4	&17\\
      1000 \GeV	&1.1&0.4	 &2.1&0.8	 &\multicolumn{1}{r@{\ $^+_-$\ }}{2.1}&{$^{2.1}_{2.1}$}	 &5.3&2.3	&1\\
      1500 \GeV	&{0.08}&{0.08}	 &0.28&0.15	 &\multicolumn{1}{r@{\ $^+_-$\ }}{0.5}&{$^{0.9}_{0.5}$}	 &\multicolumn{1}{r@{\ $^+_-$\ }}{0.8}&{$^{0.9}_{0.8}$}	&0\\
\hline
\multicolumn{10}{c}{$2 e/\mu + \geq 1 \tau$, off-$Z$, OSSF} \\
\hline
       600 \GeV	&0.59&0.19	 &0.87&0.30	 &17&5  	 &18&5  	&19\\
      1000 \GeV	&0.17&0.06	 &0.12&0.10	 &1.8&1.0	 &2.1&1.0	&2\\
      1500 \GeV	&\multicolumn{1}{r@{\ $^+_-$\ }}{0}&{$^{0.003}_{0}$}	 &\multicolumn{1}{r@{\ $^+_-$\ }}{0}&{$^{0.05}_{0}$}	 &\multicolumn{1}{r@{\ $^+_-$\ }}{0}&{$^{0.7}_{0}$}	 &\multicolumn{1}{r@{\ $^+_-$\ }}{0}&{$^{0.7}_{0}$}	&0\\
\hline
\multicolumn{10}{c}{$\geq 3 e/\mu$, on-$Z$} \\
\hline
       600 \GeV	&26&8   	 &126&29	 &9.2&3.5	 &161&31	&147\\
      1000 \GeV	&4.6&1.5	 &21&8  	 &1.1&1.0	 &27&8	&27\\
      1500 \GeV	&0.48&0.17	 &3.2&1.9	 &\multicolumn{1}{r@{\ $^+_-$\ }}{0}&{$^{0.6}_{0}$}	 &3.7&2.0	&2\\
\hline
\multicolumn{10}{c}{$2 e/\mu + \geq 1 \tau$, on-$Z$} \\
\hline
       600 \GeV	&1.4&0.5	 &8.7&2.1	 &65&19 	 &75&19 	&65\\
      1000 \GeV	&0.26&0.09	 &1.5&0.6	 &3.6&1.4	 &5.3&1.6	&11\\
      1500 \GeV	&0.02&0.02	 &0.31&0.21	 &\multicolumn{1}{r@{\ $^+_-$\ }}{0.08}&{$^{0.71}_{0.08}$}	 &\multicolumn{1}{r@{\ $^+_-$\ }}{0.4}&{$^{0.7}_{0.4}$}	&1\\
\hline
\end{tabular}
   \caption{Expected and observed event yields for the inclusive $\St$~signal regions.}
   \label{t:Result_All_ST}
\end{center}
\end{table*}

\begin{table*}[tbp]\footnotesize
 \begin{center}
  \begin{tabular}{r r@{\ $\pm$\ }r r@{\ $\pm$\ }r r@{\ $\pm$\ }r r@{\ $\pm$\ }r c}
\hline
$\St \geq$ &\multicolumn{2}{c}{$t\bar{t}+V(V)$}	 &\multicolumn{2}{c}{$VV(V)$}	 &\multicolumn{2}{c}{Reducible}	 &\multicolumn{2}{c}{Total}	&Observed\\\hline
\multicolumn{10}{c}{$\geq 3 e/\mu$, off-$Z$, no-OSSF} \\
\hline
 Inclusive	&1.7&0.6	 &0.90&0.24	 &4.0&1.5	 &6.6&1.6	&8\\
       600 \GeV	&0.71&0.27	 &0.25&0.06	 &0.8&0.8	 &1.7&0.8	&2\\
      1000 \GeV	&0.24&0.15	 &0.01&0.00	 &\multicolumn{1}{r@{\ $^+_-$\ }}{0.15}&{$^{0.70}_{0.15}$}	 &\multicolumn{1}{r@{\ $^+_-$\ }}{0.4}&{$^{0.7}_{0.4}$}	&1\\
      1200 \GeV	&0.14&0.09	 &0.01&0.00	 &\multicolumn{1}{r@{\ $^+_-$\ }}{0.15}&{$^{0.70}_{0.15}$}	 &\multicolumn{1}{r@{\ $^+_-$\ }}{0.29}&{$^{0.70}_{0.29}$}	&0\\
\hline
\multicolumn{10}{c}{$2 e/\mu + \geq 1 \tau$, off-$Z$, no-OSSF} \\
\hline
 Inclusive	&1.3&0.4	 &2.0&0.4	 &29&8  	 &32&8  	&28\\
       600 \GeV	&0.81&0.29	 &0.66&0.19	 &8.0&2.5	 &9.5&2.5	&9\\
      1000 \GeV	&\multicolumn{1}{r@{\ $^+_-$\ }}{0.16}&{$^{0.20}_{0.16}$}	 &0.06&0.05	 &1.4&0.9	 &1.6&0.9	&2\\
      1200 \GeV	&0.10&0.05	 &\multicolumn{1}{r@{\ $^+_-$\ }}{0.01}&{$^{0.05}_{0.01}$}	 &\multicolumn{1}{r@{\ $^+_-$\ }}{0.4}&{$^{0.8}_{0.4}$}	 &\multicolumn{1}{r@{\ $^+_-$\ }}{0.5}&{$^{0.8}_{0.5}$}	&2\\
\hline
\multicolumn{10}{c}{$\geq 3 e/\mu$, off-$Z$, OSSF} \\
\hline
 Inclusive	&4.6&1.4	 &12.2&1.9	 &3.6&1.4	 &20.4&2.8	&16\\
       600 \GeV	&2.8&0.9	 &4.7&1.1	 &0.8&0.8	 &8.2&1.7	&7\\
      1000 \GeV	&0.47&0.17	 &1.2&0.4	 &\multicolumn{1}{r@{\ $^+_-$\ }}{0.16}&{$^{0.71}_{0.16}$}	 &1.8&0.9	&1\\
      1200 \GeV	&0.18&0.09	 &0.39&0.19	 &\multicolumn{1}{r@{\ $^+_-$\ }}{0.16}&{$^{0.71}_{0.16}$}	 &{0.7}&{0.7}	&1\\
\hline
\multicolumn{10}{c}{$2 e/\mu + \geq 1 \tau$, off-$Z$, OSSF} \\
\hline
 Inclusive	&0.54&0.18	 &1.47&0.31	 &14&4	 &16&4	&17\\
       600 \GeV	&0.34&0.13	 &0.44&0.17	 &4.8&1.8	 &5.6&1.8	&5\\
      1000 \GeV	&0.15&0.07	 &\multicolumn{1}{r@{\ $^+_-$\ }}{0.07}&{$^{0.08}_{0.07}$}	 &\multicolumn{1}{r@{\ $^+_-$\ }}{0.6}&{$^{0.8}_{0.6}$}	 &0.8&0.8	&2\\
      1200 \GeV	&\multicolumn{1}{r@{\ $^+_-$\ }}{0}&{$^{0.006}_{0}$}	 &\multicolumn{1}{r@{\ $^+_-$\ }}{0.07}&{$^{0.08}_{0.07}$}	 &\multicolumn{1}{r@{\ $^+_-$\ }}{0.17}&{$^{0.73}_{0.17}$}	 &\multicolumn{1}{r@{\ $^+_-$\ }}{0.24}&{$^{0.73}_{0.24}$}	&2\\
\hline
\multicolumn{10}{c}{$\geq 3 e/\mu$, on-$Z$} \\
\hline
 Inclusive	&14&4   	 &148&19	 &5.7&1.9	 &167&20	&123\\
       600 \GeV	&8.7&2.7	 &41&9  	 &1.3&0.9	 &51&10	&39\\
      1000 \GeV	&2.5&0.8	 &9.1&3.2	 &\multicolumn{1}{r@{\ $^+_-$\ }}{0}&{$^{0.69}_{0}$}	 &11.6&3.5	&12\\
      1200 \GeV	&1.01&0.33	 &4.0&1.8	 &\multicolumn{1}{r@{\ $^+_-$\ }}{0}&{$^{0.69}_{0}$}	 &5.0&2.0	&4\\
\hline
\multicolumn{10}{c}{$2 e/\mu + \geq 1 \tau$, on-$Z$} \\
\hline
 Inclusive	&1.01&0.32	 &12.1&1.7	 &13.8&4.1	 &26.9&4.5	&24\\
       600 \GeV	&0.62&0.21	 &4.1&1.0	 &3.5&1.4	 &8.2&1.7	&9\\
      1000 \GeV	&0.16&0.06	 &1.2&0.4	 &\multicolumn{1}{r@{\ $^+_-$\ }}{0.4}&{$^{0.8}_{0.4}$}	 &1.7&0.9	&0\\
      1200 \GeV	&0.07&0.03	 &0.53&0.27	 &\multicolumn{1}{r@{\ $^+_-$\ }}{0.33}&{$^{0.74}_{0.33}$}	 &0.9&0.8	&0\\
\hline
\end{tabular}
   \caption{Expected and observed event yields for the high-\met, $\St$~signal regions.}
   \label{t:Result_HighMET_ST}
\end{center}
\end{table*}

\begin{table*}[tbp]\footnotesize
 \begin{center}
  \begin{tabular}{r r@{\ $\pm$\ }r r@{\ $\pm$\ }r r@{\ $\pm$\ }r r@{\ $\pm$\ }r c}
\hline
$\St \geq$ &\multicolumn{2}{c}{$t\bar{t}+V(V)$}	 &\multicolumn{2}{c}{$VV(V)$}	 &\multicolumn{2}{c}{Reducible}	 &\multicolumn{2}{c}{Total}	&Observed\\\hline
\multicolumn{10}{c}{$\geq 3 e/\mu$, on-$Z$} \\
\hline
 Inclusive	&11.2&3.5	 &174&23	 &9.0&2.7	 &194&24	&164\\
       600 \GeV	&5.5&1.7	 &22&6	 &{0.9}&{0.9}	 &28&6	&29\\
      1200 \GeV	&0.33&0.12	 &2.5&1.3	 &\multicolumn{1}{r@{\ $^+_-$\ }}{0.16}&{$^{0.71}_{0.16}$}	 &3.0&1.5	&2\\
\hline
\multicolumn{10}{c}{$2 e/\mu + \geq 1 \tau$, on-$Z$} \\
\hline
 Inclusive	&0.38&0.13	 &2.2&0.9	 &51&17	 &54&17	&46\\
       600 \GeV	&0.10&0.06	 &0.12&0.08	 &5.4&2.2	 &5.6&2.2	&8\\
       1200 \GeV	&{0.01}&{0.01}	 &0.04&0.04	 &\multicolumn{1}{r@{\ $^+_-$\ }}{0.22}&{$^{0.73}_{0.22}$}	 &\multicolumn{1}{r@{\ $^+_-$\ }}{0.27}&{$^{0.73}_{0.27}$}	&0\\
\hline
\end{tabular}
   \caption{Expected and observed event yields for the high-$m_{\mathrm{T}}^{W}$, $\St$~signal regions.}
   \label{t:Result_HighmTW_ST}
\end{center}
\end{table*}

\begin{table*}[tbp]\footnotesize
 \begin{center}
  \begin{tabular}{r r@{\ $\pm$\ }r r@{\ $\pm$\ }r r@{\ $\pm$\ }r r@{\ $\pm$\ }r c}
\hline
$\met \geq$ &\multicolumn{2}{c}{$t\bar{t}+V(V)$}	 &\multicolumn{2}{c}{$VV(V)$}	 &\multicolumn{2}{c}{Reducible}	 &\multicolumn{2}{c}{Total}	&Observed\\\hline
\multicolumn{10}{c}{$\geq 3 e/\mu$, off-$Z$, no-OSSF} \\
\hline
 Inclusive	&3.7&1.2	 &0.85&0.26	 &7.1&2.2	 &11.7&2.5	&18\\
       100 \GeV	&1.2&0.4	 &0.24&0.06	 &2.6&1.1	 &4.0&1.2	&8\\
       200 \GeV	&0.18&0.07	 &0.01&0.01	 &\multicolumn{1}{r@{\ $^+_-$\ }}{0.34}&{$^{0.71}_{0.34}$}	 &\multicolumn{1}{r@{\ $^+_-$\ }}{0.5}&{$^{0.7}_{0.5}$}	&0\\
       300 \GeV	&0.12&0.07	 &\multicolumn{1}{r@{\ $^+_-$\ }}{0}&{$^{0.03}_{0}$}	 &\multicolumn{1}{r@{\ $^+_-$\ }}{0.15}&{$^{0.70}_{0.15}$}	 &\multicolumn{1}{r@{\ $^+_-$\ }}{0.28}&{$^{0.70}_{0.28}$}	&0\\
\hline
\multicolumn{10}{c}{$2 e/\mu + \geq 1 \tau$, off-$Z$, no-OSSF} \\
\hline
 Inclusive	&2.7&0.8	 &3.7&0.7	 &71&19	 &77&19	&83\\
       100 \GeV	&1.1&0.4	 &1.18&0.29	 &19&5	 &21&5	&24\\
       200 \GeV	&0.04&0.04	 &\multicolumn{1}{r@{\ $^+_-$\ }}{0.16}&{$^{0.18}_{0.16}$}	 &1.7&0.9	 &1.9&1.0	&2\\
       300 \GeV	&{0.01}&{0.01}	 &0.05&0.03	 &\multicolumn{1}{r@{\ $^+_-$\ }}{0.25}&{$^{0.73}_{0.25}$}	 &\multicolumn{1}{r@{\ $^+_-$\ }}{0.31}&{$^{0.73}_{0.31}$}	&0\\
\hline
\multicolumn{10}{c}{$\geq 3 e/\mu$, off-$Z$, OSSF} \\
\hline
 Inclusive	&11.7&3.5	 &32&6  	 &12&4  	 &56&8  	&53\\
       100 \GeV	&3.6&1.1	 &5.0&1.2	 &1.9&1.0	 &10.5&1.9	&8\\
       200 \GeV	&0.41&0.17	 &0.86&0.24	 &\multicolumn{1}{r@{\ $^+_-$\ }}{0.21}&{$^{0.71}_{0.21}$}	 &1.5&0.8	&2\\
       300 \GeV	&\multicolumn{1}{r@{\ $^+_-$\ }}{0.04}&{$^{0.05}_{0.04}$}	 &0.28&0.12	 &\multicolumn{1}{r@{\ $^+_-$\ }}{0}&{$^{0.7}_{0}$}	 &\multicolumn{1}{r@{\ $^+_-$\ }}{0.33}&{$^{0.70}_{0.33}$}	&0\\
\hline
\multicolumn{10}{c}{$2 e/\mu + \geq 1 \tau$, off-$Z$, OSSF} \\
\hline
 Inclusive	&1.07&0.35	 &2.4&0.6	 &95&26	 &98&26	&83\\
       100 \GeV	&0.39&0.13	 &0.63&0.21	 &10.1&3.2	 &11.1&3.2	&9\\
       200 \GeV	&0.03&0.02	 &0.20&0.11	 &\multicolumn{1}{r@{\ $^+_-$\ }}{0.35}&{$^{0.74}_{0.35}$}	 &\multicolumn{1}{r@{\ $^+_-$\ }}{0.6}&{$^{0.8}_{0.6}$}	&1\\
       300 \GeV	&\multicolumn{1}{r@{\ $^+_-$\ }}{0}&{$^{0.003}_{0}$}	 &\multicolumn{1}{r@{\ $^+_-$\ }}{0.02}&{$^{0.05}_{0.02}$}	 &\multicolumn{1}{r@{\ $^+_-$\ }}{0}&{$^{0.71}_{0}$}	 &\multicolumn{1}{r@{\ $^+_-$\ }}{0.02}&{$^{0.71}_{0.02}$}	&0\\
\hline
\multicolumn{10}{c}{$\geq 3 e/\mu$, on-$Z$} \\
\hline
 Inclusive	&52&16  	 &391&70	 &40&10 	 &480&7 	&446\\
       100 \GeV	&13&4   	 &57&12  	 &2.7&1.2	 &73&13 	&53\\
       200 \GeV	&1.7&0.5	 &9.6&2.6	 &\multicolumn{1}{r@{\ $^+_-$\ }}{0.5}&{$^{0.8}_{0.5}$}	 &11.8&2.8	&13\\
       300 \GeV	&0.24&0.09	 &2.3&0.8	 &\multicolumn{1}{r@{\ $^+_-$\ }}{0}&{$^{0.69}_{0}$}	 &2.6&1.1	&1\\
\hline
\multicolumn{10}{c}{$2 e/\mu + \geq 1 \tau$, on-$Z$} \\
\hline
 Inclusive	&3.0&0.9	 &26&5  	 &640&180	 &670&180	&554\\
       100 \GeV	&0.93&0.30	 &5.3&1.3	 &8.1&2.5	 &14.3&2.9	&17\\
       200 \GeV	&\multicolumn{1}{r@{\ $^+_-$\ }}{0.13}&{$^{0.14}_{0.13}$}	 &1.2&0.4	 &\multicolumn{1}{r@{\ $^+_-$\ }}{0.4}&{$^{0.8}_{0.4}$}	 &1.7&0.9	&3\\
       300 \GeV	&0.02&0.01	 &0.35&0.16	 &\multicolumn{1}{r@{\ $^+_-$\ }}{0}&{$^{0.7}_{0}$}	 &\multicolumn{1}{r@{\ $^+_-$\ }}{0.4}&{$^{0.7}_{0.4}$}	&1\\
\hline
\end{tabular}
   \caption{Expected and observed event yields for the high-\Ht, $\met$~signal regions.}
   \label{t:Result_Strong_MET}
\end{center}
\end{table*}

\begin{table*}[tbp]\footnotesize
 \begin{center}
  \begin{tabular}{r r@{\ $\pm$\ }r r@{\ $\pm$\ }r r@{\ $\pm$\ }r r@{\ $\pm$\ }r c}
\hline
$\met \geq$ &\multicolumn{2}{c}{$t\bar{t}+V(V)$}	 &\multicolumn{2}{c}{$VV(V)$}	 &\multicolumn{2}{c}{Reducible}	 &\multicolumn{2}{c}{Total}	&Observed\\\hline
\multicolumn{10}{c}{$\geq 3 e/\mu$, off-$Z$, no-OSSF} \\
\hline
 Inclusive	&1.9&0.6	 &6.1&1.2	 &10.4&2.9	 &18.4&3.3	&18\\
       100 \GeV	&0.53&0.22	 &0.66&0.19	 &1.4&0.8	 &2.6&0.9	&0\\
       200 \GeV	&\multicolumn{1}{r@{\ $^+_-$\ }}{0}&{$^{0.003}_{0}$}	 &0.08&0.03	 &\multicolumn{1}{r@{\ $^+_-$\ }}{0}&{$^{0.7}_{0}$}	 &\multicolumn{1}{r@{\ $^+_-$\ }}{0.08}&{$^{0.69}_{0.08}$}	&0\\
       300 \GeV	&\multicolumn{1}{r@{\ $^+_-$\ }}{0}&{$^{0.003}_{0}$}	 &0.01&0.01	 &\multicolumn{1}{r@{\ $^+_-$\ }}{0}&{$^{0.7}_{0}$}	 &\multicolumn{1}{r@{\ $^+_-$\ }}{0.01}&{$^{0.69}_{0.01}$}	&0\\
\hline
\multicolumn{10}{c}{$2 e/\mu + \geq 1 \tau$, off-$Z$, no-OSSF} \\
\hline
 Inclusive	&0.63&0.24	 &19.1&2.5	 &105&25	 &125&25	&125\\
       100 \GeV	&0.25&0.12	 &0.85&0.23	 &9.7&3.0	 &10.8&3.0	&4\\
       200 \GeV	&\multicolumn{1}{r@{\ $^+_-$\ }}{0}&{$^{0.006}_{0}$}	 &0.02&0.01	 &\multicolumn{1}{r@{\ $^+_-$\ }}{0.06}&{$^{0.71}_{0.06}$}	 &\multicolumn{1}{r@{\ $^+_-$\ }}{0.08}&{$^{0.71}_{0.08}$}	&0\\
       300 \GeV	&\multicolumn{1}{r@{\ $^+_-$\ }}{0}&{$^{0.003}_{0}$}	 &\multicolumn{1}{r@{\ $^+_-$\ }}{0}&{$^{0.03}_{0}$}	 &\multicolumn{1}{r@{\ $^+_-$\ }}{0}&{$^{0.71}_{0}$}	 &\multicolumn{1}{r@{\ $^+_-$\ }}{0}&{$^{0.71}_{0}$}	&0\\
\hline
\multicolumn{10}{c}{$\geq 3 e/\mu$, off-$Z$, OSSF} \\
\hline
 Inclusive	&3.1&1.0	 &159&18	 &21&6  	 &183&19	&168\\
       100 \GeV	&0.95&0.34	 &7.2&1.0	 &1.8&1.0	 &9.9&1.4	&8\\
       200 \GeV	&0.06&0.06	 &0.71&0.16	 &\multicolumn{1}{r@{\ $^+_-$\ }}{0.09}&{$^{0.70}_{0.09}$}	 &0.9&0.7	&1\\
       300 \GeV	&\multicolumn{1}{r@{\ $^+_-$\ }}{0.05}&{$^{0.07}_{0.05}$}	 &0.11&0.06	 &\multicolumn{1}{r@{\ $^+_-$\ }}{0}&{$^{0.6}_{0}$}	 &\multicolumn{1}{r@{\ $^+_-$\ }}{0.16}&{$^{0.70}_{0.16}$}	&0\\
\hline
\multicolumn{10}{c}{$2 e/\mu + \geq 1 \tau$, off-$Z$, OSSF} \\
\hline
 Inclusive	&0.41&0.15	 &10.6&1.2	 &530&150	 &540&150	&539\\
       100 \GeV	&0.16&0.08	 &0.84&0.20	 &4.2&1.5	 &5.2&1.5	&8\\
       200 \GeV	&\multicolumn{1}{r@{\ $^+_-$\ }}{0}&{$^{0.003}_{0}$}	 &0.06&0.05	 &\multicolumn{1}{r@{\ $^+_-$\ }}{0}&{$^{0.71}_{0}$}	 &\multicolumn{1}{r@{\ $^+_-$\ }}{0.06}&{$^{0.71}_{0.06}$}	&0\\
       300 \GeV	&\multicolumn{1}{r@{\ $^+_-$\ }}{0}&{$^{0.003}_{0}$}	 &\multicolumn{1}{r@{\ $^+_-$\ }}{0}&{$^{0.03}_{0}$}	 &\multicolumn{1}{r@{\ $^+_-$\ }}{0}&{$^{0.71}_{0}$}	 &\multicolumn{1}{r@{\ $^+_-$\ }}{0}&{$^{0.71}_{0}$}	&0\\
\hline
\multicolumn{10}{c}{$\geq 3 e/\mu$, on-$Z$} \\
\hline
 Inclusive	&8.4&2.9	 &2450&290	 &142&35	 &2600&290	&2539\\
       100 \GeV	&1.2&0.4	 &90&9  	 &3.0&1.3	 &94&9	&70\\
       200 \GeV	&0.05&0.02	 &6.2&0.7	 &\multicolumn{1}{r@{\ $^+_-$\ }}{0.04}&{$^{0.70}_{0.04}$}	 &6.3&1.0	&3\\
       300 \GeV	&0.01&0.01	 &1.23&0.26	 &\multicolumn{1}{r@{\ $^+_-$\ }}{0}&{$^{0.69}_{0}$}	 &1.2&0.7	&0\\
\hline
\multicolumn{10}{c}{$2 e/\mu + \geq 1 \tau$, on-$Z$} \\
\hline
 Inclusive	&0.58&0.23	 &112&10	 &9600&2600	 &9800&2600	&9149\\
       100 \GeV	&0.08&0.04	 &6.8&1.0	 &5.7&2.0	 &12.6&2.2	&7\\
       200 \GeV	&\multicolumn{1}{r@{\ $^+_-$\ }}{0}&{$^{0.012}_{0}$}	 &0.72&0.18	 &\multicolumn{1}{r@{\ $^+_-$\ }}{0.4}&{$^{0.8}_{0.4}$}	 &1.1&0.8	&0\\
       300 \GeV	&\multicolumn{1}{r@{\ $^+_-$\ }}{0}&{$^{0.003}_{0}$}	 &0.20&0.10	 &\multicolumn{1}{r@{\ $^+_-$\ }}{0}&{$^{0.7}_{0}$}	 &\multicolumn{1}{r@{\ $^+_-$\ }}{0.20}&{$^{0.71}_{0.20}$}	&0\\
\hline
\end{tabular}
   \caption{Expected and observed event yields for the low-\Ht, $\met$~signal regions.}
   \label{t:Result_Weak_MET}
\end{center}
\end{table*}

\clearpage

\label{app:limits}

Expected limits with confidence intervals of one and two standard deviations, observed limits, and one-sided $p$-values with corresponding significance in units of $\sigma$ are provided in tables~\ref{tbl:limits_htlep}--\ref{tbl:limits_metwk}.

\renewcommand{\arraystretch}{1.3}
\begin{table}[tbp]
\centering
\begin{tabular}{l r c l l c c c}
\hline\hline
\multicolumn{2}{c}{\htlep}    &Expected &$\pm 1\sigma$ &$\pm 2\sigma$    & Observed   & $p_0$  & Significance \\
\multicolumn{2}{c}{[GeV]}  &[fb]  &\multicolumn{1}{c}{[fb]}  &\multicolumn{1}{c}{[fb]}                              &[fb]  &  &[$\sigma$] \\
\hline
\multicolumn{8}{c}{$\geq 3 e/\mu$, no-OSSF} \\
\hline
$\geq $ &200  &0.34 &$^{+0.13}_{-0.08}$ &$^{+0.30}_{-0.15}$  &0.34  & 0.46  &0.1 \\ 
$\geq $ &500  &0.22 &$^{+0.07}_{-0.04}$ &$^{+0.12}_{-0.06}$  &0.22  & 0.47  &0.1 \\ 
$\geq $ &800  &0.14 &$^{+0.01}_{-0.00}$ &$^{+0.08}_{-0.02}$  &0.14  & 0.50  &0.0 \\ 
\hline
\multicolumn{8}{c}{$2 e/\mu + \geq 1 \tau_{\mathrm{had}}$, no-OSSF} \\
\hline
$\geq $ &200  &0.60 &$^{+0.22}_{-0.17}$ &$^{+0.40}_{-0.28}$  &0.41  & 0.50  &0.0 \\ 
$\geq $ &500  &0.15 &$^{+0.01}_{-0.01}$ &$^{+0.08}_{-0.01}$  &0.14  & 0.50  &0.0 \\ 
$\geq $ &800  &0.15 &$^{+0.00}_{-0.01}$ &$^{+0.06}_{-0.02}$  &0.14  & 0.50  &0.0 \\ 
\hline
\multicolumn{8}{c}{$\geq 3 e/\mu$, OSSF} \\
\hline
$\geq $ &200  &1.2 &$^{+0.5}_{-0.3}$ &$^{+1.0}_{-0.5}$  &0.70  & 0.50  &0.0 \\ 
$\geq $ &500  &0.26 &$^{+0.03}_{-0.06}$ &$^{+0.10}_{-0.10}$  &0.18  & 0.50  &0.0 \\ 
$\geq $ &800  &0.16 &$^{+0.05}_{-0.01}$ &$^{+0.13}_{-0.02}$  &0.15  & 0.50  &0.0 \\ 
\hline
\multicolumn{8}{c}{$2 e/\mu + \geq 1 \tau_{\mathrm{had}}$, OSSF} \\
\hline
$\geq $ &200  &1.8 &$^{+0.6}_{-0.4}$ &$^{+1.2}_{-0.7}$  &1.68  & 0.50  &0.0 \\ 
$\geq $ &500  &0.16 &$^{+0.06}_{-0.02}$ &$^{+0.14}_{-0.03}$  &0.14  & 0.50  &0.0 \\ 
$\geq $ &800  &0.16 &$^{+0.00}_{-0.02}$ &$^{+0.05}_{-0.03}$  &0.14  & 0.50  &0.0 \\ 
\hline
\multicolumn{8}{c}{$\geq 3 e/\mu$, on-$Z$} \\
\hline
$\geq $ &200  &4.9 &$^{+1.7}_{-1.3}$ &$^{+3.6}_{-2.1}$  &3.58  & 0.50  &0.0 \\ 
$\geq $ &500  &0.47 &$^{+0.21}_{-0.14}$ &$^{+0.41}_{-0.23}$  &0.47  & 0.50  &0.0 \\ 
$\geq $ &800  &0.27 &$^{+0.04}_{-0.05}$ &$^{+0.09}_{-0.08}$  &0.30  & 0.15  &1.1 \\ 
\hline
\multicolumn{8}{c}{$2 e/\mu + \geq 1 \tau_{\mathrm{had}}$, on-$Z$} \\
\hline
$\geq $ &200  &3.7 &$^{+1.0}_{-0.8}$ &$^{+2.1}_{-1.5}$  &3.14  & 0.50  &0.0 \\ 
$\geq $ &500  &0.24 &$^{+0.06}_{-0.04}$ &$^{+0.08}_{-0.06}$  &0.29  & 0.30  &0.5 \\ 
$\geq $ &800  &0.15 &$^{+0.01}_{-0.00}$ &$^{+0.07}_{-0.02}$  &0.16  & 0.50  &0.0 \\ 
\hline\hline
\end{tabular}
\caption{Expected and observed limits, and corresponding $p$-values and significances (in standard deviations), for
signal regions based on cuts on \htlep.
}
\label{tbl:limits_htlep}
\end{table}
\renewcommand{\arraystretch}{1.0}

\renewcommand{\arraystretch}{1.3}
\begin{table}[tbp]
\centering
\begin{tabular}{l r c l l c c c}
\hline\hline
\multicolumn{2}{c}{\minleppt}    &Expected &$\pm 1\sigma$ &$\pm 2\sigma$    & Observed   & $p_0$  & Significance \\
\multicolumn{2}{c}{[GeV]}  &[fb]  &\multicolumn{1}{c}{[fb]}  &\multicolumn{1}{c}{[fb]}                              &[fb]        &        &[$\sigma$]\\
\hline
\multicolumn{8}{c}{$\geq 3 e/\mu$, no-OSSF} \\
\hline
$\geq $ &50  &0.25 &$^{+0.04}_{-0.04}$ &$^{+0.09}_{-0.07}$  &0.29  & 0.10  &1.3 \\ 
$\geq $ &100  &0.13 &$^{+0.02}_{-0.00}$ &$^{+0.09}_{-0.00}$  &0.15  & 0.50  &0.0 \\ 
$\geq $ &150  &0.14 &$^{+0.01}_{-0.01}$ &$^{+0.07}_{-0.03}$  &0.14  & 0.50  &0.0 \\ 
\hline
\multicolumn{8}{c}{$2 e/\mu + \geq 1 \tau_{\mathrm{had}}$, no-OSSF} \\
\hline
$\geq $ &50  &0.34 &$^{+0.13}_{-0.08}$ &$^{+0.31}_{-0.13}$  &0.31  & 0.50  &0.0 \\ 
$\geq $ &100  &0.15 &$^{+0.00}_{-0.00}$ &$^{+0.07}_{-0.02}$  &0.14  & 0.50  &0.0 \\ 
$\geq $ &150  &0.15 &$^{+0.00}_{-0.01}$ &$^{+0.07}_{-0.02}$  &0.14  & 0.50  &0.0 \\ 
\hline
\multicolumn{8}{c}{$\geq 3 e/\mu$, OSSF} \\
\hline
$\geq $ &50  &0.70 &$^{+0.30}_{-0.20}$ &$^{+0.68}_{-0.33}$  &0.50  & 0.50  &0.0 \\ 
$\geq $ &100  &0.21 &$^{+0.08}_{-0.04}$ &$^{+0.13}_{-0.07}$  &0.20  & 0.49  &0.0 \\ 
$\geq $ &150  &0.14 &$^{+0.04}_{-0.02}$ &$^{+0.12}_{-0.03}$  &0.13  & 0.50  &0.0 \\ 
\hline
\multicolumn{8}{c}{$2 e/\mu + \geq 1 \tau_{\mathrm{had}}$, OSSF} \\
\hline
$\geq $ &50  &0.71 &$^{+0.20}_{-0.20}$ &$^{+0.43}_{-0.34}$  &0.86  & 0.21  &0.8 \\ 
$\geq $ &100  &0.15 &$^{+0.01}_{-0.01}$ &$^{+0.08}_{-0.02}$  &0.16  & 0.50  &0.0 \\ 
$\geq $ &150  &0.16 &$^{+0.00}_{-0.02}$ &$^{+0.05}_{-0.03}$  &0.16  & 0.50  &0.0 \\ 
\hline
\multicolumn{8}{c}{$\geq 3 e/\mu$, on-$Z$} \\
\hline
$\geq $ &50  &2.3 &$^{+0.8}_{-0.6}$ &$^{+1.8}_{-1.0}$  &1.77  & 0.50  &0.0 \\ 
$\geq $ &100  &0.44 &$^{+0.20}_{-0.13}$ &$^{+0.42}_{-0.22}$  &0.69  & 0.09  &1.4 \\ 
$\geq $ &150  &0.31 &$^{+0.06}_{-0.01}$ &$^{+0.28}_{-0.03}$  &0.30  & 0.48  &0.1 \\ 
\hline
\multicolumn{8}{c}{$2 e/\mu + \geq 1 \tau_{\mathrm{had}}$, on-$Z$} \\
\hline
$\geq $ &50  &1.3 &$^{+0.4}_{-0.3}$ &$^{+1.0}_{-0.6}$  &1.05  & 0.50  &0.0 \\ 
$\geq $ &100  &0.16 &$^{+0.07}_{-0.01}$ &$^{+0.13}_{-0.02}$  &0.16  & 0.50  &0.0 \\ 
$\geq $ &150  &0.15 &$^{+0.00}_{-0.01}$ &$^{+0.08}_{-0.02}$  &0.15  & 0.50  &0.0 \\ 
\hline\hline
\end{tabular}
\caption{Expected and observed limits, and corresponding $p$-values and significances (in standard deviations), for
signal regions based on cuts on \minleppt.
}
\label{tbl:limits_minpt}
\end{table}
\renewcommand{\arraystretch}{1.0}

\renewcommand{\arraystretch}{1.3}
\begin{table}[tbp]
\centering
\begin{tabular}{l r c l l c c c}
\hline\hline
\multicolumn{2}{c}{$b$ tags}    &Expected &$\pm 1\sigma$ &$\pm 2\sigma$    & Observed   & $p_0$  & Significance \\
\multicolumn{2}{c}{}  &[fb]  &\multicolumn{1}{c}{[fb]}  &\multicolumn{1}{c}{[fb]}                              &[fb]          &        & [$\sigma$]\\
\hline
\multicolumn{8}{c}{$\geq 3 e/\mu$, no-OSSF} \\
\hline
$\geq $ &1  &0.59 &$^{+0.23}_{-0.17}$ &$^{+0.41}_{-0.28}$  &0.64  & 0.35  &0.4 \\ 
$\geq $ &2  &0.30 &$^{+0.09}_{-0.10}$ &$^{+0.21}_{-0.15}$  &0.32  & 0.40  &0.3 \\ 
\hline
\multicolumn{8}{c}{$2 e/\mu + \geq 1 \tau_{\mathrm{had}}$, no-OSSF} \\
\hline
$\geq $ &1  &2.4 &$^{+0.7}_{-0.6}$ &$^{+1.6}_{-1.0}$  &2.44  & 0.45  &0.1 \\ 
$\geq $ &2  &0.96 &$^{+0.35}_{-0.25}$ &$^{+0.72}_{-0.43}$  &1.07  & 0.35  &0.4 \\ 
\hline
\multicolumn{8}{c}{$\geq 3 e/\mu$, OSSF} \\
\hline
$\geq $ &1  &0.92 &$^{+0.35}_{-0.25}$ &$^{+0.79}_{-0.41}$  &0.86  & 0.50  &0.0 \\ 
$\geq $ &2  &0.45 &$^{+0.18}_{-0.10}$ &$^{+0.28}_{-0.16}$  &0.47  & 0.45  &0.1 \\ 
\hline
\multicolumn{8}{c}{$2 e/\mu + \geq 1 \tau_{\mathrm{had}}$, OSSF} \\
\hline
$\geq $ &1  &1.8 &$^{+0.6}_{-0.4}$ &$^{+1.2}_{-0.7}$  &1.57  & 0.50  &0.0 \\ 
$\geq $ &2  &0.55 &$^{+0.22}_{-0.16}$ &$^{+0.39}_{-0.26}$  &0.43  & 0.50  &0.0 \\ 
\hline
\multicolumn{8}{c}{$\geq 3 e/\mu$, on-$Z$} \\
\hline
$\geq $ &1  &3.9 &$^{+1.3}_{-1.0}$ &$^{+2.7}_{-1.7}$  &3.91  & 0.49  &0.0 \\ 
$\geq $ &2  &0.89 &$^{+0.33}_{-0.24}$ &$^{+0.73}_{-0.39}$  &0.70  & 0.50  &0.0 \\ 
\hline
\multicolumn{8}{c}{$2 e/\mu + \geq 1 \tau_{\mathrm{had}}$, on-$Z$} \\
\hline
$\geq $ &1  &9.4 &$^{+2.5}_{-2.1}$ &$^{+5.1}_{-3.7}$  &9.38  & 0.50  &0.0 \\ 
$\geq $ &2  &0.79 &$^{+0.30}_{-0.21}$ &$^{+0.68}_{-0.35}$  &0.66  & 0.50  &0.0 \\ 
\hline\hline
\end{tabular}
\caption{Expected and observed limits, and corresponding $p$-values and significances (in standard deviations), for
signal regions based on cuts on the number of $b$-tagged jets.
}
\label{tbl:limits_nbtag}
\end{table}
\renewcommand{\arraystretch}{1.0}

\renewcommand{\arraystretch}{1.3}
\begin{table}[tbp]
\centering
\begin{tabular}{l r c l l c c c}
\hline\hline
\multicolumn{2}{c}{\meff}    &Expected &$\pm 1\sigma$ &$\pm 2\sigma$    & Observed   & $p_0$  & Significance \\
\multicolumn{2}{c}{[GeV]}  &[fb]  &\multicolumn{1}{c}{[fb]}  &\multicolumn{1}{c}{[fb]}                              &[fb]    &        &[$\sigma$]\\
\hline
\multicolumn{8}{c}{$\geq 3 e/\mu$, no-OSSF} \\
\hline
$\geq $ &600  &0.27 &$^{+0.04}_{-0.06}$ &$^{+0.09}_{-0.10}$  &0.25  & 0.50  &0.0 \\ 
$\geq $ &1000  &0.18 &$^{+0.06}_{-0.02}$ &$^{+0.12}_{-0.04}$  &0.19  & 0.37  &0.3 \\ 
$\geq $ &1500  &0.15 &$^{+0.01}_{-0.01}$ &$^{+0.07}_{-0.01}$  &0.15  & 0.50  &0.0 \\ 
\hline
\multicolumn{8}{c}{$2 e/\mu + \geq 1 \tau_{\mathrm{had}}$, no-OSSF} \\
\hline
$\geq $ &600  &0.68 &$^{+0.20}_{-0.19}$ &$^{+0.41}_{-0.32}$  &0.77  & 0.34  &0.4 \\ 
$\geq $ &1000  &0.24 &$^{+0.05}_{-0.06}$ &$^{+0.10}_{-0.09}$  &0.22  & 0.50  &0.0 \\ 
$\geq $ &1500  &0.18 &$^{+0.06}_{-0.01}$ &$^{+0.11}_{-0.02}$  &0.20  & 0.26  &0.6 \\ 
\hline
\multicolumn{8}{c}{$\geq 3 e/\mu$, OSSF} \\
\hline
$\geq $ &600  &0.65 &$^{+0.21}_{-0.18}$ &$^{+0.41}_{-0.30}$  &0.49  & 0.50  &0.0 \\ 
$\geq $ &1000  &0.25 &$^{+0.11}_{-0.05}$ &$^{+0.25}_{-0.09}$  &0.16  & 0.50  &0.0 \\ 
$\geq $ &1500  &0.15 &$^{+0.05}_{-0.01}$ &$^{+0.12}_{-0.01}$  &0.14  & 0.50  &0.0 \\ 
\hline
\multicolumn{8}{c}{$2 e/\mu + \geq 1 \tau_{\mathrm{had}}$, OSSF} \\
\hline
$\geq $ &600  &0.66 &$^{+0.21}_{-0.18}$ &$^{+0.40}_{-0.30}$  &0.69  & 0.44  &0.1 \\ 
$\geq $ &1000  &0.23 &$^{+0.06}_{-0.04}$ &$^{+0.11}_{-0.06}$  &0.22  & 0.50  &0.0 \\ 
$\geq $ &1500  &0.15 &$^{+0.00}_{-0.00}$ &$^{+0.05}_{-0.02}$  &0.15  & 0.50  &0.0 \\ 
\hline
\multicolumn{8}{c}{$\geq 3 e/\mu$, on-$Z$} \\
\hline
$\geq $ &600  &3.2 &$^{+1.0}_{-0.8}$ &$^{+2.1}_{-1.3}$  &2.93  & 0.50  &0.0 \\ 
$\geq $ &1000  &0.90 &$^{+0.20}_{-0.17}$ &$^{+0.75}_{-0.33}$  &0.91  & 0.48  &0.0 \\ 
$\geq $ &1500  &0.26 &$^{+0.04}_{-0.07}$ &$^{+0.09}_{-0.10}$  &0.22  & 0.50  &0.0 \\ 
\hline
\multicolumn{8}{c}{$2 e/\mu + \geq 1 \tau_{\mathrm{had}}$, on-$Z$} \\
\hline
$\geq $ &600  &1.7 &$^{+0.6}_{-0.4}$ &$^{+1.2}_{-0.7}$  &1.49  & 0.50  &0.0 \\ 
$\geq $ &1000  &0.38 &$^{+0.14}_{-0.09}$ &$^{+0.32}_{-0.15}$  &0.64  & 0.05  &1.7 \\ 
$\geq $ &1500  &0.18 &$^{+0.06}_{-0.01}$ &$^{+0.12}_{-0.02}$  &0.20  & 0.29  &0.5 \\ 
\hline\hline
\end{tabular}
\caption{Expected and observed limits, and corresponding $p$-values and significances (in standard deviations), for
signal regions based on cuts on \meff.
}
\label{tbl:limits_stall}
\end{table}
\renewcommand{\arraystretch}{1.0}

\renewcommand{\arraystretch}{1.3}
\begin{table}[tbp]
\centering
\begin{tabular}{l r c l l c c c}
\hline\hline
\multicolumn{2}{c}{\meff}    &Expected &$\pm 1\sigma$ &$\pm 2\sigma$    & Observed   & $p_0$  & Significance \\
\multicolumn{2}{c}{[GeV]}  &[fb]  &\multicolumn{1}{c}{[fb]}  &\multicolumn{1}{c}{[fb]}                              &[fb]    &        & [$\sigma$]\\
\hline
\multicolumn{8}{c}{$\geq 3 e/\mu$, no-OSSF} \\
\hline
$\geq $ &0  &0.37 &$^{+0.14}_{-0.09}$ &$^{+0.34}_{-0.16}$  &0.41  & 0.34  &0.4 \\ 
$\geq $ &600  &0.22 &$^{+0.07}_{-0.04}$ &$^{+0.13}_{-0.06}$  &0.24  & 0.44  &0.2 \\ 
$\geq $ &1200  &0.15 &$^{+0.02}_{-0.01}$ &$^{+0.07}_{-0.02}$  &0.15  & 0.50  &0.0 \\ 
\hline
\multicolumn{8}{c}{$2 e/\mu + \geq 1 \tau_{\mathrm{had}}$, no-OSSF} \\
\hline
$\geq $ &0  &0.87 &$^{+0.33}_{-0.23}$ &$^{+0.73}_{-0.39}$  &0.77  & 0.50  &0.0 \\ 
$\geq $ &600  &0.43 &$^{+0.17}_{-0.12}$ &$^{+0.31}_{-0.17}$  &0.42  & 0.50  &0.0 \\ 
$\geq $ &1200  &0.21 &$^{+0.07}_{-0.02}$ &$^{+0.12}_{-0.02}$  &0.27  & 0.17  &1.0 \\ 
\hline
\multicolumn{8}{c}{$\geq 3 e/\mu$, OSSF} \\
\hline
$\geq $ &0  &0.57 &$^{+0.23}_{-0.17}$ &$^{+0.41}_{-0.27}$  &0.48  & 0.50  &0.0 \\ 
$\geq $ &600  &0.39 &$^{+0.16}_{-0.10}$ &$^{+0.34}_{-0.15}$  &0.35  & 0.50  &0.0 \\ 
$\geq $ &1200  &0.18 &$^{+0.06}_{-0.02}$ &$^{+0.12}_{-0.03}$  &0.19  & 0.40  &0.2 \\ 
\hline
\multicolumn{8}{c}{$2 e/\mu + \geq 1 \tau_{\mathrm{had}}$, OSSF} \\
\hline
$\geq $ &0  &0.60 &$^{+0.22}_{-0.17}$ &$^{+0.40}_{-0.28}$  &0.62  & 0.45  &0.1 \\ 
$\geq $ &600  &0.34 &$^{+0.12}_{-0.09}$ &$^{+0.30}_{-0.12}$  &0.32  & 0.50  &0.0 \\ 
$\geq $ &1200  &0.25 &$^{+0.05}_{-0.04}$ &$^{+0.10}_{-0.04}$  &0.28  & 0.16  &1.0 \\ 
\hline
\multicolumn{8}{c}{$\geq 3 e/\mu$, on-$Z$} \\
\hline
$\geq $ &0  &2.0 &$^{+0.7}_{-0.5}$ &$^{+1.6}_{-0.9}$  &1.21  & 0.50  &0.0 \\ 
$\geq $ &600  &1.1 &$^{+0.4}_{-0.3}$ &$^{+0.9}_{-0.5}$  &0.84  & 0.50  &0.0 \\ 
$\geq $ &1200  &0.29 &$^{+0.01}_{-0.06}$ &$^{+0.07}_{-0.11}$  &0.27  & 0.50  &0.0 \\ 
\hline
\multicolumn{8}{c}{$2 e/\mu + \geq 1 \tau_{\mathrm{had}}$, on-$Z$} \\
\hline
$\geq $ &0  &0.69 &$^{+0.21}_{-0.20}$ &$^{+0.43}_{-0.32}$  &0.60  & 0.50  &0.0 \\ 
$\geq $ &600  &0.39 &$^{+0.17}_{-0.11}$ &$^{+0.33}_{-0.16}$  &0.42  & 0.42  &0.2 \\ 
$\geq $ &1200  &0.16 &$^{+0.06}_{-0.02}$ &$^{+0.13}_{-0.03}$  &0.14  & 0.50  &0.0 \\ 
\hline\hline
\end{tabular}
\caption{Expected and observed limits, and corresponding $p$-values and significances (in standard deviations), for
signal regions based on cuts on \meff.
All signal regions above have an additional requirement of $\met\geq 100 \GeV$.
}
\label{tbl:limits_stmet}
\end{table}
\renewcommand{\arraystretch}{1.0}

\renewcommand{\arraystretch}{1.3}
\begin{table}[tbp]
\centering
\begin{tabular}{l r c l l c c c}
\hline\hline
\multicolumn{2}{c}{\meff}    &Expected &$\pm 1\sigma$ &$\pm 2\sigma$    & Observed   & $p_0$  & Significance \\
\multicolumn{2}{c}{[GeV]}  &[fb]  &\multicolumn{1}{c}{[fb]}  &\multicolumn{1}{c}{[fb]}                              &[fb]    &        & [$\sigma$]\\
\hline
\multicolumn{8}{c}{$\geq 3 e/\mu$, on-$Z$} \\
\hline
$\geq $ &0  &2.5 &$^{+0.9}_{-0.7}$ &$^{+1.9}_{-1.1}$  &1.83  & 0.50  &0.0 \\ 
$\geq $ &600  &0.84 &$^{+0.18}_{-0.20}$ &$^{+0.57}_{-0.36}$  &0.86  & 0.46  &0.1 \\ 
$\geq $ &1200  &0.26 &$^{+0.04}_{-0.06}$ &$^{+0.09}_{-0.07}$  &0.22  & 0.50  &0.0 \\ 
\hline
\multicolumn{8}{c}{$2 e/\mu + \geq 1 \tau_{\mathrm{had}}$, on-$Z$} \\
\hline
$\geq $ &0  &1.45 &$^{+0.45}_{-0.35}$ &$^{+0.99}_{-0.59}$  &1.31  & 0.50  &0.0 \\ 
$\geq $ &600  &0.39 &$^{+0.14}_{-0.09}$ &$^{+0.33}_{-0.15}$  &0.48  & 0.25  &0.7 \\ 
$\geq $ &1200  &0.15 &$^{+0.01}_{-0.00}$ &$^{+0.08}_{-0.02}$  &0.15  & 0.50  &0.0 \\ 
\hline\hline
\end{tabular}
\caption{Expected and observed limits, and corresponding $p$-values and significances (in standard deviations), for
signal regions based on cuts on \meff.
All signal regions above have an additional requirement of $\mtw\geq 100 \GeV$.
}
\label{tbl:limits_stmtw}
\end{table}
\renewcommand{\arraystretch}{1.0}

\renewcommand{\arraystretch}{1.3}
\begin{table}[tbp]
\centering
\begin{tabular}{l r c l l c c c}
\hline\hline
\multicolumn{2}{c}{\met}    &Expected &$\pm 1\sigma$ &$\pm 2\sigma$    & Observed   & $p_0$  & Significance \\
\multicolumn{2}{c}{[GeV]}  &[fb]  &\multicolumn{1}{c}{[fb]}  &\multicolumn{1}{c}{[fb]}                              &[fb]  &        &[$\sigma$]\\
\hline
\multicolumn{8}{c}{$\geq 3 e/\mu$, no-OSSF} \\
\hline
$\geq $ &0  &0.50 &$^{+0.22}_{-0.15}$ &$^{+0.42}_{-0.24}$  &0.77  & 0.09  &1.3 \\ 
$\geq $ &100  &0.28 &$^{+0.01}_{-0.04}$ &$^{+0.05}_{-0.07}$  &0.29  & 0.08  &1.4 \\ 
$\geq $ &200  &0.15 &$^{+0.04}_{-0.01}$ &$^{+0.10}_{-0.02}$  &0.15  & 0.50  &0.0 \\ 
$\geq $ &300  &0.15 &$^{+0.02}_{-0.01}$ &$^{+0.09}_{-0.01}$  &0.14  & 0.50  &0.0 \\ 
\hline
\multicolumn{8}{c}{$2 e/\mu + \geq 1 \tau_{\mathrm{had}}$, no-OSSF} \\
\hline
$\geq $ &0  &2.0 &$^{+0.6}_{-0.5}$ &$^{+1.4}_{-0.8}$  &2.00  & 0.40  &0.3 \\ 
$\geq $ &100  &0.73 &$^{+0.19}_{-0.20}$ &$^{+0.43}_{-0.34}$  &0.81  & 0.36  &0.4 \\ 
$\geq $ &200  &0.22 &$^{+0.07}_{-0.05}$ &$^{+0.12}_{-0.06}$  &0.23  & 0.48  &0.0 \\ 
$\geq $ &300  &0.15 &$^{+0.00}_{-0.01}$ &$^{+0.08}_{-0.02}$  &0.14  & 0.50  &0.0 \\ 
\hline
\multicolumn{8}{c}{$\geq 3 e/\mu$, OSSF} \\
\hline
$\geq $ &0  &1.2 &$^{+0.4}_{-0.3}$ &$^{+1.0}_{-0.5}$  &1.11  & 0.50  &0.0 \\ 
$\geq $ &100  &0.42 &$^{+0.16}_{-0.11}$ &$^{+0.30}_{-0.15}$  &0.35  & 0.50  &0.0 \\ 
$\geq $ &200  &0.21 &$^{+0.08}_{-0.04}$ &$^{+0.13}_{-0.05}$  &0.24  & 0.37  &0.3 \\ 
$\geq $ &300  &0.16 &$^{+0.04}_{-0.01}$ &$^{+0.12}_{-0.01}$  &0.15  & 0.50  &0.0 \\ 
\hline
\multicolumn{8}{c}{$2 e/\mu + \geq 1 \tau_{\mathrm{had}}$, OSSF} \\
\hline
$\geq $ &0  &2.2 &$^{+0.7}_{-0.5}$ &$^{+1.4}_{-0.9}$  &1.88  & 0.50  &0.0 \\ 
$\geq $ &100  &0.46 &$^{+0.16}_{-0.12}$ &$^{+0.28}_{-0.19}$  &0.41  & 0.50  &0.0 \\ 
$\geq $ &200  &0.19 &$^{+0.05}_{-0.01}$ &$^{+0.11}_{-0.03}$  &0.19  & 0.37  &0.3 \\ 
$\geq $ &300  &0.14 &$^{+0.01}_{-0.00}$ &$^{+0.06}_{-0.01}$  &0.13  & 0.50  &0.0 \\ 
\hline
\multicolumn{8}{c}{$\geq 3 e/\mu$, on-$Z$} \\
\hline
$\geq $ &0  &7.2 &$^{+2.2}_{-1.8}$ &$^{+4.7}_{-3.0}$  &6.38  & 0.50  &0.0 \\ 
$\geq $ &100  &1.3 &$^{+0.5}_{-0.4}$ &$^{+1.1}_{-0.6}$  &0.96  & 0.50  &0.0 \\ 
$\geq $ &200  &0.51 &$^{+0.22}_{-0.15}$ &$^{+0.40}_{-0.24}$  &0.55  & 0.41  &0.2 \\ 
$\geq $ &300  &0.23 &$^{+0.07}_{-0.06}$ &$^{+0.12}_{-0.10}$  &0.18  & 0.50  &0.0 \\ 
\hline
\multicolumn{8}{c}{$2 e/\mu + \geq 1 \tau_{\mathrm{had}}$, on-$Z$} \\
\hline
$\geq $ &0  &12.4 &$^{+3.3}_{-2.8}$ &$^{+6.6}_{-4.9}$  &10.66  & 0.50  &0.0 \\ 
$\geq $ &100  &0.53 &$^{+0.23}_{-0.16}$ &$^{+0.41}_{-0.26}$  &0.64  & 0.30  &0.5 \\ 
$\geq $ &200  &0.24 &$^{+0.05}_{-0.05}$ &$^{+0.10}_{-0.07}$  &0.29  & 0.25  &0.7 \\ 
$\geq $ &300  &0.22 &$^{+0.08}_{-0.04}$ &$^{+0.12}_{-0.06}$  &0.23  & 0.48  &0.0 \\ 
\hline\hline
\end{tabular}
\caption{Expected and observed limits, and corresponding $p$-values and significances (in standard deviations), for
signal regions based on cuts on \met.
All signal regions above have an additional requirement of $\htjets\geq 150 \GeV$.
}
\label{tbl:limits_metst}
\end{table}
\renewcommand{\arraystretch}{1.0}

\renewcommand{\arraystretch}{1.3}
\begin{table}[tbp]
\centering
\begin{tabular}{l r c l l c c c}
\hline\hline
\multicolumn{2}{c}{\met}    &Expected &$\pm 1\sigma$ &$\pm 2\sigma$    & Observed   & $p_0$  & Significance \\
\multicolumn{2}{c}{[GeV]}  &[fb]  &\multicolumn{1}{c}{[fb]}  &\multicolumn{1}{c}{[fb]}                              &[fb]   &       & [$\sigma$]\\
\hline
\multicolumn{8}{c}{$\geq 3 e/\mu$, no-OSSF} \\
\hline
$\geq $ &0  &0.58 &$^{+0.23}_{-0.17}$ &$^{+0.41}_{-0.27}$  &0.56  & 0.50  &0.0 \\ 
$\geq $ &100  &0.21 &$^{+0.07}_{-0.06}$ &$^{+0.13}_{-0.08}$  &0.15  & 0.50  &0.0 \\ 
$\geq $ &200  &0.16 &$^{+0.01}_{-0.01}$ &$^{+0.08}_{-0.02}$  &0.15  & 0.50  &0.0 \\ 
$\geq $ &300  &0.14 &$^{+0.01}_{-0.01}$ &$^{+0.01}_{-0.01}$  &0.14  & 0.50  &0.0 \\ 
\hline
\multicolumn{8}{c}{$2 e/\mu + \geq 1 \tau_{\mathrm{had}}$, no-OSSF} \\
\hline
$\geq $ &0  &2.5 &$^{+0.8}_{-0.6}$ &$^{+1.7}_{-1.1}$  &2.48  & 0.50  &0.0 \\ 
$\geq $ &100  &0.38 &$^{+0.15}_{-0.11}$ &$^{+0.33}_{-0.17}$  &0.22  & 0.50  &0.0 \\ 
$\geq $ &200  &0.14 &$^{+0.00}_{-0.00}$ &$^{+0.06}_{-0.01}$  &0.14  & 0.50  &0.0 \\ 
$\geq $ &300  &0.15 &$^{+0.00}_{-0.01}$ &$^{+0.05}_{-0.02}$  &0.14  & 0.50  &0.0 \\ 
\hline
\multicolumn{8}{c}{$\geq 3 e/\mu$, OSSF} \\
\hline
$\geq $ &0  &2.2 &$^{+0.8}_{-0.6}$ &$^{+1.8}_{-1.0}$  &1.80  & 0.50  &0.0 \\ 
$\geq $ &100  &0.39 &$^{+0.16}_{-0.09}$ &$^{+0.34}_{-0.17}$  &0.34  & 0.50  &0.0 \\ 
$\geq $ &200  &0.18 &$^{+0.06}_{-0.02}$ &$^{+0.12}_{-0.04}$  &0.19  & 0.45  &0.1 \\ 
$\geq $ &300  &0.14 &$^{+0.01}_{-0.00}$ &$^{+0.08}_{-0.01}$  &0.15  & 0.50  &0.0 \\ 
\hline
\multicolumn{8}{c}{$2 e/\mu + \geq 1 \tau_{\mathrm{had}}$, OSSF} \\
\hline
$\geq $ &0  &12.4 &$^{+3.3}_{-2.8}$ &$^{+6.6}_{-4.9}$  &12.32  & 0.50  &0.0 \\ 
$\geq $ &100  &0.36 &$^{+0.15}_{-0.08}$ &$^{+0.32}_{-0.13}$  &0.43  & 0.18  &0.9 \\ 
$\geq $ &200  &0.15 &$^{+0.01}_{-0.01}$ &$^{+0.08}_{-0.02}$  &0.14  & 0.50  &0.0 \\ 
$\geq $ &300  &0.13 &$^{+0.01}_{-0.01}$ &$^{+0.05}_{-0.03}$  &0.13  & 0.50  &0.0 \\ 
\hline
\multicolumn{8}{c}{$\geq 3 e/\mu$, on-$Z$} \\
\hline
$\geq $ &0  &26 &$^{+9}_{-7}$ &$^{+19}_{-11}$  &24.77  & 0.50  &0.0 \\ 
$\geq $ &100  &1.2 &$^{+0.5}_{-0.3}$ &$^{+1.1}_{-0.6}$  &0.69  & 0.50  &0.0 \\ 
$\geq $ &200  &0.31 &$^{+0.16}_{-0.10}$ &$^{+0.38}_{-0.16}$  &0.20  & 0.50  &0.0 \\ 
$\geq $ &300  &0.19 &$^{+0.07}_{-0.05}$ &$^{+0.11}_{-0.05}$  &0.14  & 0.50  &0.0 \\ 
\hline
\multicolumn{8}{c}{$2 e/\mu + \geq 1 \tau_{\mathrm{had}}$, on-$Z$} \\
\hline
$\geq $ &0  &205 &$^{+50}_{-45}$ &$^{+102}_{-79}$  &194.36  & 0.50  &0.0 \\ 
$\geq $ &100  &0.40 &$^{+0.17}_{-0.11}$ &$^{+0.33}_{-0.17}$  &0.29  & 0.50  &0.0 \\ 
$\geq $ &200  &0.17 &$^{+0.07}_{-0.02}$ &$^{+0.13}_{-0.04}$  &0.14  & 0.50  &0.0 \\ 
$\geq $ &300  &0.14 &$^{+0.03}_{-0.01}$ &$^{+0.10}_{-0.02}$  &0.13  & 0.50  &0.0 \\ 
\hline\hline
\end{tabular}
\caption{Expected and observed limits, and corresponding $p$-values and significances (in standard deviations), for
signal regions based on cuts on \met.
All signal regions above have an additional requirement of $\htjets < 150 \GeV$.
}
\label{tbl:limits_metwk}
\end{table}
\renewcommand{\arraystretch}{1.0}

\clearpage

\begin{flushleft}
{\Large The ATLAS Collaboration}

\bigskip

G.~Aad$^{\rm 85}$,
B.~Abbott$^{\rm 113}$,
J.~Abdallah$^{\rm 152}$,
S.~Abdel~Khalek$^{\rm 117}$,
O.~Abdinov$^{\rm 11}$,
R.~Aben$^{\rm 107}$,
B.~Abi$^{\rm 114}$,
M.~Abolins$^{\rm 90}$,
O.S.~AbouZeid$^{\rm 159}$,
H.~Abramowicz$^{\rm 154}$
% ...
\bigskip
\\
% ...
$^{11}$ Institute of Physics, Azerbaijan Academy of Sciences, Baku, Azerbaijan\\
% ...
$^{85}$ CPPM, Aix-Marseille Universit{\'e} and CNRS/IN2P3, Marseille, France\\
% ...
$^{90}$ Department of Physics and Astronomy, Michigan State University, East Lansing MI, United States of America\\
% ...
$^{107}$ Nikhef National Institute for Subatomic Physics and University of Amsterdam, Amsterdam, Netherlands\\
% ...
$^{113}$ Homer L. Dodge Department of Physics and Astronomy, University of Oklahoma, Norman OK, United States of America\\
$^{114}$ Department of Physics, Oklahoma State University, Stillwater OK, United States of America\\
% ...
$^{117}$ LAL, Universit{\'e} Paris-Sud and CNRS/IN2P3, Orsay, France\\
% ...
$^{152}$ Institute of Physics, Academia Sinica, Taipei, Taiwan\\
% ...
$^{154}$ Raymond and Beverly Sackler School of Physics and Astronomy, Tel Aviv University, Tel Aviv, Israel\\
% ...
$^{159}$ Department of Physics, University of Toronto, Toronto ON, Canada\\
% ...
\end{flushleft}

\end{document}